%% file: main.tex
\RequirePackage{fix-cm}
\documentclass{pckg/fac}
\usepackage{tcolorbox}
\tcbuselibrary{theorems}
\usepackage{amsmath} 
\usepackage[T1]{fontenc}
\usepackage[ngerman,american]{babel}
\usepackage{amssymb}
\usepackage{dashrule}
\usepackage{xargs}
\usepackage{amsfonts}
\usepackage{stmaryrd}
\usepackage{paralist}
\usepackage{booktabs}
\usepackage{titlecaps}
\usepackage{todonotes}
\usepackage{algorithmic}
\usepackage{algorithm}
\usepackage{ifthen}
\usepackage{multirow}
\usepackage{tikz}
\usepackage{mathtools}
\usepackage[ddmmyyyy]{datetime}
\usepackage[bottom]{footmisc}
\usetikzlibrary{shapes,decorations.pathreplacing,calc,patterns}
\usepackage{glossaries}
\usepackage[xindy]{imakeidx}
\usepackage{hyperref}
\makeglossaries
\newtcolorbox{mymathbox}[2][0mm]{leftupper=#1, ams align*, fontupper=\footnotesize, title = #2, colback=white, sharp corners, colframe=black}

\include{macros/headings}
\include{macros/macros}%
\include{macros/templates}%
\loadglsentries[main]{glossary}%
\makeglossaries
\loadglsentries{acronyms}
\title[On the Specification of Constraints for Dynamic Architectures]{On the Specification of Constraints for Dynamic Architectures}
\author[D. Marmsoler]
{Diego Marmsoler\\
	Technische Universit\"at M\"unchen, Germany}

\correspond{Diego Marmsoler, Boltzmannstr. 3, 85748 Garching bei M\"unchen, GER.
	e-mail: diego.marmsoler@tum.de}

\begin{document}
\Addlcwords{of and a}

\makecorrespond
\maketitle

\input{abstract}
\input{introduction}
\input{model}
\input{specification}
\input{interfaces}
\input{cnftrace}
\input{cnfdiag}
\input{verifying}
\input{discussion}
\input{bgrw}
\input{conclusion}
\renewcommand{\section}{\oldsec}%
\bibliographystyle{plain}
\bibliography{references}
%
\input{conventions}
\end{document}

%% file: macros/headings.tex
\let\oldsec\section
\renewcommand{\section}[1]{\oldsec{\texorpdfstring{\titlecap{#1}}{#1}}}
\let\oldssec\subsection
\renewcommand{\subsection}[1]{\oldssec{\texorpdfstring{\titlecap{#1}}{#1}}}
\let\oldsssec\subsubsection
\renewcommand{\subsubsection}[1]{\oldsssec{\texorpdfstring{\titlecap{#1}}{#1}}}

%% file: macros/macros.tex
\newcommand{\etal}{et al.}

\newcommand{\NN}{\mathbb{N}}
\newcommand{\NNP}{\mathbb{N}^+}

\newcommand{\pset}[1]{\wp{\left(#1\right)}}
\newcommand{\pFun}[2]{#1\dashrightarrow#2}
\newcommand{\domain}[1]{\mathrm{dom}\left(#1\right)}
\newcommand{\range}[1]{\mathrm{ran}\left(#1\right)}
\newcommand{\rest}[1]{\mid_{#1}}
\newcommand{\cp}[2]{\left[#1\right]^#2}
\newcommand{\defeq}{\overset{\text{def}}{\quad=\quad}}
\newcommand{\qqquad}{\qquad\quad}
\newcommand{\update}[3]{\ensuremath{#1[#2\mapsto#3]}}
\newcommand{\fupdate}[4]{\ensuremath{#1[#2\colon #3\mapsto#4]}}
\newcommand{\fmerge}[2]{\ensuremath{#1 \cup #2}}
\newcommand{\card}[1]{\ensuremath{|#1|}}
\newcommand{\cProd}[2]{\ensuremath{#1^{#2}}}
\newcommand{\inverse}[1]{\ensuremath{#1^{-1}}}

\newcommand{\mess}{\ensuremath{\mathsf{M}}}
\newcommand{\port}{\ensuremath{\mathsf{P}}}
\newcommand{\cID}{\ensuremath{\mathsf{C}_{id}}}
\newcommand{\iID}{\ensuremath{\mathsf{I}_{id}}}
\newcommand{\pID}{\ensuremath{\mathsf{P}_{id}}}
\newcommand{\val}[1]{\ensuremath{\overline{#1}}}

\newcommand{\comp}[1][]{\ensuremath{(\cmpId#1,\cmpLP#1,\cmpIP#1,\cmpOP#1,\cmpVal#1)}}
\newcommand{\cmpId}{\ensuremath{d}}
\newcommand{\cmpLP}{\ensuremath{L}}
\newcommand{\cmpIP}{\ensuremath{I}}
\newcommand{\cmpOP}{\ensuremath{O}}
\newcommand{\cmpVal}{\ensuremath{\mu}}
\newcommand{\cmpAll}{\ensuremath{\mathcal{C}}}
\newcommand{\cmpIn}[1]{\ensuremath{\mathsf{in}(#1)}}
\newcommand{\cmpOut}[1]{\ensuremath{\mathsf{out}(#1)}}
\newcommand{\cmpLoc}[1]{\ensuremath{\mathsf{loc}(#1)}}
\newcommand{\cmpPort}[1]{\ensuremath{\mathsf{port}(#1)}}
\newcommand{\cmpPin}[1]{\ensuremath{\mathsf{in}_{#1}}}
\newcommand{\cmpPout}[1]{\ensuremath{\mathsf{out}_{#1}}}
\newcommand{\cmpPloc}[1]{\ensuremath{\mathsf{loc}_{#1}}}
\newcommand{\cmpPport}[1]{\ensuremath{\mathsf{port}_{#1}}}
\newcommand{\cmpPlocVal}[1]{\ensuremath{\mathit{lV}_{#1}}}
\newcommand{\cmpName}[1]{\ensuremath{\mathsf{id}(#1)}}

\newcommand{\configuration}{\ensuremath{(\cnfComp,\cnfConn)}}
\newcommand{\cnfComp}{\ensuremath{C'}}
\newcommand{\cnfConn}{\ensuremath{N}}
\newcommand{\cnfVal}[1]{\ensuremath{\mathit{val}_{#1}}}
\newcommand{\oIP}[1]{\mathsf{in}_o(#1)}
\newcommand{\cnfAll}[1]{\ensuremath{\mathcal{K}(#1)}}

\newcommand{\allTraces}[1]{\ensuremath{\mathcal{R}(#1)}}

\newcommand{\signature}{\ensuremath{(\sigSort,\sigFun,\sigPred)}}
\newcommand{\sigSort}{\ensuremath{S}}
\newcommand{\sigFun}{\ensuremath{F}}
\newcommand{\sigPred}{\ensuremath{B}}
\newcommand{\sftype}[1][]{\ensuremath{\mathit{sort_{#1}}}}
\newcommand{\sptype}[1][]{\ensuremath{\mathit{sort_{#1}}}}
\newcommand{\fsort}[1]{\ensuremath{\mathit{F^{#1}}}}
\newcommand{\psort}[1]{\ensuremath{\mathit{B^{#1}}}}

\newcommand{\algebra}{(\algSet,\algFun,\algPred,\intSort,\intFun,\intPred)}
\newcommand{\algSet}{\ensuremath{S'}}
\newcommand{\algFun}{\ensuremath{F'}}
\newcommand{\algPred}{\ensuremath{B'}}
\newcommand{\intSort}{\ensuremath{\alpha}}
\newcommand{\intFun}{\ensuremath{\beta}}
\newcommand{\intPred}{\ensuremath{\gamma}}
\newcommand{\algAll}[1]{\ensuremath{\mathcal{A}(#1)}}

\newcommand{\dtVar}{\ensuremath{\prescript{}{d}{\mathcal{V}}}}
\newcommand{\dtVI}{\ensuremath{\iota}}
\newcommand{\dtVAll}[2]{\ensuremath{\mathcal{I}^{#1}_{#2}}}
\newcommand{\dtTerms}[3][]{\ensuremath{\prescript{#1}{d}{\mathcal{T}}(#2,#3)}}
\newcommand{\dtFormula}[2]{\ensuremath{\prescript{}{d}{\Gamma}(#1,#2)}}
\newcommand{\dtsem}[3]{\llbracket #1 \rrbracket^{#2}_{#3}}

\newcommand{\dtmod}[3][]{\ifthenelse{\equal{#1}{}}{#2 \models #3}{#2,#1 \models #3}}
\newcommand{\dteq}[2]{\ensuremath{#1 = #2}}
\newcommand{\dtneg}[1]{\ensuremath{\neg #1}}
\newcommand{\dtland}[2]{\ensuremath{#1 \land #2}}
\newcommand{\dtlor}[2]{\ensuremath{#1 \lor #2}}
\newcommand{\dtimplies}[2]{\ensuremath{#1 \implies #2}}
\newcommand{\dtiff}{\ensuremath{\iff}}
\newcommand{\dtforall}[1]{\ensuremath{\forall #1.}}
\newcommand{\dtexists}[1]{\ensuremath{\exists #1.}}
\newcommand{\dtspec}{\ensuremath{\Phi}}

\newcommand{\pspec}{\ensuremath{(\psid, \pstype)}}
\newcommand{\psid}{\ensuremath{P}}
\newcommand{\pstype}{\ensuremath{t^p}}
\newcommand{\pspecAll}[1]{\ensuremath{\mathcal{S}_p(#1)}}
\newcommand{\interface}{\ensuremath{(\ifLocP, \ifInP,\ifOutP)}}
\newcommand{\ifLocP}{\ensuremath{L_i}}
\newcommand{\ifInP}{\ensuremath{I_i}}
\newcommand{\ifOutP}{\ensuremath{O_i}}
\newcommand{\ifAll}[1]{\ensuremath{\mathcal{I}(#1)}}
\newcommand{\ifIid}{\ensuremath{I_{\mathit{id}}}}
\newcommand{\ifspec}{\ensuremath{(\isId,\isInt)}}
\newcommand{\isId}{\ensuremath{N}}
\newcommand{\isInt}{\ensuremath{Q}}
\newcommand{\IS}[1]{\mathcal{S}_i(#1)}
\newcommand{\ifLoc}[1]{\mathsf{loc}(#1)}%
\newcommand{\ifIn}[1]{\mathsf{in}(#1)}%
\newcommand{\ifOut}[1]{\mathsf{out}(#1)}%
\newcommand{\ifPort}[1]{\mathsf{port}(#1)}%
\newcommand{\ifTerms}[4][]{\ensuremath{\prescript{#1}{i}{\mathcal{T}}_{#4}(#2,#3)}}
\newcommand{\ifFormulas}[3]{\ensuremath{\prescript{}{i}{\Gamma}_{#3}(#1,#2)}}
\newcommand{\ifTsem}[4]{\llbracket #1 \rrbracket_{(#2,#3)}^{#4}}
\newcommand{\ifmod}[4][]{#2\models^{#1}_{#4} #3}
\newcommand{\ifint}{(\intCmp,\intLoc,\intIn,\intOut)}
\newcommand{\intCmp}{\ensuremath{c}}
\newcommand{\intLoc}{\ensuremath{\delta^l}}
\newcommand{\intIn}{\ensuremath{\delta^i}}
\newcommand{\intOut}{\ensuremath{\delta^o}}
\newcommand{\ifintAll}[2]{\mathcal{Q}(#1,#2)}
\newcommand{\cint}{\ensuremath{J}}
\newcommand{\cicomp}[2][]{\ensuremath{\mathit{cmp}_{#1}(#2)}}
\newcommand{\cintAll}[2]{\ensuremath{\mathcal{J}(#1,#2)}}

\newcommand{\ctAll}[2]{\ensuremath{\mathcal{T}_c(#1,#2)}}

\newcommand{\caVar}{\ensuremath{\prescript{}{c}{\mathcal{V}}}}
\newcommand{\caVI}{\ensuremath{\iota'}}
\newcommand{\caVAll}[2]{\ensuremath{\mathcal{I'}^{#1}_{#2}}}
\newcommand{\caTerms}[5][]{\ensuremath{\prescript{#1}{c}{\mathcal{T}}_{#4}^{#5}(#2,#3)}}
\newcommand{\caTVal}[2]{#1.#2}
\newcommand{\caFormulas}[4]{\ensuremath{\prescript{}{c}{\Gamma}_{#3}^{#4}(#1,#2)}}
\newcommand{\caFMin}[2]{|#1|_{#2}}
\newcommand{\caFMax}[2]{|#1|^{#2}}
\newcommand{\caFMinMax}[3]{|#1|_{#2}^{#3}}
\newcommand{\caFActive}[1]{\Vert #1 \Vert}
\newcommand{\caFConn}[4]{#1.#2\rightarrow#3.#4}
\newcommand{\caIRConn}[4]{#1.#2\rightarrow#3.#4}

\newcommand{\caforall}[1]{\ensuremath{\forall #1.}}
\newcommand{\caexists}[1]{\ensuremath{\exists #1.}}
\newcommand{\casem}[6]{\llbracket #1 \rrbracket_{(#2,#3)}^{(#4,#5)}(#6)}
\newcommandx*{\camod}[6][1=, 2=]{\ifthenelse{\equal{#1}{}}{#3 \models_{#5}^{#6} #4}{#1,#2,#3 \models_{#5}^{#6} #4}}

\newcommand{\ctaDTVar}{\ensuremath{\prescript{r}{d}{\mathcal{V}}}}
\newcommand{\ctaDTVI}{\ensuremath{\kappa}}
\newcommand{\ctaDTVAll}[2]{\ensuremath{\mathcal{K}^{#1}_{#2}}}
\newcommand{\ctaCVar}{\ensuremath{\prescript{r}{c}{\mathcal{V}}}}
\newcommand{\ctaCVI}{\ensuremath{\kappa'}}
\newcommand{\ctaCVAll}[2]{\ensuremath{\mathcal{K'}^{#1}_{#2}}}
\newcommand{\ctFormulas}[6]{\ensuremath{\prescript{}{t}{\Gamma}}^{(#3,#4)}_{(#5,#6)}(#1,#2)}
\newcommandx*{\ctmod}[6][1=, 2=]{\ifthenelse{\equal{#1}{}}{#3 \models_{#5}^{#6} #4}{#1,#2,#3 \models_{#5}^{#6} #4}}

\newcommand{\ctexistsD}[1]{\ensuremath{\exists #1.}}
\newcommand{\ctforallD}[1]{\ensuremath{\forall #1.}}
\newcommand{\ctexistsC}[1]{\ensuremath{\exists #1.}}
\newcommand{\ctforallC}[1]{\ensuremath{\forall #1.}}
\newcommand{\ctnext}[1]{\ensuremath{\bigcirc #1}}
\newcommand{\cteventually}[1]{\ensuremath{\diamondsuit #1}}
\newcommand{\ctglobally}[1]{\ensuremath{\Box #1}}
\newcommand{\ctuntilS}[2]{\ensuremath{\big(#1~\mathcal{U}~#2\big)}}
\newcommand{\ctuntilW}[2]{\ensuremath{\big(#1~\mathcal{W}~#2\big)}}

\newcommand{\acspec}{\ensuremath{(\Sigma,\dtspec,\specP,\acIf,\acIA,\acCTA)}}
\newcommand{\specP}{\ensuremath{S_p}}
\newcommand{\acIf}{\ensuremath{S_i}}
\newcommand{\acIA}{\ensuremath{\Omega}}
\newcommand{\acCTA}{\ensuremath{\Gamma}}
\newcommand{\cdMinMax}{(\mmMin, \mmMax)}
\newcommand{\mmMin}{\ensuremath{A_\mathit{min}}}
\newcommand{\mmMax}{\ensuremath{A_\mathit{max}}}
\newcommand{\cdRig}{\ensuremath{A_r}}
\newcommand{\cdRConn}{\ensuremath{A_c}}

\newcommand{\bbip}{\ensuremath{r_p}}
\newcommand{\bbis}{\ensuremath{n_s}}
\newcommand{\bbop}{\ensuremath{o_p}}
\newcommand{\bbos}{\ensuremath{c_s}}
\newcommand{\ksip}{\bbop}
\newcommand{\ksis}{\bbos}
\newcommand{\ksop}{\bbip}
\newcommand{\ksos}{\bbis}

\newtheorem{definition}{Definition}
\newtheorem{theorem}{Theorem}
\newtheorem{note}{Note}
\newtheorem{property}{Property}
\newtheorem{example}{Example}

%% file: macros/templates.tex
\newcommand{\inlineeqno}[1]{\refstepcounter{equation}\label{#1}\ (\theequation)}
\newenvironment{dtstmp}[4][250]{
	\def\func{35}
	\def\fvar{1}
	\def\awidth{#1}
	\newcommand{\symb}[2]{%
		\node[anchor=base west, text width=1.5cm] at (0,-\func-5 pt) {##1$\colon$};
		\node[anchor=base west, text width=0.5*\awidth] at (2cm,-\func-5 pt) {##2};
		\pgfmathparse{\func+10}%
		\edef\func{\pgfmathresult}
	}
	\newcommand{\sline}[1][solid]{%
		\draw[##1] (0,-\func pt) -- (\awidth pt,-\func pt);
		\pgfmathparse{\func+5}%
		\edef\func{\pgfmathresult}		
	}
	\newcommand{\var}[2]{%
		\node[anchor=base west, text width=0.8cm] at (0,-\func-5 pt) {\ifthenelse{\equal{\fvar}{1}}{\textbf{var}}{}};
		\node[anchor=base west, text width=0.5*\awidth] at (0.7cm,-\func-5 pt) {##1$\colon$ ##2};
		\pgfmathparse{\fvar+1}%
		\edef\fvar{\pgfmathresult}
		\pgfmathparse{\func+10}%
		\edef\func{\pgfmathresult}
	}
			
	\newcommand{\axiom}[2][]{%
		\node [anchor=base west, text width=\awidth pt-1cm] at (0,-\func-8 pt) {##2};
		\ifthenelse{\equal{##1}{}}{}{
			\node [anchor=base west, text width=0.7cm] at (\awidth pt-1cm,-\func-8 pt) {\inlineeqno{##1}};}
		\pgfmathparse{\func+15}		
		\edef\func{\pgfmathresult}			
	}
	\begin{tikzpicture}
	\draw [thick] (0,0) -- (\awidth pt,0);
	\node [anchor=base west, text width=0.5*\awidth] at (0,-10pt) {\textbf{DTSpec} #2};
	\node [align=right,anchor=base east, text width=0.5*\awidth] at (\awidth pt,-10pt) {\textbf{import} #3};
	\draw [thin, double,double distance=2pt] (0,-15pt) -- (\awidth pt,-15pt);
	\node [anchor=base west, text width=\awidth pt-0.5cm] at (0,-26pt) {\textbf{sort} #4};
	\draw (0,-30pt) -- (\awidth pt,-30pt);
}
{
	\draw[thick] (0,-\func pt) -- (\awidth pt,-\func pt);
	\end{tikzpicture}		
}

\newenvironment{pstmp}[3][200]{
	\def\func{20}
	\def\fvar{1}
	\def\pwidth{#1}
	\newcommand{\psport}[2]{%
		\node[anchor=base west, text width=1cm] at (0,-\func-5 pt) {##1$\colon$};
		\node[anchor=base west, text width=3cm] at (1.5cm,-\func-5 pt) {##2};
		\pgfmathparse{\func+10}%
		\edef\func{\pgfmathresult}
	}
	
	\begin{tikzpicture}
	\draw[thick] (0,0) -- (\pwidth pt,0);
	\node [anchor=north west, text width=0.5*\pwidth] at (0,0) {\textbf{PSpec} #2};
	\node [align=right,anchor=north east, text width=0.5*\pwidth] at (\pwidth pt,0) {\textbf{uses} #3};
	\draw[thin, double,double distance=2pt] (0,-15pt) -- (\pwidth pt,-15pt);
}
{
	\draw[thick] (0,-\func pt) -- (\pwidth pt,-\func pt);
	\end{tikzpicture}		
}

\newenvironment{istmp}[7][250]{
	\def\func{50}
	\def\fvar{1}
	\def\iwidth{#1}
	\newcommand{\isport}[2]{%
		\node[anchor=base west, text width=2cm] at (0,-\func-5 pt) {##1$\colon$};
		\node[anchor=base west, text width=4cm] at (2cm,-\func-5 pt) {##2};
		\pgfmathparse{\func+10}%
		\edef\func{\pgfmathresult}
	}
	\newcommand{\aline}[1][solid]{%
		\draw[##1] (0,-\func pt) -- (\iwidth pt,-\func pt);
		\pgfmathparse{\func+5}%
		\edef\func{\pgfmathresult}		
	}
	\newcommand{\var}[2]{%
		\node[anchor=base west, text width=5cm] at (0,-\func-5 pt) {\ifthenelse{\equal{\fvar}{1}}{\textbf{var}}{}};
		\node[anchor=base west, text width=5cm] at (1cm,-\func-5 pt) {##1$\colon$};
		\node[anchor=base west, text width=4cm] at (2cm,-\func-5 pt) {##2};		
		\pgfmathparse{\fvar+1}%
		\edef\fvar{\pgfmathresult}
		\pgfmathparse{\func+10}%
		\edef\func{\pgfmathresult}
	}
	\newcommand{\axiom}[2][]{%
		\node [anchor=base west, text width=5cm] at (0,-\func-8 pt) {##2};
		\ifthenelse{\equal{##1}{}}{}{
			\node [anchor=base west, text width=1cm] at (\iwidth pt-1cm,-\func-8 pt) {\inlineeqno{##1}};}
		\pgfmathparse{\func+15}
		\edef\func{\pgfmathresult}			
	}
	
	\begin{tikzpicture}	
	\draw[thick] (0,0) -- (\iwidth pt,0);
	\node [anchor=base west, text width=0.3*\iwidth] at (0,-10pt) {\textbf{ISpec} #2};
	\node [align=right, anchor=base east, text width=0.7*\iwidth] at (\iwidth pt,-10 pt) {\textbf{based on} #6
		\ifthenelse{\equal{#7}{}}{}{\textbf{uses} #7}};
	\draw [double,double distance=2pt] (0,-15pt) -- (\iwidth pt,-15pt);
	\node [anchor=base west, text width=4cm] at (0,-25pt) {\textbf{loc}: #3};
	\node [anchor=base west, text width=4cm] at (0,-35pt) {\textbf{in}: #4};
	\node [anchor=base west, text width=4cm] at (0,-45pt) {\textbf{out}: #5};
}
{
	\draw[thick] (0,-\func pt) -- (\iwidth pt,-\func pt);
	\end{tikzpicture}		
}

\newenvironment{ctstmp}[3][200]{
	\def\func{20}
	\def\cwidth{#1}
	\def\bfirst{0}
	\newcommand{\ctvar}[2]{%
		\node [anchor=base west, text width=1cm] at (0,-\func-5 pt) {\ifthenelse{\bfirst=0}{\textbf{var}}{}};
		\node[anchor=base west, text width=3cm] at (1cm,-\func-5 pt) {##1$\colon$};
		\node[anchor=base west, text width=3cm] at (3.5cm,-\func-5 pt) {##2};
		\edef\bfirst{1}	
		\pgfmathparse{\func+10}	
		\edef\func{\pgfmathresult}
	}
	
	\newcommand{\ctline}[1][solid]{%
		\draw[##1] (0,-\func pt) -- (\cwidth pt,-\func pt);
		\pgfmathparse{\func+5}
		\edef\func{\pgfmathresult}		
	}	
	\newcommand{\ctspace}[1][10]{%
		\pgfmathparse{\func+##1}		
		\edef\func{\pgfmathresult}	
	}		
	\newcommand{\ctaxiom}[2][]{%
		\node [anchor=base west, text width=\cwidth pt-1cm] at (0,-\func-8 pt) {##2};
		\ifthenelse{\equal{##1}{}}{}{
			\node [anchor=base west, text width=1cm] at (\cwidth pt-1cm,-\func-8 pt) {\inlineeqno{##1}};}
		\pgfmathparse{\func+15}		
		\edef\func{\pgfmathresult}			
	}
	\begin{tikzpicture}	
 	\draw [thick] (0,0) -- (\cwidth pt,0);
	\node [anchor=base west, text width=0.6*\cwidth] at (0,-10pt) {\textbf{Spec} #2};
	\node [align=right, anchor=base east, text width=0.4*\cwidth] at (\cwidth pt,-10pt) {\textbf{uses} \textit{#3}};	
	\draw [thin, double,double distance=2pt] (0,-15pt) -- (\cwidth pt,-15pt);
}
{
	\draw [thick] (0,-\func pt) -- (\cwidth pt,-\func pt);
	\end{tikzpicture}		
}

%% file: glossary.tex
\newglossaryentry{gls:mess}
{	name=message,
	description=atomic data entity,
	symbol=$\mess$}

\newglossaryentry{gls:port}
{	name=port,
	description=means to exchange messages,
	symbol=$\port$}

\newglossaryentry{gls:val}
{	name=port-valuation,
	description=assignment of messages to set of ports $P$,
	symbol=$\val P$}

\newglossaryentry{gls:cmp:id}
{	name=component identifier,
	description={identifier for components},
	symbol=$\cID$}

\newglossaryentry{gls:comp}
{	name=component,
	description={identifier, local/input/output ports, and port-valuations},
	symbol=$\cmpAll$}

\newglossaryentry{gls:cmp:healthy}
{	name=healthy,
	description=set of components with interface and valuations of local ports determined by component identifiers}

\newglossaryentry{gls:cmp:val}
{	name=component-port-valuation,
	description=component port valuation,
	plural=component-port-valuations}

\newglossaryentry{gls:aconf}
{	name=architecture configuration,
	description={set of active components and connections between their ports over a \glsentrytext{gls:cmp:healthy} set of components $C$},
	symbol=$\cnfAll C$}

\newglossaryentry{gls:cnf:trace}
{	name=configuration trace,
	description=sequence of \glsentryplural{gls:aconf} over a set of \glsentrytext{gls:cmp:healthy} components $C$,
	symbol=\allTraces{C}}

\newglossaryentry{gls:bhv:trace}
{	name=behavior trace,
	description=behavior trace,
	plural=behavior traces}

\newglossaryentry{gls:cnf:oin}
{	name=open input port,
	description=open input port,
	plural=open input ports}

\newglossaryentry{gls:datatype}
{	name=datatype,
	description=datatype,
	plural=datatypes}

\newglossaryentry{gls:dt:spec:temp}
{	name=datatype specification template,
	description=structured technique to specify \glsentryplural{gls:datatype},
	symbol=graphical}

\newglossaryentry{gls:dt:assert}
{	name=datatype assertion,
	description=formula over \glsentryplural{gls:dt:term} with corresponding signature $\Sigma$ and \glsentryplural{gls:dt:var} $\dtVar$,
	symbol=$\dtFormula \Sigma \dtVar$}

\newglossaryentry{gls:dt:spec}
{	name=datatype specification,
	description=set of \glsentryplural{gls:dt:assert},
	plural=datatype specifications,
	symbol=\dtspec}

\newglossaryentry{gls:dt:term}
{	name=datatype term,
	description=term over a signature $\Sigma$ and \glsentryplural{gls:dt:var} $\dtVar$,
	symbol={$\dtTerms[s] \Sigma \dtVar$, $\dtTerms \Sigma \dtVar$}}

\newglossaryentry{gls:dt:var}
{	name=datatype variable,
	description=variable for \glsentrytext{gls:datatype} elements of sort $s$,
	symbol=$\dtVar_s$}

\newglossaryentry{gls:dt:va}
{	name=datatype variable assignment,
	description=assignment of elements of an algebra $A$ to a set of \glsentryplural{gls:dt:var} $\dtVar$,
	symbol=$\dtVAll{\dtVar}{A}$}

\newglossaryentry{gls:dt:sem}
{	name=datatype semantic function,
	description=assigns elements of an algebra $A$ to \glsentryplural{gls:dt:term} under a certain \glsentrytext{gls:dt:va} $\dtVI$,
	symbol=$\dtsem{\_}{\dtVI}{A}$}

\newglossaryentry{gls:dt:ass:sem}
{	name=datatype models relation,
	description=relates \glsentryplural{gls:dt:assert} with \glsentryplural{gls:alg} and corresponding \glsentrytext{gls:dt:va} $\dtVI$,
	symbol=$\dtmod[\dtVI]{\_}{\_}$ / $\dtmod{\_}{\_}$}

\newglossaryentry{gls:if:pid}
{	name=port identifier,
	description=identifier for ports,
	symbol=$\pID$}

\newglossaryentry{gls:if:pspec}
{	name=port specification,
	description=\glsentryplural{gls:if:pid} and corresponding typing w.r.t. a signature $\Sigma$,
	symbol=$\pspecAll \Sigma$}

\newglossaryentry{gls:if:ptype}
{	name=port type,
	description=mapping assigning a sort to each port identifier,
	symbol=$\pID$}

\newglossaryentry{gls:portint}
{	name=port-interpretation,
	description=port interpretation,
	plural=port interpretations}

\newglossaryentry{gls:if:term}
{	name=interface term,
	description={term over the ports of a certain interface $F$, signature $\Sigma$, and \glsentryplural{gls:dt:var} $\dtVar$},
	symbol=$\ifTerms{\Sigma}{F}{\dtVar}$}

\newglossaryentry{gls:if:t:sem}
{	name=interface semantic function,
	description=assigns elements of an algebra $A$ to \glsentryplural{gls:if:term} under \glsentrytext{gls:dt:va} $\dtVI$ and \glsentrytext{gls:if:int} $J$,
	symbol=$\ifTsem{\_}{A}{\dtVI}{J}$}

\newglossaryentry{gls:if:assert}
{	name=interface assertion,
	description={formula over \glsentryplural{gls:if:term} with signature $\Sigma$, interface $F$, and \glsentryplural{gls:dt:var} $\dtVar$},
	symbol=$\ifFormulas{\Sigma}{F}{\dtVar}$}

\newglossaryentry{gls:if:a:sem}
{	name=interface models relation,
	description={relates \glsentryplural{gls:if:assert} $\varphi$ with \glsentrytext{gls:if:int} $J$ under algebra $A$ and \glsentrytext{gls:dt:va} $\dtVI$},
	symbol={$\ifmod[\dtVI]{J}{\varphi}{A}$/$\ifmod{J}{\varphi}{A}$}}

\newglossaryentry{gls:interface}
{	name=interface,
	description={set of local, input, and output port identifiers of a port specification $P$},
	symbol=$\ifAll P$}

\newglossaryentry{gls:if:id}
{	name=interface identifier,
	description=identifier for interfaces,
	symbol=$\ifIid$}

\newglossaryentry{gls:ifspec}
{	name=interface specification,
	description={interface-identifiers and corresponding interfaces over \glsentrytext{gls:sig} $\Sigma$},
	symbol=$\IS{\Sigma}$}

\newglossaryentry{gls:ifspec:int}
{	name=interface specification interpretation,
	description={set of \glsentryplural{gls:if:int} for \glsentryplural{gls:interface} of an \glsentrytext{gls:ifspec} $S_i$ under algebra $A$},
	symbol=$\cintAll{S_i}{A}$}

\newglossaryentry{gls:ifspec:int:c}
{	name=component interpretation,
	description={bijection between interface identifiers and component identifiers},
	symbol=$\ciInt$}

\newglossaryentry{gls:ifspec:temp}
{	name=interface specification template,
	description=structured technique to specify \glsentryplural{gls:ctype},
	symbol=graphical}

\newglossaryentry{gls:ifspec:ptemp}
{	name=port specification template,
	description=structured technique to specify ports,
	symbol=graphical}

\newglossaryentry{gls:if:int}
{	name=interface interpretation,
	description={components and corresponding \glsentryplural{gls:if:int:p} for an interface $F$},
	symbol=$\ifintAll{F}$}

\newglossaryentry{gls:if:int:p}
{	name=port interpretation,
	description={bijection between port identifiers and ports of a component},
	symbol={$\intLoc$, $\intIn$ ,$\intOut$}}

\newglossaryentry{gls:templog}
{	name=temporal logic,
	description=temporal logic,
	plural=temporal logic}

\newglossaryentry{gls:cnfdiag}
{	name=configuration diagram,
	description=graphical notation to specify interfaces as well as common activation and connection constraints,
	symbol=graphical,
	plural=configuration diagrams}

\newglossaryentry{gls:cnfassert}
{	name=configuration assertion,
	description={formula over \glsentryplural{gls:ca:term} with corresponding \glsentrytext{gls:sig} $\Sigma$, \glsentrytext{gls:ifspec} $S_i$, \glsentryplural{gls:dt:var} $\dtVar$, and \glsentryplural{gls:ca:var} $\caVar$},
	symbol=$\caFormulas{\Sigma}{S_i}{\dtVar}{\caVar}$,
	plural=configuration assertions}

\newglossaryentry{gls:ca:term:sem}
{	name=configuration semantic function,
	description={assigns elements of an algebra $A$ to \glsentryplural{gls:ca:term} under a certain \glsentrytext{gls:dt:va} $\dtVI$, \glsentrytext{gls:ifspec:int} $J$ and corresponding \glsentrytext{gls:ca:va} $\caVI$, and \glsentrytext{gls:aconf} $k$},
	symbol=$\casem{\_}{A}{\dtVI}{J}{\caVI}{k}$}

\newglossaryentry{gls:ca:var}
{	name=component variable,
	description=variable for \glsentryplural{gls:cmp:id},
	symbol=$\caVar$}

\newglossaryentry{gls:ca:term}
{	name=configuration term,
	description={term over \glsentrytext{gls:sig} $\Sigma$, \glsentrytext{gls:ifspec} $S_i$, \glsentryplural{gls:dt:var} $\dtVar$, and \glsentryplural{gls:ca:var} $\caVar$},
	symbol=$\caTerms{\Sigma}{S_i}{\dtVar}{\caVar}$,
	plural=configuration terms}

\newglossaryentry{gls:ca:va}
{	name=component variable assignment,
	description={assignment of \glsentryplural{gls:cmp:id} of \glsentrytext{gls:ifspec:int} $J$ to \glsentryplural{gls:ca:var} $\caVar$},
	symbol=$\caVAll{\caVar}{J}$}

\newglossaryentry{gls:ca:sem}
{	name=configuration models relation,
	description={relates \glsentryplural{gls:cnfassert} with \glsentryplural{gls:aconf} under an \glsentrytext{gls:alg} $A$ and corresponding \glsentrytext{gls:dt:va} $\dtVI$, and \glsentrytext{gls:ifspec:int} $J$ and corresponding \glsentrytext{gls:ca:va} $\caVI$},
	symbol=$\camod[\dtVI][\caVI]{\_}{\_}{A}{J}$}

\newglossaryentry{gls:cnftraceassert}
{	name=configuration trace assertion,
	description={formula over \glsentryplural{gls:cnfassert} with corresponding \glsentrytext{gls:sig} $\Sigma$, \glsentrytext{gls:ifspec} $S_i$, \glsentryplural{gls:dt:var} $\dtVar$, \glsentryplural{gls:ca:var} $\caVar$, \glsentryplural{gls:ct:dtvar} $\ctaDTVar$, and \glsentryplural{gls:ct:cvar} $\ctaCVar$},
	symbol=$\ctFormulas{\Sigma}{S_i}{\dtVar}{\caVar}{\ctaDTVar}{\ctaCVar}$,
	plural=configuration trace assertions}

\newglossaryentry{gls:cta:sem}
{	name=configuration trace models relation,
	description={relates \glsentryplural{gls:cnftraceassert} with \glsentryplural{gls:cnf:trace} under an \glsentrytext{gls:alg} $A$ and corresponding \glsentrytext{gls:ct:dtva} $\ctaDTVI$, and \glsentrytext{gls:ifspec:int} $J$ and corresponding \glsentrytext{gls:ct:cva} $\ctaCVI$},
	symbol=$\ctmod[\ctaDTVI][\ctaCVI]{\_}{\_}{A}{J}$}

\newglossaryentry{gls:const:spec}
{	name=architecture constraint specification,
	description={a \glsentrytext{gls:sig} $\Sigma$, \glsentrytext{gls:dtspect} $\dtspec$, \glsentrytext{gls:ifspec} $\acIf$, and a set of \glsentryplural{gls:cnftraceassert} $\acCTA$},
	symbol=$\acspec$}

\newglossaryentry{gls:cta:temp}
{	name=configuration trace specification template,
	description={structured technique to specify \glsentryplural{gls:cnf:trace}},
	symbol=graphical}

\newglossaryentry{gls:connassert}
{	name=connection assertion,
	description=connection assertion,
	plural=configuration trace assertions}

\newglossaryentry{gls:actassert}
{	name=activation assertion,
	description=activation assertion,
	plural=activation assertions}

\newglossaryentry{gls:bhvassert}
{	name=behavior assertion,
	description=behavior assertion,
	plural=behavior assertions}

\newglossaryentry{gls:btassert}
{	name=behavior trace assertion,
	description=behavior trace assertion,
	plural=behavior trace assertions}

\newglossaryentry{gls:ct:dtvar}
{	name=rigid datatype variable,
	description=\glsentrytext{gls:dt:var} with fixed assignment for the whole execution,
	symbol=$\ctaDTVar$,
	plural=rigid datatype variables}

\newglossaryentry{gls:ct:cvar}
{	name=rigid component variable,
	description=\glsentrytext{gls:ca:var} with fixed assignment for the whole execution,
	symbol=$\ctaCVar$,	
	plural=rigid component variables}

\newglossaryentry{gls:ct:dtva}
{	name=rigid datatype variable assignment,
	description={assignment of elements of an algebra $A$ to a set of \glsentryplural{gls:ct:dtvar} $\dtVar$,},
	symbol=$\ctaDTVAll{\ctaDTVar}{A}$,
	plural=rigid datatype variable assignments}

\newglossaryentry{gls:ct:cva}
{	name=rigid component variable assignment,
	description={assignment of \glsentryplural{gls:cmp:id} of \glsentrytext{gls:ifspec:int} $J$ to \glsentryplural{gls:ct:cvar} $\caVar$},
	symbol=$\ctaCVAll{\ctaCVar}{J}$,
	plural=rigid component variable assignments}

\newglossaryentry{gls:archprop}
{	name=architecture property,
	description=a set of configuration traces,
	plural=architecture properties}

\newglossaryentry{gls:prop:spec}
{	name=architecture property specification,
	description=architecture property specification,
	plural=architecture property specifications}

\newglossaryentry{gls:ctype}
{	name=component type,
	description={an interface and corresponding \glsentryplural{gls:if:assert} over \glsentrytext{gls:sig} $\Sigma$ and \glsentryplural{gls:dt:var} $\dtVar$},
	symbol={$\ctAll \Sigma \dtVar$},
	plural=component types}

\newglossaryentry{gls:cns:arch}
{	name=architecture constraint,
	description=a general constraint over the architecture,
	plural=architecture constraints}

\newglossaryentry{gls:cns:bhv}
{	name=behavioral constraint,
	description=behavioral constraint,
	plural=behavioral contstraints}

\newglossaryentry{gls:cns:act}
{	name=activation constraint,
	description=constraint on activation and deactivation of components,
	plural=activation contstraints}

\newglossaryentry{gls:act:minmax}
{	name=min-max annotation,
	description={specifies the minimal($\mathit{min}$)/maximal($\mathit{max}$) number of components of a certain type},
	symbol={$[\mathit{min}..\mathit{max}]$}}

\newglossaryentry{gls:act:rigid}
{	name=rigid annotation,
	description={set of \glsentryplural{gls:ct:cvar} denoting the set of all possible active components of an interface $\mathit{If}$},
	symbol={$c_1, c_2\colon \mathit{If}$}}

\newglossaryentry{gls:cns:conn}
{	name=connection constraint,
	description=constraint on the connection between components}

\newglossaryentry{gls:conn:feasible}
{	name=possible connection annotation,
	description=specification of possible port-connections}

\newglossaryentry{gls:conn:req}
{	name=required connection annotation,
	description=specification of required port-connection}

\newglossaryentry{gls:bb:p}
{	name=Blackboard,
	description=blackboard,
	plural=Blackboards}

\newglossaryentry{gls:bb:bb}
{	name=Blackboard,
	description=Blackboard,
	plural=Blackboards}

\newglossaryentry{gls:bb:ks}
{	name=Knowledge-source,
	description=Knowledge-source,
	plural=Knowledge-sources}

\newglossaryentry{gls:sig}
{	name=signature,
	description={sorts, function/predicate symbols},
	symbol=$\Sigma$}

\newglossaryentry{gls:alg}
{	name=algebra,
	description={sets, functions, predicates, and corresponding mappings for a signature $\Sigma$},
	symbol=$\algAll{\Sigma}$}

%% file: acronyms.tex
\newacronym[see={[Glossary:]{gls:datatype}}]{DT}{DT}{datatype\glsadd{gls:datatype}}
\newacronym[see={[Glossary:]{gls:cmpif}}]{CI}{CI}{component interface\glsadd{gls:cmpif}}
\newacronym[see={[Glossary:]{gls:val}}]{PV}{PV}{port-valuation\glsadd{gls:val}}
\newacronym[see={[Glossary:]{gls:cmp:val}}]{CPV}{CPV}{component-port-valuation\glsadd{gls:cmp:val}}
\newacronym[see={[Glossary:]{gls:aconf}}]{AC}{AC}{architecture configuration\glsadd{gls:aconf}}
\newacronym[see={[Glossary:]{gls:cnf:trace}}]{CT}{CT}{configuration trace\glsadd{gls:cnf:trace}}
\newacronym[see={[Glossary:]{gls:bhv:trace}}]{BT}{BT}{behavior trace\glsadd{gls:bhv:trace}}
\newacronym[see={[Glossary:]{gls:cnfdiag}}]{CD}{CD}{configuration diagram\glsadd{gls:cnfdiag}}
\newacronym[see={[Glossary:]{gls:cnftraceassert}}]{CTA}{CTA}{configuration trace assertion\glsadd{gls:cnftraceassert}}
\newacronym[see={[Glossary:]{gls:bhvassert}}]{BA}{BA}{behavior assertion\glsadd{gls:bhvassert}}
\newacronym[see={[Glossary:]{gls:btassert}}]{BTA}{BTA}{behavior trace assertion\glsadd{gls:btassert}}
\newacronym[see={[Glossary:]{gls:actassert}}]{AA}{AA}{activation assertion\glsadd{gls:actassert}}
\newacronym[see={[Glossary:]{gls:connassert}}]{CA}{CA}{connection assertion\glsadd{gls:connassert}}
\newacronym[see={[Glossary:]{gls:dt:spec}}]{DTS}{DTS}{datatype specification\glsadd{gls:dt:spec}}
\newacronym[see={[Glossary:]{gls:ctype}}]{CPT}{CPT}{component type\glsadd{gls:ctype}}
\newacronym[see={[Glossary:]{gls:bb:bb}}]{BB}{BB}{Blackboard\glsadd{gls:bb:bb}}
\newacronym[see={[Glossary:]{gls:bb:ks}}]{KS}{KS}{Knowledge-source\glsadd{gls:bb:ks}}
\newacronym[see={[Glossary:]{gls:ifspec}}]{IS}{IS}{interface specification\glsadd{gls:ifspec}}
\newacronym[see={[Glossary:]{gls:dt:va}}]{VA}{VA}{variable assignment\glsadd{gls:dt:va}}
\newacronym[see={[Glossary:]{gls:portint}}]{PI}{PI}{port-interpretation\glsadd{gls:portint}}
\newacronym[see={[Glossary:]{gls:cmp:port}}]{CP}{CP}{component-port\glsadd{gls:cmp:port}}

\newacronym{ITP}{ITP}{Interactive Theorem Proving}

%% file: abstract.tex
\begin{abstract}
In dynamic architectures, component activation and connections between components may vary over time.
With the emergence of mobile computing such architectures became increasingly important and several techniques emerged to support in their specification.
%
These techniques usually allow for the specification of concrete architecture instances.
Sometimes, however, it is desired to focus on the specification of constraints, rather than concrete architectures.
Especially specifications of architecture patterns usually focus on a few, important constraints, leaving out the details of the concrete architecture implementing the pattern.
%
With this article we introduce an approach to specify such constraints for dynamic architectures.
To this end, we introduce the notion of \emph{configuration traces} as an abstract model for dynamic architectures.
Then, we introduce the notion of \emph{configuration trace assertions} as a formal language based on linear temporal logic to specify constraints for such architectures.
In addition, we also introduce the notion of \emph{configuration diagrams} to specify interfaces and certain common activation and connection constraints in one single, graphical notation.
%
The approach is well-suited to specify patterns for dynamic architectures and verify them by means of formal analyses.
%
This is demonstrated by applying the approach to specify and verify the Blackboard pattern for dynamic architectures.
\end{abstract}

\begin{keywords}
	Dynamic Architectures; Algebraic Specifications; Architecture Constraints
\end{keywords}

%% file: introduction.tex
\section{Introduction}\label{sec:intro}
\begin{sloppypar}
	A systems \emph{architecture} describes the components of a system as well as connections between those components.
	\emph{Dynamic architectures} are architectures in which component activation as well as connections can change over time~\cite{Wermelinger2001}.
	Specifying such architectures remains an active topic of research~\cite{Fiadeiro2013,Broy2014} and over the last years, several approaches emerged to support in this endeavor~\cite{Luckham1995,Allen1997,Dormoy2010,Broy2014}.
	These approaches usually focus on the specification of concrete architecture instances characterized by the following properties:
	\begin{itemize}
		\item \emph{Concrete model of execution}: Component behavior is either specified using some notion of state-machine~\cite{Allen1998}, guarded commands~\cite{Wermelinger2001}, or even stream-processing functions~\cite{Broy2014}.
		\item \emph{Concrete model of component interaction}: Interaction between components is either message-synchronous (for approaches based on CSP~\cite{Hoare1978}), time-synchronous~\cite{Broy2014}, or also action-synchronous~\cite{Wermelinger2001}.
		\item \emph{Concrete model of component activation/deactivation}: Component activation/deactivation is either specified by arbitrary components~\cite{Luckham1995} or by a designated component~\cite{Allen1998}.
		\item \emph{Concrete model of reconfiguration}: Similarly, \emph{connection changes} are either implemented by each single component~\cite{Luckham1995} or again by a designated component~\cite{Allen1998}.
	\end{itemize}
\end{sloppypar}

While these aspects are important when specifying concrete architecture instances, they play only a secondary role when specifying architectural constraints as it is the case when specifying architectural patterns, for example.

Consider, for example, the specification of the \glsentrytext{gls:bb:p} architecture pattern. In this pattern, a set of experts (aka. \glsentryplural{gls:bb:ks}) collaborate through a so-called \glsentrytext{gls:bb:bb} component to collaboratively solve a complex problem~\cite{Buschmann1996,Shaw1996,Taylor2009}.
The pattern requires that for each problem provided by a \glsentrytext{gls:bb:bb} component, eventually an expert exists which is able to handle this problem.
In turn, the pattern guarantees that a complex problem can be collaboratively solved, even if no single expert exists which can solve the problem on its own.

If we look at this specification, several observations can be made:
\begin{inparaenum}[(i)]
	\item The specification does not prescribe how a component is implemented as long as it satisfies certain behavioral constraints.	
	\item Neither does it constrain how components communicate as long as they do.
	\item Moreover, it is not specified who creates a component as long as it is created somehow.
	\item Finally, it is not specified how the connections were established as long as they are.
\end{inparaenum}
Thus, we argue, that traditional architecture specification techniques meet its limits when it comes to the specification of such constraints and that new, more abstract techniques are required.

To address this problem, we introduce an abstract model of dynamic architectures.
Thereby, an architecture is modeled as a set of so-called configuration traces which are sequences over architecture configurations.
An architecture configuration, on the other hand, consists of a set of active components, connections between component ports and (important) a valuation of the component ports with messages.

\begin{sloppypar}
	Based on this model, we provide a model-theoretic approach to formally specify properties of dynamic architectures by means of architecture constraints:
	\begin{inparaenum}[(i)]
		\item First, abstract \glsentryplural{gls:datatype} are specified by means of algebraic specifications.
		\item Then, \glsentryplural{gls:interface} and \glsentryplural{gls:ctype} are specified over these \glsentryplural{gls:datatype} by means of \glsentryplural{gls:ifspec}.
		\item Finally, properties can be specified over the interfaces by means of \glsentryplural{gls:cnftraceassert}.
	\end{inparaenum}
	To facilitate the specification process we also introduce the notion of \glsentryplural{gls:cnfdiag} as a graphical means to specify interfaces as well as certain common activation and connection constraints.
\end{sloppypar}

The approach allows to specify properties of dynamic architectures by means of \glsentryplural{gls:cns:arch} and is thus well suited for the specification of patterns for this kind of architectures.
To demonstrate this, we evaluate the approach by specifying and verifying the \emph{\glsentrytext{gls:bb:p}} pattern~\cite{Buschmann1996,Shaw1996,Taylor2009} for dynamic architectures.
Thereby, we identify and formalize the patterns key \glsentryplural{gls:cns:arch} as a set of \glsentryplural{gls:cnftraceassert} and prove its guarantee from the specification.

\subsection{Specifying Properties of Dynamic Architectures}
\begin{figure}\centering
	\begin{tikzpicture}[phase1/.style={align=center, rectangle, draw, rounded corners=0.2cm, minimum width=2.5cm, text width=2cm}]

	\node[phase1, text width=6.2cm, align=left, minimum width=7cm] (sig) at (0,4) {Signature:\\ Sorts, Function, and Predicate Symbols};
	\node[phase1] (dt) at (-3,2) {Datatype \\ Specification};
	\node[phase1] (if) at (3,2) {Interface \\ Specification};
	\node[phase1] (ac) at (3,0) {Architecture \\ Constraints \\ Specification};
	
	\draw[-latex] ($(sig.south west) + (2,0)$)--(dt.north);
	\draw[-latex] ($(sig.south east) - (2,0)$)--(if.north);	
	\draw[-latex] (if.south)--(ac.north);	
	\end{tikzpicture}
	\caption{Approach to specify properties of dynamic architectures.}
	\label{fig:approach}
\end{figure}
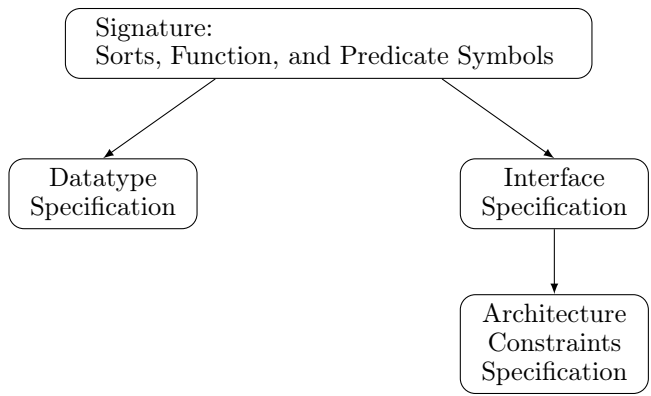

Fig.~\ref{fig:approach} provides an overview of our approach to specify properties of dynamic architectures.
As a first step, a suitable \gls{gls:sig} is specified to introduce symbols for sets, functions, and predicates. These symbols form the primitive entities of the whole specification process: \glsentryplural{gls:dt:spec} and \glsentryplural{gls:ifspec} as well as \glsentryplural{gls:const:spec} are based on these symbols.

Then, \glspl{gls:datatype} are specified over the \glsentrytext{gls:sig}. Therefore, so-called \glspl{gls:dt:assert} are build over \glspl{gls:dt:term} to assert the \glsentryplural{gls:datatype} characteristic properties and provide meaning for the symbols introduced in the \glsentrytext{gls:sig}.

Interfaces are also directly specified over the \glsentrytext{gls:sig}. Therefore, a set of ports is typed by sorts of the corresponding \glsentrytext{gls:sig}. Then, an interface is specified by assigning an interface identifier with three sets of ports: local, input, and output ports. Finally, a set of \glspl{gls:if:assert} is associated with each interface identifier to specify component types, i.e., interfaces with associated global invariants.

Finally, \glspl{gls:cns:arch} can be specified by means of \glspl{gls:cnftraceassert} over the interfaces.
\Glspl{gls:cnftraceassert} are a special kind of linear-temporal formulas build over \glspl{gls:cnfassert}, i.e., assertions over an \gls{gls:aconf}.

\subsection{Running example: \glsentrytext{gls:bb:p} architectures}
In the following we introduce the \glsentrytext{gls:bb:p} architecture pattern. It is used as a running example throughout the text to illustrate the main concepts and ideas.

The \gls{gls:bb:p} pattern (as described, for example, by Shaw and Garlan~\cite{Shaw1996}, Buschmann \etal~\cite{Buschmann1996}, and Taylor \etal~\cite{Taylor2009}), is a pattern for dynamic architectures used for collaborative problem solving.
In a \glsentrytext{gls:bb:p} architecture, several experts (aka.\ \glsentryplural{gls:bb:ks}) collaborate through a central component (aka. \glsentrytext{gls:bb:bb}) to solve a complex problem consisting of several sub-problems.

\begin{sloppypar}
	Although the pattern is not too complex (consisting of only two types of components), it incorporates several aspects of dynamic architectures:
	\begin{inparaenum}[(i)]
		\item \glsentrytext{gls:bb:ks} components can be activated and deactivated over time,
		\item connections between the various \glsentryplural{gls:bb:ks} and the \glsentrytext{gls:bb:bb} component can also change over time.
	\end{inparaenum}
\end{sloppypar}

\subsection{Overview}
The remainder of this article is structured as follows:
In Sect.~\ref{sec:model} we discuss our model of dynamic architectures.
Then, the different modeling techniques to specify \glsentryplural{gls:datatype} (Sect.~\ref{sec:ds}), interfaces (Sect.~\ref{sec:ifspec}), and configuration traces (Sect.~\ref{sec:cnft}) are introduced. For each technique we provide a formal description of its syntax as well as its semantics in terms of the model introduced in Sect.~\ref{sec:model}.
While these techniques already suffice to specify all kinds of architecture properties, in Sect.~\ref{sec:cnfd} we introduce the notion of configuration diagrams as a graphical notation to support in the specification of interfaces and certain common activation and connection constraints.
In Sect.~\ref{sec:analysis} we demonstrate how a specification in our language can be used to formally reason about the specification. To this end, we verify the Blackboard architecture pattern by proving one of its characteristic properties from its specification.
Finally, we discuss our approach and possible limitations thereof in Sect.~\ref{sec:discussion}.
In Sect.~\ref{sec:bgrw} we provide related work and conclude with a brief summary in Sect.~\ref{sec:conc}.

To not lose track of the various concepts and notations, Sect.~\ref{sec:model} $-$ Sect.~\ref{sec:cnfd} conclude with a tabular overview of all the new concepts and notations introduced in the corresponding section.

Sect.~\ref{sec:conv} provides our notation for some general, mathematical concepts.
If at any time during the reading a symbol was used but not properly introduced, its definition can be found in this section.

%% file: model.tex
\section{A Model of Dynamic Architectures}\label{sec:model}
In the following, we describe our model of dynamic architectures introduced in~\cite{Marmsoler2016}. It is based on Broy's \textsc{Focus} theory~\cite{Broy2010} and an adaptation of its dynamic extension~\cite{Broy2014}.
An \glsentrytext{gls:archprop} is thereby modeled as a set of \glsentryplural{gls:cnf:trace} which are sequences of \glsentryplural{gls:aconf} that, in turn, consist of a set of active components, valuations of their ports with messages, and connections between their ports.

\subsection{Foundations}\label{sec:mod:foundations}
In our model, components communicate by exchanging \glsentryplural{gls:mess} over ports. Thus, we assume the existence of sets $\mess$ and $\port$ containing all \glsentryplural{gls:mess} and ports, respectively.

\subsubsection{\glsentryplural{gls:val}}
Components communicate by sending and receiving messages through ports. This is achieved through the notion of \glsentrytext{gls:val}. Roughly speaking, a valuation for a set of ports is an assignment of messages to each port.
\begin{definition}[\Glsentrytext{gls:val}]
	For a set of ports $P\subseteq \port$, we denote by $\val{P}$ the set of all possible \emph{\glspl{gls:val}}, formally:
	\begin{equation*}
	\val P\defeq (P\to\pset{\mess})\enspace.
	\end{equation*}
\end{definition}
Note that in our model, ports can be valuated by a \emph{set} of messages, meaning that a component can send/receive no message, a single message, or multiple messages at each point in time.

\subsubsection{\glsentryplural{gls:comp}}
In our model, the basic unit of computation is a \glsentrytext{gls:comp}. It consists of an identifier, a set of ports and a valuation of ports with messages. Indeed it is rather a snapshot of a component at a certain point in time with concrete messages on its ports.

Thus, we assume the existence of set \glsentrysymbol{gls:cmp:id} containing all \glsentryplural{gls:cmp:id}.
\begin{definition}[\Glsentrytext{gls:comp}]\label{def:cmp}
	A \emph{\gls{gls:comp}} is a 5-tuple $\comp$ consisting of:
	\begin{itemize}
		\item a \glsentrytext{gls:cmp:id} $\cmpId\in\cID$\enspace;
		\item \emph{disjoint} sets of local ports $\cmpLP\subseteq\port$, input ports $\cmpIP\subseteq\port$, and output ports $\cmpOP\subseteq\port$; and
		\item a valuation of its ports $\cmpVal\in\val{\cmpLP\cup\cmpIP\cup\cmpOP}$.
	\end{itemize}
	The set of all \glsentryplural{gls:comp} is denoted by $\cmpAll$.

	For a set of \glsentryplural{gls:comp} $C\subseteq\cmpAll$, we denote by:
	\begin{itemize}
		\item $\cmpLoc{C}\defeq\bigcup_{\comp\in C} (\{\cmpId\}\times\cmpLP)$ the set of \emph{component local ports},
		\item $\cmpIn{C}\defeq\bigcup_{\comp\in C} (\{\cmpId\}\times\cmpIP)$ the set of \emph{component input ports},
		\item $\cmpOut{C}\defeq\bigcup_{\comp\in C} (\{\cmpId\}\times\cmpOP)$ the set of \emph{component output ports},
		\item $\cmpPort{C}\defeq \cmpLoc{C}\cup\cmpIn{C}\cup\cmpOut{C}$ the set of all \emph{component ports}, and
		\item $\cmpName{C}\defeq \bigcup_{\comp\in C} (\{\cmpId\})$ the set of all \emph{component identifiers}.
	\end{itemize}	
	
	A set of components $C\subseteq\cmpAll$ is called \emph{\gls{gls:cmp:healthy}} if for each $\comp$, $\comp[']$ $\in C$ the following conditions are fulfilled:
	\begin{itemize}
		\item a component's interface is determined by its identifier:
		\begin{equation}\label{eq:healthy}
			\cmpId=\cmpId'\implies \cmpLP=\cmpLP'\land \cmpIP=\cmpIP'\land\cmpOP=\cmpOP'\enspace,
		\end{equation}
		\item the valuation of the local ports is also determined by a component's identifier:
		\begin{equation}\label{eq:healthy2}
			\cmpId=\cmpId'\implies \forall p\in\cmpLP\colon \cmpVal(p)=\cmpVal'(p) \enspace.
		\end{equation}
	\end{itemize}
	\begin{note}[Well-definedness of Eq.~\eqref{eq:healthy2}]
		Due to Eq.~\eqref{eq:healthy} $\cmpLP=\cmpLP'$ which is why Eq.~\eqref{eq:healthy2} is indeed well-defined.
	\end{note}	
	A \glsentrytext{gls:cmp:healthy} set of \glsentryplural{gls:comp} $C\subseteq\cmpAll$ induces mappings to assigning local, input, output, and all ports to component identifier $d\in \cmpName{C}$:
	\begin{itemize}
		\item $\cmpPloc{C}(d) = O \iff \exists I,O\subseteq \port\colon (d,L,I,O)\in C$,
		\item $\cmpPin{C}(d) = I \iff \exists O,L\subseteq \port\colon (d,L,I,O)\in C$,
		\item $\cmpPout{C}(d) = O \iff \exists I,L\subseteq \port\colon (d,L,I,O)\in C$, and
		\item $\cmpPport{C}(d) \defeq \cmpPloc{C}(d)\cup\cmpPin{C}(d)\cup\cmpPout{C}(d)$\enspace.
	\end{itemize}
	Moreover, it induces a mapping $\cmpPlocVal{C}$ to access the valuation of a component's local ports:
	\begin{align}
		\cmpPlocVal{C}(d)(p) = M & \iff \exists I,O,L\subseteq \port, \cmpVal\in\val{L\cup I\cup O}\colon\nonumber\\
		&(d,L,I,O,\cmpVal)\in C \land \cmpVal(p)=M\enspace.
	\end{align}

	\begin{note}[Well-definedness of $\cmpPloc{C}, \cmpPin{C}, \cmpPout{C}$ and $\cmpPlocVal{C}(d)$]
		While Eq.~\eqref{eq:healthy} guarantees that $\cmpPloc{C}$, $\cmpPin{C}$, and $\cmpPout{C}$ are well-defined,
		Eq.~\eqref{eq:healthy2} guarantees that $\cmpPlocVal{C}(d)$ is well-defined.
	\end{note}
\end{definition}

\begin{sloppypar}
	Note that a \glsentrytext{gls:cmp:healthy} set of \glsentryplural{gls:comp} does restrict only the interface of \glsentryplural{gls:comp} and the \glsentryplural{gls:val} of its local ports. However, there may be several \glsentryplural{gls:comp} with the same identifier but different valuations of its input and output ports.
	Thus, it is indeed possible to have two \emph{different} \glsentryplural{gls:comp} $\comp, (\cmpId,\cmpLP,\cmpIP,\cmpOP,\cmpVal')$ in a \glsentrytext{gls:cmp:healthy} set of components, as long as there exists a port $p\in\cmpIP\cup\cmpOP$, such that $\cmpVal(p)\neq\cmpVal'(p)$.
\end{sloppypar}

An important property of \glsentrytext{gls:cmp:healthy} is, that it is preserved under the subset relation.
\begin{property}[Subset preserves healthiness]
	For a \glsentrytext{gls:cmp:healthy} set of components $C$, each subset $C'\subseteq C$ is again \glsentrytext{gls:cmp:healthy}.
\end{property}
\begin{proof}
	Assume $C$ is \glsentrytext{gls:cmp:healthy} and let $C'\subseteq C$.
	Moreover, let $c=\comp, c'=\comp[']\in C'$.
	We first show $\cmpId=\cmpId'\implies \cmpLP=\cmpLP'\land \cmpIP=\cmpIP'\land\cmpOP=\cmpOP'$.
	Thus, assume $\cmpId=\cmpId'$ and have $\cmpLP=\cmpLP'\land \cmpIP=\cmpIP'\land\cmpOP=\cmpOP'$ by Eq.~\eqref{eq:healthy}, since $c,c'\in C$.
	Now we show $\cmpId=\cmpId'\implies \forall p\in\cmpLP\colon \cmpVal(p)=\cmpVal'(p)$.
	Thus, assume $\cmpId=\cmpId'$, let $p\in \cmpLP$ and have $\cmpVal(p)=\cmpVal'(p)$ by Eq.~\eqref{eq:healthy2}, since $c,c'\in C$.
\end{proof}

\begin{example}[\Glsentrytext{gls:comp}]\label{ex:cmp}
	\begin{figure}
		\centering
		\input{img/cmp.tex}
		\caption{Conceptual representation of a \glsentrytext{gls:comp} with identifier $c_2$, local ports $l_0, l_1$, input ports $i_0,i_1,i_2$, output ports $o_0$, and corresponding valuations $\{4\}, \{C\}, \{Z\}, \{A\}, \{8\}, \{9\}$.}
		\label{fig:cmp}
	\end{figure}
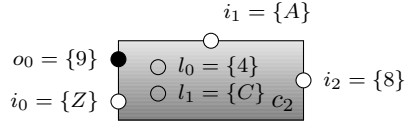
	\begin{sloppypar}
		Assuming $\mess$ consists of all characters and numbers, $c_2\in \cID$, and $l_0,l_1,i_0,i_1,i_2,o_0\in\port$.
		Figure~\ref{fig:cmp} shows a conceptual representation of a component $\comp$, with:
	\end{sloppypar}
	\begin{itemize}
		\item identifier: $\cmpId=c_2$,
		\item local ports: $\cmpLP=\{l_0,l_1\}$,
		\item input ports: $\cmpIP=\{i_0,i_1,i_2\}$,
		\item output ports: $\cmpOP=\{o_0\}$, and
		\item valuation $\cmpVal$ defined as follows:
		\begin{itemize}
			\item $\cmpVal(l_0)=\{4\}$ and $\cmpVal(l_1)=\{C\}$,
			\item $\cmpVal(i_0)=\{Z\}$, $\cmpVal(i_1)=\{A\}$, and $\cmpVal(i_2)=\{8\}$; and
			\item $\cmpVal(o_0)=\{9\}$.
		\end{itemize}
	\end{itemize}
\end{example}

\subsection{Architecture Configurations and Configuration Traces}\label{sec:mod:config}
Architecture properties are modeled as sets of \emph{\glsentryplural{gls:cnf:trace}} which are sequences over \emph{\glsentryplural{gls:aconf}}.

\subsubsection{\glsentryplural{gls:aconf}}
In our model, an \glsentrytext{gls:aconf} \emph{connects} ports of \emph{active} components.
\begin{definition}[\Glsentrytext{gls:aconf}]\label{def:aconf}
	An \emph{\gls{gls:aconf}} over a \emph{\glsentrytext{gls:cmp:healthy}} set of components $C\subseteq\cmpAll$ is a pair $\configuration$, consisting of:
	\begin{itemize}
		\item a set of active components $\cnfComp\subseteq C$ and
		\item a connection of their ports $\cnfConn\colon \cmpIn{\cnfComp}\to\pset{\cmpOut{\cnfComp}}$\enspace.
	\end{itemize}
	We require the valuation of active components of an architecture configuration to be determined by a component's identifier:
	\begin{equation}\label{eq:ac:val}
		\begin{split}
			&\forall \comp, \comp['] \in \cnfComp\colon\\
			&\quad\cmpId=\cmpId'\implies \cmpVal=\cmpVal'.
		\end{split}
	\end{equation}
	Thus, we can define a function to obtain the valuation for a component in an \glsentrytext{gls:aconf} by means of its identifier, characterized by the following equation:
	\begin{equation}\label{eq:ac:fv}\thinmuskip=0mu\medmuskip=0mu\thickmuskip=1mu
		\cnfVal{\configuration}(d)=\mu \iff \exists \cmpLP, \cmpIP, \cmpOP \subseteq\port\colon \comp\in \cnfComp\enspace.
	\end{equation}
	\begin{note}[Well-definedness of function $\cnfVal{\configuration}$]
		Function $\cnfVal{\configuration}$ is well-defined by Eq.~\eqref{eq:ac:fv}, due to Eq.~\eqref{eq:ac:val}.
	\end{note}

	For an \gls{gls:aconf} $k=\configuration$, we denote by:
	\begin{equation}
		\oIP{k} \defeq \big\{p\in \cmpIn{\cnfComp}\mid \cnfConn(p)=\emptyset\big\}\enspace,
	\end{equation}
	the set of \emph{\glspl{gls:cnf:oin}}.
		
	Moreover, we require connected ports to be consistent in their valuation, that is, if a component provides messages at its output ports, these messages are transferred to the corresponding, connected input ports:
	\begin{align}
		&\forall (d_i,p_i)\in \oIP{\configuration}\colon\\
		&\cnfVal{\configuration}(d_i)(p_i)=\bigcup_{(d_o,p_o)\in \cnfConn(d_i,p_i)} \cnfVal{\configuration}(d_o)(p_o)\enspace.\nonumber
	\end{align}
	The set of all possible \glsentryplural{gls:aconf} over a \glsentrytext{gls:cmp:healthy} set of components $C\subseteq\cmpAll$ is denoted by $\cnfAll C$.
\end{definition}
Note that a connection is modeled as a set-valued function from component input ports to component output ports, meaning that
input/output ports can be connected to several output/input ports, respectively, and
not every input/output port needs to be connected to an output/input port.
\begin{example}[\Glsentrytext{gls:aconf}]\label{ex:config}
	\begin{figure}
		\centering
		\input{img/config.tex}
		\caption{\Glsentrytext{gls:aconf} consisting of 3 components and a connection between ports $(c_2,i_1)$ and $(c_1,o_1)$, $(c_2,i_2)$ and $(c_3,o_1)$, and $(c_3,i_1)$ and $(c_1,o_2)$.}
		\label{fig:config}
	\end{figure}
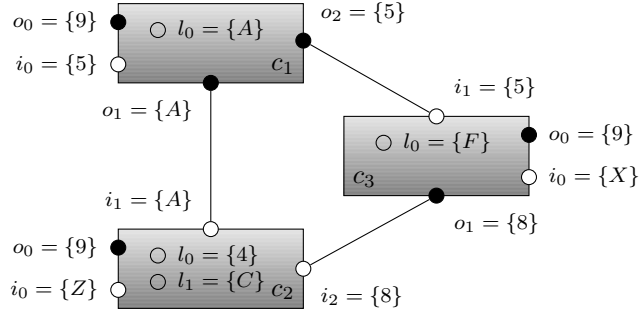	
	\begin{sloppypar}
		Assuming $\mess$ consists of all characters and numbers, $c_1,c_2,c_3\in \cID$, and $l_0,l_1,i_0,i_1,i_2,o_0,o_1,o_2\in\port$.
		Figure~\ref{fig:config} shows an architecture configuration $\configuration$, with:
	\end{sloppypar}			
	\begin{itemize}
		\item active components $\cnfComp=\{C_1,C_2,C_3\}$, where $C_2$, for example, is shown in Ex.~\ref{ex:cmp}, and
		\item connection $\cnfConn$ defined as follows:
		\begin{itemize}
			\item $\cnfConn((c_2,i_1))=\{(c_1,o_1)\}$,
			\item $\cnfConn((c_3,i_1))=\{(c_1,o_2)\}$,
			\item $\cnfConn((c_2,i_2))=\{(c_3,o_1)\}$, and
			\item $\cnfConn((c_1,i_0))=\cnfConn((c_1,o_0))=\cnfConn((c_2,i_0))=\cnfConn((c_2,o_0))=\cnfConn((c_3,i_0))=\cnfConn((c_3,o_0))=\emptyset$.
		\end{itemize}
	\end{itemize}
\end{example}

Moreover, due to the healthiness condition, an active component within an \glsentrytext{gls:aconf} is fully determined by its identifier.
\begin{property}
	For a healthy set of components $C\subseteq\cmpAll$, we have
	\begin{equation}
		\forall \configuration\in \cnfAll{C}, \forall c,c'\in\cnfComp \colon \cp{c}{1}=\cp{c'}{1}\implies c=c'\enspace.
	\end{equation}
\end{property}
\begin{proof}
	Assume $c=\comp$ and $c=\comp[']$ and $\cmpId=\cmpId'$. Then, by Eq.~\eqref{eq:healthy}, we have that $\cmpLP=\cmpLP'$, $\cmpIP=\cmpIP'$, and $\cmpOP=\cmpOP'$. Moreover, by Eq.~\eqref{eq:ac:val}, we have $\cmpVal=\cmpVal'$. Thus, we can conclude $c=c'$.
\end{proof}

\subsubsection{Configuration Traces}
A \glsentrytext{gls:cnf:trace} consists of a series of configuration snapshots of an architecture during system execution.
\begin{definition}[\Glsentrytext{gls:cnf:trace}]\label{def:atrace}
	A \emph{\gls{gls:cnf:trace}} over a \glsentrytext{gls:cmp:healthy} set of components $C\subseteq \cmpAll$ is a mapping $\NN \to \cnfAll C$.
	The set of all \glsentryplural{gls:cnf:trace} over $C$ is denoted by $\allTraces C$.
\end{definition}

\begin{example}[\Glsentrytext{gls:cnf:trace}]
	Figure~\ref{fig:trace} shows a conceptual representation of a \glsentrytext{gls:cnf:trace} $t \in \allTraces C$ with corresponding \glsentryplural{gls:aconf} $t(0)=k_0$, $t(1)=k_1$, and $t(2)=k_2$.
	\Glsentrytext{gls:aconf} $t_0$, e.g., is shown in Ex.~\ref{ex:config}.
	\begin{figure*}
		\centering
		\input{img/trace.tex}
		\caption{First three \glsentryplural{gls:aconf} $k_0$, $k_1$, and $k_2$, of a \glsentrytext{gls:cnf:trace}.}
		\label{fig:trace}
	\end{figure*}
\end{example}

Note that an architecture property is modeled as a \emph{set} of \glsentryplural{gls:cnf:trace} rather than just one single trace. This is due to the fact that input to an architecture is usually nondeterministic and the appearance and disappearance of components, as well as the reconfiguration of an architecture, may indeed depend on the input provided to it.

Moreover, note that our notion of architecture is dynamic in the following sense:
\begin{inparaenum}[(i)]
	\item \emph{components} may appear and disappear over time and
	\item \emph{connections} may change over time.
\end{inparaenum}

\subsection{Summary}
Table~\ref{tab:model} provides a brief overview of the main concepts introduced in this section.
For each concept it provides a brief description thereof and related notation.
\begin{table*}
	\centering
	\caption{Overview of concepts for dynamic architectures.\label{tab:model}}
	\begin{tabular}{r@{\hspace{10pt}}p{7cm}@{\hspace{15pt}}p{4.5cm}}
		\toprule
		\textbf{Concept} & \textbf{\textit{Description}} & \textbf{\textit{Related Notation}} \\
		\midrule
		\glsentrytext{gls:mess} & \glsentrydesc{gls:mess} & \glsentrysymbol{gls:mess}\\
		\cmidrule{2-3}
		\glsentrytext{gls:port} & \glsentrydesc{gls:port} & \glsentrysymbol{gls:port}\\
		\cmidrule{2-3}		
		\glsentrytext{gls:val} & \glsentrydesc{gls:val} & \glsentrysymbol{gls:val}\\
		\cmidrule{2-3}		
		\glsentrytext{gls:cmp:id} & \glsentrydesc{gls:cmp:id} & \glsentrysymbol{gls:cmp:id}\\
		\cmidrule{2-3}		
		\glsentrytext{gls:comp} & \glsentrydesc{gls:comp} & \glsentrysymbol{gls:comp}\\
			& \emph{local, input, output, all ports of components $C$} & \cmpLoc{C}, \cmpIn{C}, \cmpOut{C}, \cmpPort{C}\\
			& \emph{identifiers of components $C$} & \cmpName{C}\\
		\cmidrule{2-3}		
		\glsentrytext{gls:cmp:healthy} set of components & \glsentrydesc{gls:cmp:healthy} & \\
		& \emph{local, input, output, and all ports of component with identifier $d$ of healthy set $C$} & $\cmpPloc{C}(d), \cmpPin{C}(d), \cmpPout{C}(d), \cmpPport{C}(d)$\\
		& \emph{valuation of local ports of component with identifier $d$ of healthy set $C$} & $\cmpPlocVal{C}(d)$\\
		\cmidrule{2-3}
		\glsentrytext{gls:aconf} & \glsentrydesc{gls:aconf} & \glsentrysymbol{gls:aconf}\\
			& \emph{valuation of component with identifier $d$ of \glsentrytext{gls:aconf} $k$} & $\cnfVal{k}(d)$\\
		\cmidrule{2-3}		
		\glsentrytext{gls:cnf:trace} & \glsentrydesc{gls:cnf:trace} & \glsentrysymbol{gls:cnf:trace}\\
		\bottomrule
	\end{tabular}
\end{table*}

%% file: img/cmp.tex
\begin{tikzpicture}[lport/.style={circle,draw,inner sep=2pt}, iport/.style={circle,draw,,fill=white,inner sep=2pt}, oport/.style={circle,draw,fill,inner sep=2pt}, component/.style={draw,minimum height=30pt,minimum width=70pt, bottom color=black!50, top color=black!10}, port/.style={font=\small}]
		iffalse Component C2------------------------------------------------------------- fi
		\node [component] (c2) at (-1.5,-1.5) {};
		\node [above left] at (c2.south east) {$c_2$};		
		\node [lport] (c2l0) at ($(c2)+(-20pt,5pt)$) {};
		\node [port, right = 1pt of c2l0]{$l_0=\{4\}$};		
		\node [lport] (c2l1) at ($(c2)+(-20pt,-5pt)$) {};
		\node [port, right = 1pt of c2l1]{$l_1=\{C\}$};		
		\node [iport] (c2i0) at ($(c2)-(35pt,8pt)$) {};
		\node [port, left = 1pt of c2i0]{$i_0=\{Z\}$};		
		\node [oport] (c2o0) at ($(c2)+(-35pt,8pt)$) {};
		\node [port, left = 1pt of c2o0]{$o_0=\{9\}$};		
		\node [iport] (c2i1) at ($(c2)+(0pt,15pt)$) {};
		\node [port, above right = 1pt and -1pt of c2i1]{$i_1=\{A\}$};
		\node [iport] (c2i2) at ($(c2)+(35pt,0pt)$) {};
		\node [port, right = 1pt and 1pt of c2i2]{$i_2=\{8\}$};
\end{tikzpicture}

%% file: img/config.tex
\begin{tikzpicture}[lport/.style={circle,draw,inner sep=2pt}, oport/.style={circle,draw,fill,inner sep=2pt}, iport/.style={circle,draw,fill=white,inner sep=2pt}, component/.style={draw,minimum height=30pt,minimum width=70pt, bottom color=black!50, top color=black!10}, port/.style={font=\small}]
		iffalse Component C1------------------------------------------------------------- fi
		\node [component] (c1) at (-1.5,1.5) {};
		\node [above left] at (c1.south east) {$c_1$};
		\node [lport] (c1l0) at ($(c1)+(-20pt,5pt)$) {};
		\node [port, right = 1pt of c1l0]{$l_0=\{A\}$};		
		\node [oport] (c1o0) at ($(c1)+(-35pt,8pt)$) {};
		\node [port, left = 1pt of c1o0]{$o_0=\{9\}$};
		\node [iport] (c1i0) at ($(c1)+(-35pt,-8pt)$) {};
		\node [port, left = 1pt of c1i0]{$i_0=\{5\}$};
		\node [oport] (c1o1) at ($(c1)-(0pt,15pt)$) {};
		\node [port, below left = 1pt and 1pt of c1o1]{$o_1=\{A\}$};
		\node [oport] (c1o2) at ($(c1)+(35pt,1pt)$) {};
		\node [port, above right = 1pt and 1pt of c1o2]{$o_2=\{5\}$};
		iffalse Component C2------------------------------------------------------------- fi
		\node [component] (c2) at (-1.5,-1.5) {};
		\node [above left] at (c2.south east) {$c_2$};		
		\node [lport] (c2l0) at ($(c2)+(-20pt,5pt)$) {};
		\node [port, right = 1pt of c2l0]{$l_0=\{4\}$};		
		\node [lport] (c2l1) at ($(c2)+(-20pt,-5pt)$) {};
		\node [port, right = 1pt of c2l1]{$l_1=\{C\}$};		
		\node [iport] (c2i0) at ($(c2)-(35pt,8pt)$) {};
		\node [port, left = 1pt of c2i0]{$i_0=\{Z\}$};		
		\node [oport] (c2o0) at ($(c2)+(-35pt,8pt)$) {};
		\node [port, left = 1pt of c2o0]{$o_0=\{9\}$};		
		\node [iport] (c2i1) at ($(c2)+(0pt,15pt)$) {};
		\node [port, above left = 1pt and 1pt of c2i1]{$i_1=\{A\}$};
		\node [iport] (c2i2) at ($(c2)+(35pt,0pt)$) {};
		\node [port, below right = 1pt and 1pt of c2i2]{$i_2=\{8\}$};
		\draw (c1o1) edge (c2i1) ;
		iffalse Component C3------------------------------------------------------------- fi
		\node [component] (c3) at (1.5,0) {};
		\node [above right] at (c3.south west) {$c_3$};
		\node [lport] (c3l0) at ($(c3)+(-20pt,5pt)$) {};
		\node [port, right = 1pt of c3l0]{$l_0=\{F\}$};		
		\node [iport] (c3i0) at ($(c3)+(35pt,-8pt)$) {};
		\node [port, right = 1pt of c3i0]{$i_0=\{X\}$};		
		\node [oport] (c3o0) at ($(c3)+(35pt,8pt)$) {};
		\node [port, right = 1pt of c3o0]{$o_0=\{9\}$};		
		\node [iport] (c3i1) at ($(c3)+(0pt,15pt)$) {};
		\node [port, above right = 1pt and 1pt of c3i1]{$i_1=\{5\}$};
		\node [oport] (c3o1) at ($(c3)+(0pt,-15pt)$) {};
		\node [port, below right = 1pt and 1pt of c3o1]{$o_1=\{8\}$};
		\draw (c1o2) edge (c3i1) ;
		\draw (c3o1) edge (c2i2) ;
\end{tikzpicture}

%% file: img/trace.tex
\begin{tikzpicture}[scale=0.7, every node/.style={transform shape}, lport/.style={circle,draw,inner sep=2pt}, oport/.style={circle,fill,inner sep=2pt},iport/.style={circle,draw,fill=white,inner sep=2pt}, component/.style={draw,minimum height=30pt,minimum width=70pt, bottom color=black!50, top color=black!10}, port/.style={font=\small}]
	iffalse Conf 1--------------------------------------------------------------- fi
		iffalse Component C1------------------------------------------------------------- fi
		\node [component] (c1) at (-8.5,1.5) {};
		\node [above left] at (c1.south east) {$c_1$};
		\node [lport] (c1l0) at ($(c1)+(-20pt,5pt)$) {};
		\node [port, right = 1pt of c1l0]{$l_0=\{A\}$};		
		\node [oport] (c1o0) at ($(c1)+(-35pt,8pt)$) {};
		\node [port, left = 1pt of c1o0]{$o_0=\{9\}$};
		\node [iport] (c1i0) at ($(c1)+(-35pt,-8pt)$) {};
		\node [port, left = 1pt of c1i0]{$i_0=\{5\}$};
		\node [oport] (c1o1) at ($(c1)-(0pt,15pt)$) {};
		\node [port, below left = 1pt and 1pt of c1o1]{$o_1=\{A\}$};
		\node [oport] (c1o2) at ($(c1)+(35pt,1pt)$) {};
		\node [port, above right = 1pt and 1pt of c1o2]{$o_2=\{5\}$};
		iffalse Component C2------------------------------------------------------------- fi
		\node [component] (c2) at (-8.5,-1.5) {};
		\node [above left] at (c2.south east) {$c_2$};		
		\node [lport] (c2l0) at ($(c2)+(-20pt,5pt)$) {};
		\node [port, right = 1pt of c2l0]{$l_0=\{4\}$};		
		\node [lport] (c2l1) at ($(c2)+(-20pt,-5pt)$) {};
		\node [port, right = 1pt of c2l1]{$l_1=\{C\}$};		
		\node [iport] (c2i0) at ($(c2)-(35pt,8pt)$) {};
		\node [port, left = 1pt of c2i0]{$i_0=\{Z\}$};		
		\node [oport] (c2o0) at ($(c2)+(-35pt,8pt)$) {};
		\node [port, left = 1pt of c2o0]{$o_0=\{9\}$};		
		\node [iport] (c2i1) at ($(c2)+(0pt,15pt)$) {};
		\node [port, above left = 1pt and 1pt of c2i1]{$i_1=\{A\}$};
		\node [iport] (c2i2) at ($(c2)+(35pt,0pt)$) {};
		\node [port, below right = 1pt and 1pt of c2i2]{$i_2=\{8\}$};
		\draw (c1o1) edge (c2i1) ;
		iffalse Component C3------------------------------------------------------------- fi
		\node [component] (c3) at (-6.5,0) {};
		\node [above right] at (c3.south west) {$c_3$};
		\node [lport] (c3l0) at ($(c3)+(-20pt,5pt)$) {};
		\node [port, right = 1pt of c3l0]{$l_0=\{F\}$};		
		\node [iport] (c3i0) at ($(c3)+(35pt,-8pt)$) {};
		\node [port, right = 1pt of c3i0]{$i_0=\{X\}$};		
		\node [oport] (c3o0) at ($(c3)+(35pt,8pt)$) {};
		\node [port, right = 1pt of c3o0]{$o_0=\{9\}$};		
		\node [iport] (c3i1) at ($(c3)+(0pt,15pt)$) {};
		\node [port, above right = 1pt and 1pt of c3i1]{$i_1=\{5\}$};
		\node [oport] (c3o1) at ($(c3)+(0pt,-15pt)$) {};
		\node [port, below right = 1pt and 1pt of c3o1]{$o_1=\{8\}$};
		\draw (c1o2) edge (c3i1) ;
		\draw (c3o1) edge (c2i2) ;
		\node [above left = 10pt and 20pt of c1] {\huge $k_0$};
		\node (comma) at (-3.5,0) {\Huge ,};
	iffalse Conf 2--------------------------------------------------------------- fi
		iffalse Comp C1-------------------------------------------------------------- fi
		\node [component] (c1) at (-0.5,1.5) {};
		\node [above left] at (c1.south east) {$c_1$};		
		\node [lport] (c1l0) at ($(c1)+(-20pt,5pt)$) {};
		\node [port, right = 1pt of c1l0]{$l_0=\{A\}$};		
		\node [oport] (c1o0) at ($(c1)+(-35pt,8pt)$) {};
		\node [port, left = 1pt of c1o0]{$o_0=\{1\}$};
		\node [iport] (c1i0) at ($(c1)+(-35pt,-8pt)$) {};
		\node [port, left = 1pt of c1i0]{$i_0=\{8\}$};
		\node [oport] (c1o1) at ($(c1)-(0pt,15pt)$) {};
		\node [port, below left = 1pt and 1pt of c1o1]{$o_1=\{G\}$};
		\node [oport] (c1o2) at ($(c1)+(35pt,1pt)$) {};
		\node [port, above right = 1pt and 1pt of c1o2]{$o_2=\{7\}$};
		iffalse Comp C2--------------------------------------------------------------- fi
		\node [component] (c2) at (-0.5,-1.5) {};
		\node [above left] at (c2.south east) {$c_2$};
		\node [lport] (c2l0) at ($(c2)+(-20pt,5pt)$) {};
		\node [port, right = 1pt of c2l0]{$l_0=\{4\}$};		
		\node [lport] (c2l1) at ($(c2)+(-20pt,-5pt)$) {};
		\node [port, right = 1pt of c2l1]{$l_1=\{C\}$};		
		\node [iport] (c2i0) at ($(c2)-(35pt,8pt)$) {};
		\node [port, left = 1pt of c2i0]{$i_0=\{G\}$};		
		\node [oport] (c2o0) at ($(c2)+(-35pt,8pt)$) {};
		\node [port, left = 1pt of c2o0]{$o_0=\{1\}$};		
		\node [iport] (c2i1) at ($(c2)+(0pt,15pt)$) {};
		\node [port, above left = 1pt and 1pt of c2i1]{$i_1=\{F\}$};
		\node [iport] (c2i2) at ($(c2)+(35pt,0pt)$) {};
		\node [port, below right = 1pt and 1pt of c2i2]{$i_2=\{8\}$};
		\draw (c1o1) edge (c2i1) ;

		\node [above left = 10pt and 20pt of c1] {\huge $k_1$};
		\node (comma) at (2.5,0) {\Huge ,};		
	iffalse Conf 3--------------------------------------------------------------- fi
		iffalse Comp C1--------------------------------------------------------------- fi
		\node [component] (c1) at (7,1.5) {};
		\node [above left] at (c1.south east) {$c_1$};		
		\node [lport] (c1l0) at ($(c1)+(-20pt,5pt)$) {};
		\node [port, right = 1pt of c1l0]{$l_0=\{A\}$};		
		\node [oport] (c1o0) at ($(c1)+(-35pt,8pt)$) {};
		\node [port, left = 1pt of c1o0]{$o_0=\{4\}$};
		\node [iport] (c1i0) at ($(c1)+(-35pt,-8pt)$) {};
		\node [port, below right = 3pt and -20pt of c1i0]{$i_0=\{6\}$};
		\node [oport] (c1o1) at ($(c1)-(0pt,15pt)$) {};
		\node [port, below right = 1pt and 1pt of c1o1]{$o_1=\{K\}$};
		\node [oport] (c1o2) at ($(c1)+(35pt,1pt)$) {};
		\node [port, above right = 1pt and 1pt of c1o2]{$o_2=\{9\}$};
		iffalse Comp C2--------------------------------------------------------------- fi
		\node [component] (c2) at (7,-1.5) {};
		\node [above left] at (c2.south east) {$c_2$};
		\node [lport] (c2l0) at ($(c2)+(-20pt,5pt)$) {};
		\node [port, right = 1pt of c2l0]{$l_0=\{4\}$};		
		\node [lport] (c2l1) at ($(c2)+(-20pt,-5pt)$) {};
		\node [port, right = 1pt of c2l1]{$l_1=\{C\}$};		
		\node [iport] (c2i0) at ($(c2)-(35pt,8pt)$) {};
		\node [port, left = 1pt of c2i0]{$i_0=\{R\}$};		
		\node [oport] (c2o0) at ($(c2)+(-35pt,8pt)$) {};
		\node [port, above right = 3pt and -20pt of c2o0]{$o_0=\{2\}$};		
		\node [iport] (c2i1) at ($(c2)+(0pt,15pt)$) {};
		\node [port, above right = 1pt and 1pt of c2i1]{$i_1=\{W\}$};
		\node [iport] (c2i2) at ($(c2)+(35pt,0pt)$) {};
		\node [port, below right = 1pt and 1pt of c2i2]{$i_2=\{6\}$};
		iffalse Comp C3--------------------------------------------------------------- fi
		\node [component] (c3) at (4,0) {};
		\node [above right] at (c3.south west) {$c_4$};
		\node [lport] (c3l0) at ($(c3)+(-20pt,5pt)$) {};
		\node [port, right = 1pt of c3l0]{$l_0=\{4\}$};		
		\node [iport] (c3i0) at ($(c3)+(35pt,-8pt)$) {};
		\node [port, right = 1pt of c3i0]{$i_0=\{T\}$};		
		\node [oport] (c3o0) at ($(c3)+(35pt,8pt)$) {};
		\node [port, right = 1pt of c3o0]{$o_0=\{4\}$};		
		\node [iport] (c3i1) at ($(c3)+(0pt,15pt)$) {};
		\node [port, above left = 1pt and 1pt of c3i1]{$i_1=\{5\}$};
		\node [oport] (c3o1) at ($(c3)+(0pt,-15pt)$) {};
		\node [port, below left = 1pt and 1pt of c3o1]{$o_1=\{B\}$};
		\draw (c1o0) edge (c3i1);
		\draw (c3o1) edge (c2i0);

		\node [above left = 10pt and 60pt of c1] {\huge $k_2$};
		\node (comma) at (9.5,0) {\Huge ,};		
	iffalse dots--------------------------------------------------------------- fi
		\node [circle,fill] (dot) at (10,0) {};
		\node [circle,fill] (dot) at (11,0) {};
		\node [circle,fill] (dot) at (12,0) {};		
\end{tikzpicture}

%% file: specification.tex
\section{Datatype Specifications}\label{sec:ds}
\Glsentryplural{gls:datatype} are specified by means of algebraic specifications~\cite{Broy1996,Wirsing1991}.

Thus, a \glsentrytext{gls:dt:spec} is expressed over a signature by means of a set of so-called \glsentryplural{gls:dt:assert}, i.e., predicate-logic formulas over \glsentryplural{gls:dt:term}.
Meaning is provided in terms of a corresponding algebra, i.e., concrete mathematical structures for the sorts and functions of the corresponding signature.

\subsection{\glsentryplural{gls:sig}}
A \glsentrytext{gls:sig} determines the symbols used throughout the specification.
Sorts are symbols representing certain sets of messages while function symbols and predicate symbols, represent functions, and predicates over those sets.
\begin{definition}[\Glsentrytext{gls:sig}]
	A \emph{\gls{gls:sig}} is a triple $\Sigma=\signature$, consisting of:
	\begin{itemize}
		\item a set of sorts $\sigSort$,
		\item a set of function symbols $\sigFun$ and predicate symbols $\sigPred$ with corresponding assignments $\sftype\colon \sigFun\to\cProd{\sigSort}{n}$ and $\sftype\colon \sigPred\to\cProd{\sigSort}{n}$, with:
		\begin{itemize}
			\item \fsort{n}/\psort{n} denoting the set of function/predicate symbols with arity $n\in\NN$,
			\item $\sftype[n](f)$/$\sptype[n](b)$ denoting the sort of the $n$-th parameter (with $n\in\NNP$) of function symbol $f\in\sigFun$ / predicate symbol $b\in\sigPred$, and
			\item $\sftype[0](f)$ denoting the sort of the return value of function symbol $f\in\sigFun$.
		\end{itemize}
	\end{itemize}
\end{definition}

\subsection{\glsentryplural{gls:alg}}
The meaning of the symbols introduced by a \glsentrytext{gls:sig} is determined by an \glsentrytext{gls:alg}. An \glsentrytext{gls:alg} consists of concrete sets for each sort-symbol and corresponding functions and predicates for the function-symbols and predicate-symbols, respectively. Moreover, mappings associate each symbol with the corresponding interpretation.
\begin{sloppypar}
\begin{definition}[\Glsentrytext{gls:alg}]%
	An \emph{\gls{gls:alg}} for a \glsentrytext{gls:sig} $\signature$ is a $6$-tuple $\algebra$, consisting of:
	\begin{itemize}
		\item a set of \emph{non-empty} sets of messages $\algSet\subseteq \pset{\mess}$\enspace;
		\item a set of functions $\algFun$ and predicates $\algPred$ for symbols $\sigFun$ and $\sigPred$, respectively; and
		\item interpretations $\alpha\colon\sigSort\to\algSet$, $\beta\colon\sigFun\to\algFun$, and $\gamma\colon\sigPred\to\algPred$\enspace.
	\end{itemize}
	The set of all possible \glsentryplural{gls:alg} over \glsentrytext{gls:sig} $\Sigma$ is denoted by $\algAll \Sigma$.
\end{definition}
\end{sloppypar}

\subsection{\glsentryplural{gls:dt:term}}
Terms of \glsentryplural{gls:dt:spec} are build over a given \glsentrytext{gls:sig} and corresponding \emph{\glspl{gls:dt:var}} (a family of disjoint sets of variables $\dtVar=(\dtVar_s)_{s\in\sigSort}$ with $\dtVar_s$ denoting a set of variables of sort $s\in\sigSort$).
\begin{sloppypar}
\begin{definition}[\Glsentrytext{gls:dt:term}]
	The set of all \emph{\glspl{gls:dt:term}} of sort $s\in\sigSort$ over a \glsentrytext{gls:sig} $\Sigma=\signature$ and \glsentryplural{gls:dt:var} $\dtVar$ is the smallest set $\dtTerms[s] \Sigma \dtVar$, satisfying the equations in Fig.~\ref{fig:DS:syn:t}.
	\begin{figure*}
		\begin{mymathbox}{\Glsentryplural{gls:dt:term}: syntax}
			v\in \dtVar_s &\implies v\in \dtTerms[s] \Sigma \dtVar\enspace,\\
			f\in \sigFun^0 &\implies f\in \dtTerms[s] \Sigma \dtVar~\textrm{[for $\sftype[0](f)=s$]}\enspace,\\
			f\in \sigFun^{n+1} \land
			t_1\in\dtTerms[s_1]\Sigma \dtVar, \cdots, t_{n+1} \in \dtTerms[s_{n+1}]\Sigma \dtVar &\implies f(t_1, \cdots, t_{n+1})\in \dtTerms[s] \Sigma \dtVar~\textrm{[for $n\in\NN$, $\sftype[0](f)=s$,}\\&\qqquad \textrm{and $\sftype[1](f)=s_1,\cdots, \sftype[n+1](f)=s_{n+1}$]}\enspace.
		\end{mymathbox}
		\caption{Inductive definition of \glsentryplural{gls:dt:term} $\dtTerms[s]{\Sigma}{\dtVar}$ of sort $s\in\sigSort$ over signature $\Sigma=\signature$ and \glsentryplural{gls:dt:var} $\dtVar=(\dtVar_s)_{s\in \sigSort}$\enspace.}\label{fig:DS:syn:t}
	\end{figure*}
	The set of all \glsentryplural{gls:dt:term} (of all sorts) over a signature $\Sigma=\signature$ and \glsentryplural{gls:dt:var} $\dtVar$ is denoted by $\dtTerms \Sigma \dtVar$.	
\end{definition}
\end{sloppypar}

Roughly speaking, a \glsentrytext{gls:dt:term} is the application of function symbols of a \glsentrytext{gls:sig} to other function symbols or variables. Thereby, the sorts of the parameters have to be consistent with the sorts of the corresponding function symbol. 

The semantics of \glsentryplural{gls:dt:term} is defined over an \glsentrytext{gls:alg} $A=\algebra$ and corresponding \emph{\gls{gls:dt:va}} (a family of mappings $\dtVI=(\dtVI_s)_{s\in\sigSort}$ with $\dtVI_s\colon \dtVar_s\to \alpha(s)$ for each sort $s\in\sigSort$).
In the following we denote with $\dtVAll{\dtVar}{A}$ the set of all \glsentryplural{gls:dt:va} for \glsentryplural{gls:dt:var} $\dtVar$ in \glsentrytext{gls:alg} $A$.

\begin{sloppypar}
\begin{definition}[\Glsentrytext{gls:dt:sem}]
	The \emph{\gls{gls:dt:sem}} for \glsentryplural{gls:dt:term} $\dtTerms{\Sigma}{\dtVar}$ with \glsentrytext{gls:sig} $\Sigma=\signature$ and \glsentryplural{gls:dt:var} $\dtVar$,  over \glsentrytext{gls:alg} $A\in\algAll{\Sigma}$ and \glsentrytext{gls:dt:va} $\dtVI\in\dtVAll{\dtVar}{A}$ is the mapping $\dtsem{\_}{\dtVI}{A}\colon\dtTerms[s]{\Sigma}{\dtVar}\to\intSort(s)$ (for each sort $s\in\sigSort$), characterized by the equations in Fig.~\ref{fig:DS:sem:t}.
	\begin{figure*}
		\begin{mymathbox}[3cm]{\Glsentryplural{gls:dt:term}: semantics}
			\dtsem{v}{\dtVI}{A} &\defeq \dtVI_s(v)~\textrm{[for $v\in \dtVar_s$]}\enspace,\\
			\dtsem{f}{\dtVI}{A} &\defeq \beta(f)~\textrm{[for function symbol $f\in\sigFun^0$]}\enspace,\\		
			\dtsem{f(t_1, \cdots, t_n)}{\dtVI}{A} &\defeq \beta(f)(\dtsem{t_1}{\dtVI}{A}, \cdots, \dtsem{t_n}{\dtVI}{A})~\textrm{[for function symbol $f\in F^{n+1}$]}\enspace.
		\end{mymathbox}
		\caption{Recursive definition of \glsentrytext{gls:dt:sem} for \glsentryplural{gls:dt:term} $\dtTerms[s]{\Sigma}{\dtVar}$ of sort $s\in\sigSort$ with signature $\Sigma=\signature$, algebra $A=\algebra\in\algAll{\Sigma}$, and \glsentrytext{gls:dt:va} $\dtVI=(\dtVI_s)_{s\in\sigSort}$.}\label{fig:DS:sem:t}
	\end{figure*}	
\end{definition}
\end{sloppypar}
Thus, the semantics of a \glsentrytext{gls:dt:term} is given by a function assigning a value of the corresponding \glsentrytext{gls:alg} to each term.

\subsection{\glsentryplural{gls:dt:assert}}
\Glsentryplural{gls:dt:assert} are build over \glsentryplural{gls:dt:term} by the common logical operators.
\begin{definition}[\Glsentrytext{gls:dt:assert}]
	The set of all \emph{\glspl{gls:dt:assert}} over a signature $\Sigma$ and \glsentryplural{gls:dt:var} $\dtVar$ is the smallest set $\dtFormula \Sigma \dtVar$ satisfying the equations in Fig.~\ref{fig:DS:syn:f}.
	\begin{figure*}
		\begin{mymathbox}{\Glsentryplural{gls:dt:assert}: syntax}
			b\in \sigPred^0 &\implies b\in \dtFormula \Sigma \dtVar\enspace,\\
			b\in \sigPred^{n+1} \land t_1\in\dtTerms[s_1]\Sigma \dtVar, \cdots, t_{n+1} \in \dtTerms[s_{n+1}]\Sigma \dtVar &\implies b(t_1, \cdots, t_{n+1})\in \dtFormula \Sigma \dtVar\\&\qqquad \textrm{[for $n\in\NN$ and $\sftype[1](b)=s_1,\cdots, \sftype[n+1](b)=s_{n+1}$]}\enspace,\\
			t,t'\in \dtTerms[s] \Sigma \dtVar &\implies \dteq{t}{t'}\in \dtFormula \Sigma \dtVar\enspace,\\
			E\in \dtFormula \Sigma \dtVar &\implies \dtneg{E} \in \dtFormula \Sigma \dtVar\enspace,\\	
			E, E'\in \dtFormula \Sigma \dtVar &\implies \dtland{E}{E'},~\dtlor{E}{E'},~\dtimplies{E}{E'},~E \dtiff E' \in \dtFormula \Sigma \dtVar\enspace,\\
			E\in \dtFormula \Sigma \dtVar \land x\in\dtVar_s &\implies \dtforall{x}~E\in\dtFormula \Sigma \dtVar\enspace\land\\
			&\qqquad\dtexists{x}~E\in\dtFormula \Sigma \dtVar~\textrm{[for some $s\in S$]}\enspace.
		\end{mymathbox}
		\caption{Inductive definition of \glsentryplural{gls:dt:assert} $\dtFormula{\Sigma}{\dtVar}$ over signature $\Sigma=\signature$ and \glsentryplural{gls:dt:var} $\dtVar=(\dtVar_s)_{s\in \sigSort}$\enspace.}\label{fig:DS:syn:f}
	\end{figure*}
\end{definition}

Thus, a \glsentrytext{gls:dt:assert} is obtained by applying the common logical operators to \glsentryplural{gls:dt:term} or predicate symbols.
The semantics of \glsentryplural{gls:dt:assert} is defined over an \glsentrytext{gls:alg}.

\begin{sloppypar}
\begin{definition}[\Glsentrytext{gls:dt:ass:sem}]
	The \emph{\gls{gls:dt:ass:sem}} for \glsentryplural{gls:dt:assert} $\dtFormula \Sigma \dtVar$ with \glsentrytext{gls:sig} $\Sigma$, \glsentryplural{gls:dt:var} \dtVar, and \glsentrytext{gls:alg} $A\in\algAll{\Sigma}$ is the relation $\dtmod[\_]{A}{\_}\subseteq\dtVAll{\dtVar}{A}\times\dtFormula \Sigma \dtVar$ characterized by the equations in Fig.~\ref{fig:DS:sem:f}.		
	A \glsentrytext{gls:dt:assert} $\varphi\in\dtFormula \Sigma \dtVar$ is \emph{valid} for an \glsentrytext{gls:alg} $A\in\algAll{\Sigma}$ iff there exists a \glsentrytext{gls:dt:va} $\dtVI\in\dtVAll{\dtVar}{A}$, such that $\dtmod[\dtVI]{A}{\varphi}$.
	An \glsentrytext{gls:alg} $A\in\algAll{\Sigma}$ is a \emph{model} for \glsentrytext{gls:dt:assert} $\varphi\in\dtFormula \Sigma \dtVar$, written $\dtmod{A}{\varphi}$ iff $\dtmod[\dtVI]{A}{\varphi}$ for each $\dtVI\in\dtVAll{\dtVar}{A}$.
	An algebra $A\in\algAll{\Sigma}$ is a model for a set of \glsentryplural{gls:dt:assert} $\Phi\subseteq\dtFormula \Sigma \dtVar$, written $\dtmod{A}{\Phi}$ iff $\dtmod{A}{\varphi}$ for each $\varphi\in\Phi$.
	\begin{figure*}
		\begin{mymathbox}[3cm]{\Glsentryplural{gls:dt:assert}: semantics}
			\dtmod[\dtVI]{A}{b} &\iff \gamma(b)~\textrm{[for $b\in\sigPred^0$]}\enspace,\\	
			\dtmod[\dtVI]{A}{b(t_1,\cdots,t_n)} &\iff \gamma(b)(\dtsem{t_1}{\dtVI}{A}, \cdots, \dtsem{t_n}{\dtVI}{A})~\textrm{[for $b\in\sigPred^{n+1}$]}\enspace,\\
			\dtmod[\dtVI]{A}{\dteq{t}{t'}} &\iff \dtsem{t}{\dtVI}{A}=\dtsem{t'}{\dtVI}{A}\enspace,\\
			\dtmod[\dtVI]{A}{\dtland{E}{E'}} &\iff \dtmod[\dtVI]{A}{E} \land \dtmod[\dtVI]{A}{E'}\enspace,\\
			\dtmod[\dtVI]{A}{\dtlor{E}{E'}} &\iff
			\dtmod[\dtVI]{A}{E} \lor \dtmod[\dtVI]{A}{E'}\enspace,\\
			\dtmod[\dtVI]{A}{\dtimplies{E}{E'}} &\iff
			\dtmod[\dtVI]{A}{E} \implies \dtmod[\dtVI]{A}{E'}\enspace,\\
			\dtmod[\dtVI]{A}{E \dtiff E'} &\iff
			\dtmod[\dtVI]{A}{E} \iff \dtmod[\dtVI]{A}{E'}\enspace,\\
			\dtmod[\dtVI]{A}{\dtexists{x}~E} &\iff \exists x'\in \alpha(s)\colon \dtmod[\fupdate{\dtVI}{s}{x}{x'}]{A}{E}~\textrm{[for $s\in\sigSort$ and $x\in\dtVar_s$]}\enspace,\\
			\dtmod[\dtVI]{A}{\dtforall{x}~E} &\iff \forall x'\in \alpha(s)\colon \dtmod[\fupdate{\dtVI}{s}{x}{x'}]{A}{E}~\textrm{[for $s\in\sigSort$ and $x\in\dtVar_s$]}\enspace.
		\end{mymathbox}
		\caption{Recursive definition of models relation for \glsentryplural{gls:dt:assert} $\dtFormula \Sigma \dtVar$ with signature $\Sigma=\signature$, algebra $A=\algebra\in\algAll{\Sigma}$, and \glsentrytext{gls:dt:va} $\dtVI=(\dtVI_s)_{s\in\sigSort}$.}\label{fig:DS:sem:f}
	\end{figure*}
\end{definition}
\end{sloppypar}
Thus, the semantics is given by a relation over \glsentryplural{gls:dt:assert} and \glsentryplural{gls:alg} with a corresponding \glsentrytext{gls:dt:va} satisfying the assertions.

\subsection{Specifying \glsentryplural{gls:datatype}}
\Glsentryplural{gls:sig} introduce the basic symbols used throughout the whole specification process and \glsentryplural{gls:dt:spec} provide meaning to these symbols.
\begin{sloppypar}
	\begin{definition}[\Glsentrytext{gls:dt:spec}]
	A \gls{gls:dt:spec} over a \glsentrytext{gls:sig} $\Sigma=\signature$ and a family of \glsentryplural{gls:dt:var} $\dtVar=(\dtVar_s)_{s\in \sigSort}$ is a set of \glsentryplural{gls:dt:assert} $\dtspec\subseteq\dtFormula \Sigma \dtVar$.
\end{definition}
\end{sloppypar}

\Glsentryplural{gls:sig} and corresponding \glsentryplural{gls:dt:spec} can be expressed by means of \glspl{gls:dt:spec:temp} (Fig.~\ref{fig:dtspec}).
\begin{figure}
	\centering
	\begin{dtstmp}{Name}{dtSpec}{$\mathtt{Sort1}$, $\mathtt{Sort2}$}
		\symb{$\mathit{symbol1}$}{$\mathtt{Sort1}$}
		\symb{$\mathit{symbol2}$}{$\mathtt{Sort1} \to \mathtt{Sort2}$}
		\sline
		\var{$\mathit{var1}$, $\mathit{var2}$}{$\mathtt{Sort1}$}
		\var{$\mathit{var3}$}{$\mathtt{Sort2}$}			
		\sline[dashed]
		\axiom{$\mathit{assertion1}(\mathit{symbol1}, \mathit{var1}, \mathit{var2}, \mathit{var4})$}
		\axiom{$\mathit{assertion2}(\mathit{symbol1}, \mathit{symbol2}, \mathit{var1}, \mathit{var4})$}
	\end{dtstmp}
	\caption{Datatype specification template with corresponding sorts, \glsentryplural{gls:dt:var}, symbols for functions and predicates, and \glsentryplural{gls:dt:assert}.}
	\label{fig:dtspec}
\end{figure}
Each template has a name and can import other \glsentryplural{gls:dt:spec:temp} by means of their name.
Sorts are introduced by a list of names at the beginning of the template.
Then, a list of variables for the different sorts are defined and function/predicate symbols are introduced with the corresponding types.
Finally, a list of \glsentryplural{gls:dt:assert} is specified to describe the characteristic properties of a \glsentrytext{gls:datatype}.

\subsection{\glsentrytext{gls:bb:p}: \glsentrytext{gls:dt:spec}}
\Glsentrytext{gls:bb:p} architectures work with \emph{problems} and \emph{solutions} for these problems.
Figure~\ref{fig:bb:dt} provides the corresponding \glsentrytext{gls:dt:spec:temp}.
We denote with $\mathtt{PROB}$ the set of all problems and with $\mathtt{SOL}$ the set of all solutions.
Complex problems consist of \emph{subproblems} which can be complex themselves. To solve a problem, its subproblems have to be solved first. Therefore, we assume the existence of a \emph{subproblem relation} $\prec~\subseteq~\mathtt{PROB}\times\mathtt{PROB}$.
For complex problems, this relation may not be known in advance. Indeed, one of the benefits of a Blackboard architecture is that a problem can be solved also without knowing this relation in advance.
However, the subproblem relation has to be well-founded\footnote{A partial order is well-founded iff it does not contain any infinite decreasing chains. A detailed definition can be found e.g. in~\cite{Hungerford2012}.} (Eq.~\eqref{eq:bb:dt:1}) for a problem to be solvable. In particular, we do not allow cycles in the transitive closure of $\prec$.
While there may be different approaches to solve a certain problem (i.e.\ several ways to split a problem into subproblems), we assume (without loss of generality) that the final solution for a problem is always unique. 
Thus, we assume the existence of a function $\mathit{solve}\colon \mathtt{PROB}\to \mathtt{SOL}$ which assigns the \emph{correct} solution to each problem. Note, however, that this function is not known in advance and it is one of the reasons of using this pattern to calculate this function.
\begin{figure}
	\centering
	\begin{dtstmp}[200]{ProbSol}{SET}{$\mathtt{PROB}$, $\mathtt{SOL}$}
		\symb{$\prec$}{$\mathtt{PROB}\times\mathtt{PROB}$}
		\symb{$\mathit{solve}$}{$\mathtt{PROB}\to \mathtt{SOL}$}
		\sline
		\axiom[eq:bb:dt:1]{$\mathit{well-founded}(\prec)$}
	\end{dtstmp}
	\caption{Blackboard \glsentrytext{gls:dt:spec:temp} introducing sorts, function symbols, and predicate symbols for \glsentrytext{gls:bb:p} architectures.}
	\label{fig:bb:dt}
\end{figure}

\subsection{Summary}
To conclude, Tab.~\ref{tab:dt} provides a brief overview of the main concepts introduced in this section.
For each concept it provides a brief description and related notation.
\begin{table*}
	\centering
	\caption{Overview of concepts for \glsentryplural{gls:dt:spec}.\label{tab:dt}}
	\begin{tabular}{r@{\hspace{10pt}}p{8cm}@{\hspace{15pt}}p{3cm}}
		\toprule
		\textbf{Concept} & \textbf{\textit{Description}} & \textbf{\textit{Related Notation}} \\
		\midrule
		\glsentrytext{gls:sig} & \glsentrydesc{gls:sig} & \glsentrysymbol{gls:sig}\\
			&\emph{$n$-ary function/predicate symbols} &  $\fsort{n}/\psort{n}$\\
			&\emph{sort of $n$-th parameter of function symbol $f$ / predicate symbol $b$} &$\sftype[n](f)$ / $\sptype[n](b)$\\
			&\emph{sort of return value of function symbol $f$} & $\sftype[0](f)$\\
		\cmidrule{2-3}
		\glsentrytext{gls:alg} & \glsentrydesc{gls:alg} & \glsentrysymbol{gls:alg}\\
		\cmidrule{2-3}
		\glsentrytext{gls:dt:var} & \glsentrydesc{gls:dt:var} & \glsentrysymbol{gls:dt:var}\\
		\cmidrule{2-3}
		\glsentrytext{gls:dt:va} & \glsentrydesc{gls:dt:va} & \glsentrysymbol{gls:dt:va}\\
		\cmidrule{2-3}
		\glsentrytext{gls:dt:term} & \glsentrydesc{gls:dt:term} & \glsentrysymbol{gls:dt:term}\\
		\cmidrule{2-3}
		\glsentrytext{gls:dt:sem} & \glsentrydesc{gls:dt:sem} & \glsentrysymbol{gls:dt:sem}\\
		\cmidrule{2-3}		
		\glsentrytext{gls:dt:assert} & \glsentrydesc{gls:dt:assert} & \glsentrysymbol{gls:dt:assert}\\
		\cmidrule{2-3}		
		\glsentrytext{gls:dt:ass:sem} & \glsentrydesc{gls:dt:ass:sem} & \glsentrysymbol{gls:dt:ass:sem}\\
		\cmidrule{2-3}		
		\glsentrytext{gls:dt:spec} & \glsentrydesc{gls:dt:spec} & \glsentrysymbol{gls:dt:spec}\\
		\cmidrule{2-3}		
		\glsentrytext{gls:dt:spec:temp} & \glsentrydesc{gls:dt:spec:temp} & \glsentrysymbol{gls:dt:spec:temp}\\		
		\bottomrule
	\end{tabular}
\end{table*}

%% file: interfaces.tex
\section{\glsentryplural{gls:ifspec}}\label{sec:ifspec}
Interfaces are specified over a given signature and declare a set of local, input, and output ports for a set of \glsentryplural{gls:if:id}.
Moreover, an interface specification allows to specify valuations of local ports by means of interface assertions formulated over interface terms.

Thus, in the following, we postulate the existence of the set of all \emph{\glspl{gls:if:pid}} $\pID$.

\subsection{\glsentryplural{gls:if:pspec} and \glsentryplural{gls:interface}}
Ports are specified by means of \glsentryplural{gls:if:pspec} which declare a set of \glsentryplural{gls:if:pid} and a corresponding typing.

\begin{definition}[\Glsentrytext{gls:if:pspec}]
	A \emph{\gls{gls:if:pspec}} over \glsentrytext{gls:sig} $\Sigma=\signature$ is a pair $\pspec$, consisting of:
	\begin{itemize}
		\item a set of port identifiers $\psid\subseteq\pID$ and
		\item a mapping $\pstype\colon \psid\to \sigSort$ assigning a sort to each \glsentrytext{gls:if:pid}.
	\end{itemize}
	The set of all \glsentryplural{gls:if:pspec} over \glsentrytext{gls:sig} $\Sigma$ is denoted by $\pspecAll \Sigma$.
\end{definition}

\Glsentryplural{gls:interface} are build over a given \glsentrytext{gls:if:pspec}.
They consist of a set of local, input, and output port identifiers.
\begin{definition}[\Glsentrytext{gls:interface}]
	An \emph{\gls{gls:interface}} over \glsentrytext{gls:if:pspec} $\pspec\in\pspecAll{\Sigma}$ is a triple $\interface$, consisting of \emph{disjoint} sets for:
	\begin{itemize}
		\item local port identifiers $\ifLocP\subseteq \psid$,
		\item input port identifiers $\ifInP\subseteq \psid$, and
		\item output port identifiers $\ifOutP\subseteq \psid$.
	\end{itemize}
	The set of all \glsentryplural{gls:interface} over \glsentrytext{gls:if:pspec} $S_p$ is denoted by $\ifAll {S_p}$.
\end{definition}

An \glsentrytext{gls:interface} can be interpreted by a component, relating port identifiers of the \glsentrytext{gls:interface} with concrete ports of the component. 
\begin{definition}[\Glsentrytext{gls:if:int}]
	An \emph{\gls{gls:if:int}} for an interface $\interface\in\ifAll{S_p}$ over \glsentrytext{gls:if:pspec} $S_p=\pspec\in\pspecAll \Sigma$ with signature $\Sigma$ in an algebra $A=\algebra$ is a 4-tuple $\ifint$, consisting of:
	\begin{itemize}
		\item a component $\intCmp=\comp\in\cmpAll$, and
		\item \glspl{gls:if:int:p} $\intLoc\colon \cmpLP \leftrightarrow \ifLocP$, $\intIn\colon \cmpIP \leftrightarrow \ifInP$, and $\intOut\colon \cmpOP \leftrightarrow \ifOutP$, for local, input, and output ports, respectively.
	\end{itemize}
	
	Thereby, we require that the valuations of the component ports satisfy the typing constraints induced by the corresponding \glsentrytext{gls:if:pspec}:
	\begin{align}
		\qquad&\forall p\in\cmpLP \colon \cmpVal(p)\in\intSort(\pstype(\intLoc(p))),\nonumber\\	
		\qquad&\forall p\in\cmpIP \colon \cmpVal(p)\in\intSort(\pstype(\intIn(p))),\textrm{and}\nonumber\\
		&\forall p\in\cmpOP \colon \cmpVal(p)\in\intSort(\pstype(\intOut(p)))\enspace.
	\end{align}	
	
	The set of all \glsentryplural{gls:if:int} of interface $Q$ under algebra $A$ is denoted by $\ifintAll{Q}{A}$.
\end{definition}

\subsection{\glsentryplural{gls:if:term}}
\Glsentryplural{gls:if:term} are build over a given \glsentrytext{gls:interface}, corresponding \glsentrytext{gls:sig}, and \glsentryplural{gls:dt:var}.

\begin{definition}[\Glsentrytext{gls:if:term}]
	The set of all \emph{\glspl{gls:if:term}} of sort $s\in\sigSort$ of \glsentrytext{gls:sig} $\Sigma=\signature$ over an \glsentrytext{gls:interface} $Q\in\ifAll {S_p}$ with $S_p\in\pspecAll {\Sigma}$ and \glsentryplural{gls:dt:var} $\dtVar$ is the smallest set $\ifTerms[s]{\Sigma}{Q}{\dtVar}$ satisfying the equations of Fig.~\ref{fig:if:t:syn}.
	The set of all \glsentryplural{gls:if:term} of all sorts is denoted by $\ifTerms{\Sigma}{Q}{\dtVar}$.
	\begin{figure*}
		\begin{mymathbox}[-2mm]{\Glsentryplural{gls:if:term}: syntax}
				v\in \dtVar_s &\implies v\in \ifTerms[s] {\Sigma}{Q}{\dtVar}\enspace,\\
				f\in \sigFun^0 &\implies f\in \ifTerms[s] {\Sigma}{Q}{\dtVar}~\textrm{[for $\sftype[0](f)=s$]}\enspace,\\
				f\in \sigFun^{n+1} \land
				t_1\in\ifTerms[s_1]{\Sigma}{Q}{\dtVar}, \cdots, t_{n+1} \in \ifTerms[s_{n+1}]{\Sigma}{Q}{\dtVar} &\implies f(t_1, \cdots, t_{n+1})\in \ifTerms[s] \Sigma F \dtVar~\textrm{[for $n\in\NN$, $\sftype[0](f)=s$,}\\&\qqquad \textrm{and $\sftype[1](f)=s_1,\cdots, \sftype[n+1](f)=s_{n+1}$]}\enspace,\\	
				p\in \cp{Q}{1} \cup \cp{Q}{2} \cup \cp{Q}{3} &\implies p \in\ifTerms[s]{\Sigma}{Q}{\dtVar}~[\textrm{if $\pstype(p)=s$}]\enspace.
		\end{mymathbox}
		\caption{Inductive definition of \glsentryplural{gls:if:term} $\ifTerms[s]{\Sigma}{Q}{\dtVar}$ of sort $s\in\sigSort$ over signature $\Sigma=\signature$, \glsentrytext{gls:interface} $Q$, and \glsentryplural{gls:dt:var} $\dtVar$\enspace.}\label{fig:if:t:syn}
	\end{figure*}
\end{definition}

The semantics of \glsentryplural{gls:if:term} is defined over a given \glsentrytext{gls:alg}, corresponding \glsentrytext{gls:dt:va}, and \glsentrytext{gls:if:int}.
It is given in terms of a function assigning a value of the corresponding algebra to each \glsentrytext{gls:if:term}.

\begin{sloppypar}
\begin{definition}{\Glsentrytext{gls:if:t:sem}}
	The \emph{\gls{gls:if:t:sem}} for \emph{\glsentryplural{gls:if:term}} $\ifTerms{\Sigma}{Q}{\dtVar}$ over \glsentrytext{gls:sig} $\Sigma=\signature$ and \glsentrytext{gls:interface} $Q$ in \glsentrytext{gls:alg} $A=\algebra\in\algAll{\Sigma}$ with corresponding \glsentrytext{gls:dt:va} $\dtVI\in\dtVAll{\dtVar}{A}$, and \glsentrytext{gls:if:int} $j=\ifint\in\ifintAll{Q}{A}$ is the function $\ifTsem{\_}{A}{\dtVI}{j}\colon\ifTerms[s]{\Sigma}{Q}{\dtVar}\to\intSort(s)$ (for each sort $s\in\sigSort$), characterized by the equations in Fig.~\ref{fig:if:t:sem}.
	\begin{figure*}
		\begin{mymathbox}[2cm]{\Glsentryplural{gls:if:term}: semantics}
				\ifTsem{v}{A}{\dtVI}{j} &\defeq \dtVI_s(v)~\textrm{[for $v\in \dtVar_s$]}\enspace,\\
				\ifTsem{f}{A}{\dtVI}{j} &\defeq \beta(f)~\textrm{[for function symbol $f\in\sigFun^0$]}\enspace,\\		
				\ifTsem{f(t_1, \cdots, t_n)}{A}{\dtVI}{j} &\defeq \beta(f)(\dtsem{t_1}{\dtVI}{A}, \cdots, \dtsem{t_n}{\dtVI}{A})~\textrm{[for function symbol $f\in F^{n+1}$]}\enspace,\\
				\ifTsem{p}{A}{\dtVI}{j} &\defeq \cmpVal(\inverse{(\intIn)}(p))~\textrm{[for $p\in \ifInP$]}\enspace,\\
				\ifTsem{p}{A}{\dtVI}{j} &\defeq \cmpVal(\inverse{(\intOut)}(p))~\textrm{[for $p\in \ifOutP$]}\enspace.
		\end{mymathbox}
		\caption{Recursive definition of \glsentrytext{gls:if:t:sem} for \glsentryplural{gls:if:term} $\ifTerms[s]{\Sigma}{Q}{\dtVar}$ of sort $s\in\sigSort$ with signature $\Sigma=\signature$, \glsentrytext{gls:interface} $Q=\interface$, \glsentryplural{gls:dt:var} $\dtVar$, algebra $A=\algebra\in\algAll{\Sigma}$ and corresponding \glsentryplural{gls:dt:va} $\dtVI\in\dtVAll{\dtVar}{A}$, and \glsentrytext{gls:if:int} $j=\ifint\in\ifintAll{Q}{A}$\enspace.}\label{fig:if:t:sem}
	\end{figure*}	
\end{definition}
\end{sloppypar}

\subsection{\glsentryplural{gls:if:assert}}
Interface assertions are build by the common logical operators over interface terms.
They are formulated over a given \glsentrytext{gls:interface} and \glsentryplural{gls:dt:var}.
\begin{definition}[\Glsentrytext{gls:if:assert}]
	The set of all \glspl{gls:if:assert} over a \glsentrytext{gls:sig} $\Sigma$, \glsentrytext{gls:interface} $Q$, and \glsentryplural{gls:dt:var} $\dtVar$ is the smallest set $\ifFormulas{\Sigma}{Q}{\dtVar}$ satisfying the equations in Fig.~\ref{fig:if:f:syn}.
	\begin{figure*}
		\begin{mymathbox}[-3mm]{\Glsentryplural{gls:if:assert}: syntax}
			b\in \sigPred^0 &\implies b\in \ifFormulas{\Sigma}{Q}{\dtVar}\enspace,\\
			b\in \sigPred^{n+1} \land t_1\in\ifTerms[s_1]{\Sigma}{Q}{\dtVar}, \cdots, t_{n+1} \in \ifTerms[s_{n+1}]{\Sigma}{Q}{\dtVar} &\implies b(t_1, \cdots, t_{n+1})\in \ifFormulas{\Sigma}{Q}{\dtVar}\\&\qqquad\textrm{[for $n\in\NN$ and $\sftype[1](b)=s_1,\cdots, \sftype[n+1](b)=s_{n+1}$]}\enspace,\\
			t,t'\in \caTerms{\Sigma}{S_i}{\dtVar}{\caVar} &\implies \dteq{t}{t'}\in \ifFormulas{\Sigma}{Q}{\dtVar}\enspace,\\
			E\in \ifFormulas{\Sigma}{Q}{\dtVar} &\implies \dtneg{E} \in \ifFormulas{\Sigma}{Q}{\dtVar}\enspace,\\	
			E, E'\in \ifFormulas{\Sigma}{Q}{\dtVar} &\implies \dtland{E}{E'},~\dtlor{E}{E'},~\dtimplies{E}{E'},~E \dtiff E' \in \ifFormulas{\Sigma}{Q}{\dtVar}\enspace,\\
			E\in \ifFormulas{\Sigma}{Q}{\dtVar}\land x\in\dtVar_s &\implies \dtforall{x}~E\in \ifFormulas{\Sigma}{Q}{\dtVar}\enspace\land\\
			&\qqquad\dtexists{x}~E\in\ifFormulas{\Sigma}{Q}{\dtVar}~\textrm{[for some $s\in S$]}\enspace.
		\end{mymathbox}
		\caption{Inductive definition of \glsentryplural{gls:if:assert} $\ifFormulas{\Sigma}{Q}{\dtVar}$ over signature $\Sigma=\signature$, \glsentrytext{gls:interface} $Q$, and \glsentryplural{gls:dt:var} $\dtVar$\enspace.}\label{fig:if:f:syn}
	\end{figure*}
\end{definition}

The semantics of \glsentryplural{gls:if:assert} is given by relating \glsentryplural{gls:if:assert} with corresponding \glsentryplural{gls:if:int} satisfying the assertions.
\begin{sloppypar}
\begin{definition}[\Glsentrytext{gls:if:a:sem}]
	The \emph{\gls{gls:if:a:sem}} for \emph{\glsentryplural{gls:if:assert}} $\ifFormulas{\Sigma}{Q}\dtVar$ over \glsentrytext{gls:sig} $\Sigma$, \glsentryplural{gls:dt:var} \dtVar, and interface $Q\in\ifAll{S_p}$ with $S_p\in\pspecAll \Sigma$ in \glsentrytext{gls:alg} $A\in\algAll{\Sigma}$ with corresponding \glsentrytext{gls:dt:va} $\dtVI\in\dtVAll{\dtVar}{A}$ is the relation $\ifmod[\dtVI]{\_}{\_}{A}\subseteq \ifintAll{Q}{A}\times\ifFormulas{\Sigma}{Q}\dtVar$ characterized by the equations in Fig.~\ref{fig:if:f:sem}.
	\begin{figure*}
		\begin{mymathbox}[2.2cm]{\Glsentryplural{gls:if:assert}: semantics}
			\ifmod[\dtVI]{j}{b}{A} &\iff \gamma(b)~\textrm{[for $b\in\sigPred^0$]}\enspace,\\	
			\ifmod[\dtVI]{j}{b(t_1,\cdots,t_n)}{A} &\iff \gamma(b)(\dtsem{t_1}{\dtVI}{A}, \cdots, \dtsem{t_n}{\dtVI}{A})~\textrm{[for $b\in\sigPred^{n+1}$]}\enspace,\\
			\ifmod[\dtVI]{j}{\dteq{t}{t'}}{A} &\iff \dtsem{t}{\dtVI}{A}=\dtsem{t'}{\dtVI}{A}\enspace,\\
			\ifmod[\dtVI]{j}{\dtland{E}{E'}}{A} &\iff \ifmod[\dtVI]{j}{E}{A} \land \ifmod[\dtVI]{j}{E'}{A}\enspace,\\
			\ifmod[\dtVI]{j}{\dtlor{E}{E'}}{A} &\iff \ifmod[\dtVI]{j}{E}{A} \lor \ifmod[\dtVI]{j}{E'}{A}\enspace,\\
			\ifmod[\dtVI]{j}{\dtimplies{E}{E'}}{A} &\iff \ifmod[\dtVI]{j}{E}{A} \implies \ifmod[\dtVI]{j}{E'}{A}\enspace,\\
			\ifmod[\dtVI]{j}{E \dtiff E'}{A} &\iff \ifmod[\dtVI]{j}{E}{A} \iff \ifmod[\dtVI]{j}{E'}{A}\enspace,\\				
			\ifmod[\dtVI]{j}{\dtexists{x}~E}{A} &\iff \exists x'\in \alpha(s)\colon \ifmod[\fupdate{\dtVI}{s}{x}{x'}]{j}{E}{A}~\textrm{[for $s\in\sigSort$ and $x\in\dtVar_s$]}\enspace,\\
			\ifmod[\dtVI]{j}{\dtforall{x}~E}{A} &\iff \forall x'\in \alpha(s)\colon \ifmod[\fupdate{\dtVI}{s}{x}{x'}]{j}{E}{A}~\textrm{[for $s\in\sigSort$ and $x\in\dtVar_s$]}\enspace.
		\end{mymathbox}
		\caption{Recursive definition of models relation for \glsentryplural{gls:if:assert} $\ifFormulas{\Sigma}{Q}{\dtVar}$ over \glsentrytext{gls:sig} $\Sigma=\signature$, \glsentrytext{gls:interface} $Q$, and \glsentryplural{gls:dt:var} $\dtVar$ over algebra $A\in\algAll{\Sigma}$ with corresponding \glsentrytext{gls:dt:va} $\dtVI=(\dtVI_s)_{s\in\sigSort}$\enspace.}\label{fig:if:f:sem}
	\end{figure*}
	An \glsentrytext{gls:if:assert} $\gamma\in\ifFormulas{\Sigma}{Q}\dtVar$ is \emph{valid} for \glsentrytext{gls:alg} $A\in\algAll{\Sigma}$ and \glsentrytext{gls:if:int} $j\in\cintAll{Q}{A}$ iff there exists a corresponding \glsentrytext{gls:dt:va} $\dtVI\in\dtVAll{\dtVar}{A}$, such that $\ifmod[\dtVI]{j}{\gamma}{A}$.
	\Glsentrytext{gls:if:int} $j\in\cintAll{Q}{A}$ is a \emph{model} for $\gamma\in\ifFormulas{\Sigma}{Q}\dtVar$, written $\ifmod{j}{\gamma}{A}$ iff for each corresponding \glsentrytext{gls:dt:va} $\dtVI\in\dtVAll{\dtVar}{A}$ we have $\ifmod[\dtVI]{j}{\gamma}{A}$.
	\Glsentrytext{gls:if:int} $j\in\cintAll{Q}{A}$ is a \emph{model} for a set of \glsentryplural{gls:if:assert} $\Gamma\subseteq\ifFormulas{\Sigma}{Q}\dtVar$, written $\ifmod{j}{\Gamma}{A}$ iff $\ifmod{j}{\gamma}{A}$ for each $\gamma\in\Gamma$.
\end{definition}
\end{sloppypar}

\subsection{Specifying \glsentryplural{gls:interface}}
\Glsentryplural{gls:interface} are specified by providing a set of interface identifiers.
Then, each identifier is associated with an \glsentrytext{gls:interface} (i.e., sets of local, input and output ports).
Finally, a sets of \glsentryplural{gls:if:assert} is specified for each interface identifier.

Thus, in the following, we postulate the existence of the set of all \emph{\glspl{gls:if:id}} $\iID$.
\begin{definition}[\Glsentrytext{gls:ifspec}]
	An \emph{\gls{gls:ifspec}} over port specification $S_p\in\pspecAll \Sigma$ with signature $\Sigma$ is a pair $\ifspec$, consisting of:
	\begin{itemize}
		\item a set of \glsentryplural{gls:if:id} $\isId\subseteq\iID$,
		\item a family of corresponding \glsentryplural{gls:interface} $(\isInt_i)_{i\in\isId}$ with interface $\isInt_i\in\ifAll{S_p}$ for each interface identifier $i\in\isId$.
	\end{itemize}
	The set of all \glsentryplural{gls:ifspec} over port specification $S_p\in\pspecAll{\Sigma}$ is denoted by $\IS {S_p}$.
	
	For an \glsentrytext{gls:ifspec} $S_i=\ifspec$, we denote by:
	\begin{itemize}
		\item $\ifLoc{S_i}\defeq\bigcup_{i\in \isId} (i\times\cp{\isInt_{i}}{1})$ the set of \emph{interface local ports},		
		\item $\ifIn{S_i}\defeq\bigcup_{i\in \isId} (i\times\cp{\isInt_{i}}{2})$ the set of \emph{interface input ports},
		\item $\ifOut{S_i}\defeq\bigcup_{i\in \isId} (i\times\cp{\isInt_{i}}{3})$ the set of \emph{interface output ports}, and
		\item $\ifPort{S_i}\defeq \ifLoc{S_i}\cup\ifIn{S_i}\cup\ifOut{S_i}$ the set of all \emph{interface ports}.
	\end{itemize}	
\end{definition}
Note that an \glsentrytext{gls:ifspec} actually specifies a set of interfaces, rather than just one single interface. Moreover, it allows for reuse of ports through several interfaces. Thus, if a port is specified once, it can be used to specify several, different interfaces.

\Glsentryplural{gls:ifspec} can be interpreted by a set of components with corresponding interfaces.
\begin{sloppypar}
\begin{definition}[\Glsentrytext{gls:ifspec:int}]
	An \emph{\gls{gls:ifspec:int}} for an \glsentrytext{gls:ifspec} $\ifspec\in\IS{S_p}$ over port specification $S_p\in\pspecAll{\Sigma}$ under an algebra $A=\algebra\in\algAll{\Sigma}$ is a family $\cint=(\cint_i)_{i\in\isId}$, with $\cint_i\subseteq\ifintAll{\isInt_i}{A}$ being the biggest set of \glsentryplural{gls:if:int} for interface identifier $i\in\isId$, such that 
	\begin{itemize}
		\item components with the same identifier belong only to one interface:
		\begin{equation}
			\Big(\bigcap_{i\in\isId}\bigcup_{\ifint\in\cint_i} \{\cp{\intCmp}{1}\}\Big)=\emptyset\enspace.
		\end{equation}
		\item the set of all components $\bigcup_{i\in\isId}\bigcup_{\ifint\in\cint_i} \{c\}$ is a healthy set of components.
	\end{itemize}	
	We introduce a function to return the set of all components for a certain interface identifier:
	\begin{equation}
		\cicomp[i]{\cint}=\bigcup_{\ifint\in\cint_i} \{c\}\enspace.
	\end{equation}
	With $\cicomp{\cint}=\bigcup_{i\in\isId}\cicomp[i]{\cint}$ we denote the set of all components of all interface identifiers.
	
	Note that $\cicomp{\cint}$ is a healthy set of component which allows us to apply all the functions for healthy sets of components introduced in Def.~\ref{def:cmp} to $\cicomp{\cint}$.
	The set of all possible \glsentryplural{gls:ifspec:int} for \glsentrytext{gls:ifspec}  $S_i$ over algebra $A$ is denoted by $\cintAll{S_i}{A}$\enspace.
\end{definition}
\end{sloppypar}

\subsubsection{From interfaces to component types}
Remember the healthiness condition (Eq.~\eqref{eq:healthy2}) requiring local port valuations not to change for components.
Once fixed, those ports do not change their value during the execution of an architecture and are thus a way to parametrize components.
The annotated interfaces become \glsentryplural{gls:ctype} since they enrich an interface with certain semantic constraints.

The parametrization step is done during interface specification since \glsentryplural{gls:if:assert} are the only way to determine/use local port valuations.
It is demonstrated in the running example for \glsentrytext{gls:bb:ks} components which are parametrized by a set of problems they are able to solve.

\subsubsection{Specification using templates}
\Glsentryplural{gls:ctype} can be specified by means of \glspl{gls:ifspec:ptemp} and \glspl{gls:ifspec:temp}.
Fig.~\ref{fig:pspec}, for example, shows a \glspl{gls:ifspec:ptemp} declaring 3 ports and corresponding sorts.
\begin{figure}
	\centering
	\begin{pstmp}{Name}{dtSpec}
		\psport{$p$}{$\mathtt{Sort1}$}
		\psport{$q$}{$\mathtt{Sort2}$}
		\psport{$c$}{$\mathtt{Sort3}$}
	\end{pstmp}
	\caption{\Glsentrytext{gls:ifspec:ptemp} defining three ports and corresponding sorts.}
	\label{fig:pspec}
\end{figure}
Fig.~\ref{fig:ifspec} shows a corresponding \glsentrytext{gls:ifspec:temp}.
\begin{figure}
	\centering
	\begin{istmp}{Name}{$c$}{$p$}{$q$}{pSpec}{dtSpec}
		\aline[dashed]
		\isport{$p$}{$\mathtt{Sort1}$}
		\aline
		\var{$\mathit{aVar}$}{$\mathtt{Sort4}$}
		\aline[dashed]
		\axiom{$\mathit{someAxiom}(p,q,\mathit{c},\mathit{aVar})$}
	\end{istmp}
	\caption{\Glsentrytext{gls:ifspec:temp} consisting of a port specification and definitions for variables as well as \glsentryplural{gls:if:assert}.}
	\label{fig:ifspec}
\end{figure}
Each specification has a name with corresponding input, output, and local ports and may use some \glsentryplural{gls:dt:spec}.
Then, a list of variables and corresponding sorts is specified.
Finally, a list of \glsentryplural{gls:if:assert} can be specified over the ports and variables.

Sometimes it is convenient to combine a port and interface specification into a single \glsentrytext{gls:ifspec:temp}.
Fig.~\ref{fig:ifspec}, e.g., declares an additional port $p$ and corresponding sort.

\subsection{\glsentrytext{gls:bb:p}: \glsentrytext{gls:ifspec}}\label{sec:runexa:ifspec}
A \glsentrytext{gls:bb:p} architecture consists of a \glsentrytext{gls:bb:bb} component and several \glsentrytext{gls:bb:ks} components.
\subsubsection{\glsentrytext{gls:bb:p} interface}
A \gls{BB} component is used to capture the current state on the way to a solution of the original problem. Its state consists of all currently open subproblems and solutions for already solved subproblems. 

A \glsentrytext{BB} expects two types of input:
\begin{inparaenum}
	\item a problem $p\in\mathtt{PROB}$ which a \glsentrytext{gls:bb:ks} is able to solve, together with a set of subproblems $P\subseteq \mathtt{PROB}$ the \glsentrytext{gls:bb:ks} requires to be solved before solving the original problem $p$,
	\item a problem $p\in\mathtt{PROB}$ solved by a \glsentrytext{gls:bb:ks}, together with the corresponding solution $s \in \mathtt{SOL}$.
\end{inparaenum}
A \glsentrytext{BB} returns two types of output:
\begin{inparaenum}
	\item a set $P\subseteq\mathtt{PROB}$ which contains all the problems to be solved,
	\item a set of pairs $\mathit{PS}\subseteq\mathtt{PROB}\times\mathtt{SOL}$ containing solved problems and the corresponding solutions.
\end{inparaenum}

\subsubsection{\glsentrytext{gls:bb:ks} interface}
A \gls{KS} component is a domain expert able to solve problems in that domain. It may lack expertise of other domains. Moreover, it can recognize problems which it is able to solve and subproblems which have to be solved first by other \glsentryplural{KS}.

A \glsentrytext{KS} expects two types of input:
\begin{inparaenum}
	\item a set $P\subseteq\mathtt{PROB}$ which contains all the problems to be solved,
	\item a set of pairs $\mathit{PS}\subseteq\mathtt{PROB}\times\mathtt{SOL}$ containing solutions for already solved problems.
\end{inparaenum}
A \glsentrytext{KS} returns one of two types of output:
\begin{inparaenum}
	\item a problem $p\in\mathtt{PROB}$ which it is able to solve together with a set of subproblems $P\subseteq \mathtt{PROB}$ which it requires to be solved before solving the original problem,
	\item a problem $p\in\mathtt{PROB}$ which it was able to solve together with the corresponding solution $s \in \mathtt{SOL}$.
\end{inparaenum}

Figure~\ref{fig:BB:pspec} shows a \glsentrytext{gls:ifspec:ptemp} of the \glsentrytext{gls:bb:p} pattern.
\begin{figure}
	\centering
	\begin{pstmp}{Blackboard}{ProbSol}
		\psport{$\bbip$}{$\mathtt{PROB}\times\pset{\mathtt{PROB}}$}
		\psport{$\bbis$}{$\mathtt{PROB}\times\mathtt{SOL}$}
		\psport{$\bbop$}{$\mathtt{PROP}$}
		\psport{$\bbos$}{$\mathtt{PROB}\times\mathtt{SOL}$}
	\end{pstmp}
	\caption{Blackboard port specification.}
	\label{fig:BB:pspec}
\end{figure}

Based on the induced port specification, Fig.~\ref{fig:BB:ifspec} shows the corresponding \glsentrytext{gls:ifspec} for the pattern.
\begin{figure}
	\centering
	\begin{istmp}[200]{BB}{}{$\bbip,\bbis$}{$\bbop\bbos$}{Blackboard}{}
	\end{istmp}
	\caption{Blackboard interface specification.}
	\label{fig:BB:ifspec}
\end{figure}

\begin{figure}
	\centering
	\begin{istmp}{KS}{$\mathit{prob}$}{$\ksip,\ksis$}{$\ksop,\ksos$}{Blackboard}{ProbSol}
		\aline[dashed]		
		\isport{$\mathit{prob}$}{$\pset{\mathtt{PROB}}$}
		\aline
		\var{p}{$\mathtt{PROP}$}
		\var{P}{$\pset{\mathtt{PROP}}$}
		\aline[dashed]
		\axiom[eq:bb:i:1]{$\ksop=(p,P)\implies p\in\mathit{prob}$}
	\end{istmp}
	\caption{Knowledge source interface specification.}
	\label{fig:KS:ifspec}
\end{figure}

A \glsentrytext{KS} can only solve certain types of problems which is why we assume the existence of a local port $\mathit{prob}$ for each knowledge source which is typed by the set of problems a certain \glsentrytext{KS} can solve.
In Eq.~\eqref{eq:bb:i:1} we require for each \glsentrytext{KS} that it only solves problems given by this mapping.

\subsection{Summary}
To conclude this section, Tab.~\ref{tab:if} provides a brief overview of the main concepts introduced in this section.
For each concept it provides a brief description and related notation.
\begin{table*}
	\centering
	\caption{Overview of concepts for \glsentryplural{gls:ifspec}.\label{tab:if}}
	\begin{tabular}{r@{\hspace{10pt}}p{7.5cm}@{\hspace{15pt}}p{3cm}}
		\toprule
		\textbf{Concept} & \textbf{\textit{Description}} & \textbf{\textit{Related Notation}} \\
		\midrule
		\glsentrytext{gls:if:pid}&\glsentrydesc{gls:if:pid}&\glsentrysymbol{gls:if:pid}\\
		\cmidrule{2-3}
		\glsentrytext{gls:if:pspec}&\glsentrydesc{gls:if:pspec}&\glsentrysymbol{gls:if:pspec}\\
		\cmidrule{2-3}
		\glsentrytext{gls:interface}&\glsentrydesc{gls:interface}&\glsentrysymbol{gls:interface}\\
		\cmidrule{2-3}
		\glsentrytext{gls:if:int}&\glsentrydesc{gls:if:int}&\glsentrysymbol{gls:if:int}\\
		\cmidrule{2-3}
		\glsentrytext{gls:if:int:p}&\glsentrydesc{gls:if:int:p}&\glsentrysymbol{gls:if:int:p}\\
		\cmidrule{2-3}
		\glsentrytext{gls:if:term}&\glsentrydesc{gls:if:term}&\glsentrysymbol{gls:if:term}\\
		\cmidrule{2-3}		
		\glsentrytext{gls:if:t:sem}&\glsentrydesc{gls:if:t:sem}&\glsentrysymbol{gls:if:t:sem}\\
		\cmidrule{2-3}		
		\glsentrytext{gls:if:assert}&\glsentrydesc{gls:if:assert}&\glsentrysymbol{gls:if:assert}\\
		\cmidrule{2-3}		
		\glsentrytext{gls:if:a:sem}&\glsentrydesc{gls:if:a:sem}&\glsentrysymbol{gls:if:a:sem}\\
		\cmidrule{2-3}		
		\glsentrytext{gls:if:id}&\glsentrydesc{gls:if:id}&\glsentrysymbol{gls:if:id}\\		
		\cmidrule{2-3}		
		\glsentrytext{gls:ifspec} & \glsentrydesc{gls:ifspec} &\glsentrysymbol{gls:ifspec}\\
			& \emph{local, input, output, and all ports of interface spec $S_i$} & $\ifLoc{S_i}$, $\ifIn{S_i}$,\\
			&&$\ifOut{S_i}$, $\ifPort{S_i}$\\
		\cmidrule{2-3}		
		\glsentrytext{gls:ifspec:int} & \glsentrydesc{gls:ifspec:int} & \glsentrysymbol{gls:ifspec:int}\\
		\cmidrule{2-3}		
		\glsentrytext{gls:ifspec:ptemp} & \glsentrydesc{gls:ifspec:ptemp} & \glsentrysymbol{gls:ifspec:ptemp}\\
		\cmidrule{2-3}		
		\glsentrytext{gls:ifspec:temp} & \glsentrydesc{gls:ifspec:temp} & \glsentrysymbol{gls:ifspec:temp}\\
		\bottomrule
	\end{tabular}
\end{table*}

%% file: cnftrace.tex
\section{\glsentrytext{gls:const:spec}}\label{sec:cnft}
Architecture constraints are specified as \gls{gls:templog} formulas over \glsentryplural{gls:aconf}.
Thus, we first introduce the notion of \emph{\glsentrytext{gls:cnfassert}} to specify \glsentryplural{gls:aconf}.
Then, we introduce \emph{\glsentryplural{gls:cnftraceassert}} as an extension of \glsentryplural{gls:cnfassert} to specify \glsentryplural{gls:cnf:trace}.
\subsection{\glsentryplural{gls:cnfassert}}
\Glsentryplural{gls:aconf} can be specified by so-called \glsentryplural{gls:cnfassert} formulated over \glsentryplural{gls:ca:term}.
\subsubsection{\glsentryplural{gls:ca:term}}
Terms of \glsentryplural{gls:cnfassert} are build over an \glsentrytext{gls:ifspec}, corresponding \glsentrytext{gls:sig}, \glsentryplural{gls:dt:var} and \emph{\glspl{gls:ca:var}} (a family of disjoint sets of variables $\caVar=(\caVar_i)_{i\in\isId}$ with $\caVar_i$ denoting a set of variables for interface identifier $i\in \isId$).
\begin{definition}[\Glsentrytext{gls:ca:term}]
	The set of all \emph{\glspl{gls:ca:term}} of sort $s\in\sigSort$ over a \glsentrytext{gls:sig} $\Sigma=\signature$, \glsentrytext{gls:ifspec} $S_i\in\IS{S_p}$ over \glsentrytext{gls:if:pspec} $S_p\in\pspecAll{\Sigma}$, \glsentryplural{gls:dt:var} $\dtVar$, and \glsentryplural{gls:ca:var} $\caVar$ is the smallest set $\caTerms[s]{\Sigma}{S_i}{\dtVar}{\caVar}$ satisfying the equations of Fig.~\ref{fig:CA:syn:t}.
	The set of all \glsentryplural{gls:ca:term} of all sorts is denoted by $\caTerms{\Sigma}{S_i}{\dtVar}{\caVar}$.
	\begin{figure*}
		\begin{mymathbox}[-3mm]{\Glsentryplural{gls:ca:term}: syntax}
			v\in \dtVar_s &\implies v\in \caTerms[s] {\Sigma}{S_i}{\dtVar}{\caVar}\enspace,\\
			f\in \sigFun^0 &\implies f\in \caTerms[s] {\Sigma}{S_i}{\dtVar}{\caVar}~\textrm{[for $\sftype[0](f)=s$]}\enspace,\\
			f\in \sigFun^{n+1} \land
			t_1\in\caTerms[s_1]{\Sigma}{S_i}{\dtVar}{\caVar}, \cdots, t_{n+1} \in \caTerms[s_{n+1}]{\Sigma}{S_i}{\dtVar}{\caVar} &\implies f(t_1, \cdots, t_{n+1})\in \caTerms[s] {\Sigma}{S_i}{\dtVar}{\caVar}~\textrm{[for $n\in\NN$, $\sftype[0](f)=s$,}\\&\qqquad \textrm{and $\sftype[1](f)=s_1,\cdots, \sftype[n+1](f)=s_{n+1}$]}\enspace,\\	
			v\in \caVar_i \land p\in \cp{\isInt_i}{2} \cup \cp{\isInt_i}{3}&\implies \caTVal{v}{p} \in\caTerms[s]{\Sigma}{S_i}{\dtVar}{\caVar}~[\textrm{for $i\in \isId$ and $\pstype(p)=s$}]\enspace.
		\end{mymathbox}
		\caption{Inductive definition of \glsentryplural{gls:ca:term} $\caTerms[s]{\Sigma}{S_i}{\dtVar}{\caVar}$ of sort $s\in\sigSort$ over signature $\Sigma=\signature$, \glsentrytext{gls:ifspec} $S_i=\ifspec$, \glsentryplural{gls:dt:var} $\dtVar$, and \glsentryplural{gls:ca:var} $\caVar=(\caVar_i)_{i\in \isId}$\enspace.}\label{fig:CA:syn:t}
	\end{figure*}
\end{definition}

Note the use of function symbols of the corresponding \glsentrytext{gls:dt:spec} to build \glsentryplural{gls:ca:term}.
This allows one to reuse these function symbols when specifying \glsentryplural{gls:aconf}.
Moreover, note the use of port identifiers in the specification of \glsentryplural{gls:ca:term} to denote the valuation of the corresponding port in an \glsentrytext{gls:aconf}. For example, with $\caTVal{c}{p}$ we denote the current valuation of port $p$ of a component $c$.

The semantics of \glsentryplural{gls:ca:term} is defined over an \glsentrytext{gls:alg} and corresponding \glsentryplural{gls:dt:va}, an \glsentrytext{gls:ifspec:int} and corresponding \emph{\glspl{gls:ca:va}} (a family of mappings $\caVI=(\caVI_i)_{i\in\isId}$ with $\caVI_i\colon \caVar_i\to \cmpName{\cicomp[i]{J}}$ for each interface identifier $i\in\isId$, where $J$ is the corresponding \glsentrytext{gls:ifspec:int}), and an \glsentrytext{gls:aconf}.
In the following we denote with $\caVAll{\caVar}{J}$ the set of all \glsentryplural{gls:ca:va} for \glsentryplural{gls:ca:var} and \glsentrytext{gls:ifspec:int} $J$.
\begin{definition}[\Glsentrytext{gls:ca:term:sem}]
	The \emph{\gls{gls:ca:term:sem}} for \emph{\glsentryplural{gls:ca:term}} $\caTerms[s]{\Sigma}{S_i}{\dtVar}{\caVar}$ over \glsentrytext{gls:alg} $A\in\algAll{\Sigma}$ with corresponding \glsentrytext{gls:dt:va} $\dtVI\in\dtVAll{\dtVar}{A}$, \glsentrytext{gls:ifspec:int} $\cint\in\cintAll{S_i}{A}$ with corresponding \glsentrytext{gls:ca:va} $\caVI\in\caVAll{\caVar}{\cint}$, and \glsentrytext{gls:aconf} $k\in\cnfAll{\cicomp{\cint}}$ is the function $\casem{\_}{A}{\dtVI}{\cint}{\caVI}{k}\colon\caTerms[s]{\Sigma}{S_i}{\dtVar}{\caVar}\to\intSort(s)$ characterized by the equations in Fig.~\ref{fig:CA:sem:t}.
	\begin{figure*}
		\begin{mymathbox}[2cm]{\Glsentryplural{gls:ca:term}: semantics}
			\casem{v}{A}{\dtVI}{J}{\caVI}{k} &\defeq \dtVI_s(v)~\textrm{[for $v\in \dtVar_s$]}\enspace,\\
			\casem{f}{A}{\dtVI}{J}{\caVI}{k} &\defeq \beta(f)~\textrm{[for function symbol $f\in\sigFun^0$]}\enspace,\\		
			\casem{f(t_1, \cdots, t_n)}{A}{\dtVI}{J}{\caVI}{k} &\defeq \beta(f)(\dtsem{t_1}{\dtVI}{A}, \cdots, \dtsem{t_n}{\dtVI}{A})~\textrm{[for function symbol $f\in F^{n+1}$]}\enspace,\\
			\casem{\caTVal{v}{p}}{A}{\dtVI}{J}{\caVI}{k} &\defeq \cnfVal{k}(\caVI_i(v))(p)~\textrm{[for $i\in \isId$ and $v\in \caVar_i$]}\enspace.
		\end{mymathbox}
		\caption{Recursive definition of semantic function for \glsentryplural{gls:ca:term} $\caTerms[s]{\Sigma}{S_i}{\dtVar}{\caVar}$ with signature $\Sigma$, \glsentrytext{gls:ifspec} $S_i=\ifspec\in\IS{S_p}$ over \glsentrytext{gls:if:pspec} $S_p\in\pspecAll{\Sigma}$, \glsentryplural{gls:dt:var} $\dtVar$, \glsentryplural{gls:ca:var} $\caVar=(\caVar_i)_{i\in\isId}$, algebra $A\in\algAll{\Sigma}$ and corresponding \glsentrytext{gls:dt:va} $\dtVI\in\dtVAll{\dtVar}{A}$, \glsentrytext{gls:ifspec:int} $\cint\in\cintAll{S_i}{A}$ and corresponding \glsentrytext{gls:ca:va} $\caVI=(\caVI_i)_{i\in\isId}\in\caVAll{\caVar}{J}$, and \glsentrytext{gls:aconf} $k=\configuration\in\cnfAll{\cicomp{\cint}}$\enspace.}\label{fig:CA:sem:t}
	\end{figure*}	
\end{definition}
Again, the semantics of a term is given by a function assigning a value of a corresponding set of the underlying algebra to each term.

\subsubsection{\glsentrytext{gls:cnfassert}}
Assertions for \glsentryplural{gls:aconf} are built over the corresponding \glsentryplural{gls:ca:term} by the common logical operators and some dedicated predicates for the specification of so-called activation and connection constraints.
\begin{definition}[\Glsentrytext{gls:cnfassert}]
	The set of all \emph{\glspl{gls:cnfassert}} over a \glsentrytext{gls:sig} $\Sigma$, \glsentrytext{gls:ifspec} $S_i\in\IS{S_p}$ over \glsentrytext{gls:if:pspec} $S_p\in\pspecAll{\Sigma}$, \glsentryplural{gls:dt:var} $\dtVar$, and \glsentryplural{gls:ca:var} $\caVar$ is the smallest set $\caFormulas{\Sigma}{S_i}{\dtVar}{\caVar}$ satisfying the equations in Fig.~\ref{fig:CA:syn:f}.
	\begin{figure*}
		\begin{mymathbox}[-4mm]{\Glsentryplural{gls:cnfassert}: syntax}
			b\in \sigPred^0 &\implies b\in \caFormulas{\Sigma}{S_i}{\dtVar}{\caVar}\enspace,\\
			b\in \sigPred^{n+1} \land t_1\in\caTerms[s_1]{\Sigma}{S_i}{\dtVar}{\caVar}, \cdots, t_{n+1} \in \caTerms[s_{n+1}]{\Sigma}{S_i}{\dtVar}{\caVar} &\implies b(t_1, \cdots, t_{n+1})\in \caFormulas{\Sigma}{S_i}{\dtVar}{\caVar}\\&\qqquad\textrm{[for $n\in\NN$ and $\sftype[1](b)=s_1,\cdots, \sftype[n+1](b)=s_{n+1}$]}\enspace,\\
			t,t'\in \caTerms{\Sigma}{S_i}{\dtVar}{\caVar} &\implies \dteq{t}{t'}\in \caFormulas{\Sigma}{S_i}{\dtVar}{\caVar}\enspace,\\
			E\in \caFormulas{\Sigma}{S_i}{\dtVar}{\caVar} &\implies \dtneg{E} \in \caFormulas{\Sigma}{S_i}{\dtVar}{\caVar}\enspace,\\	
			E, E'\in \caFormulas{\Sigma}{S_i}{\dtVar}{\caVar} &\implies \dtland{E}{E'},~\dtlor{E}{E'},~\dtimplies{E}{E'},~E \dtiff E' \in \caFormulas{\Sigma}{S_i}{\dtVar}{\caVar}\enspace,\\
			E\in \caFormulas{\Sigma}{S_i}{\dtVar}{\caVar}\land x\in\dtVar_s &\implies \dtforall{x}~E\in \caFormulas{\Sigma}{S_i}{\dtVar}{\caVar}\enspace\land\\
			&\qqquad\dtexists{x}~E\in\caFormulas{\Sigma}{S_i}{\dtVar}{\caVar}~\textrm{[for $s\in S$]}\enspace,\\
			E\in \caFormulas{\Sigma}{S_i}\dtVar\caVar\land x\in \caVar_i &\implies \caforall{x}~E\in\caFormulas{\Sigma}{S_i}\dtVar{\caVar}\enspace\land\\
			&\qqquad\caexists{x}~E\in\caFormulas{\Sigma}{S_i}\dtVar\caVar~\textrm{[for $i\in \isId$]}\enspace,\\
			n,m\in\NN \land n\le m&\implies\caFMin{i}{n}\in\caFormulas{\Sigma}{S_i}\dtVar\caVar\land\caFMax{i}{m}\in\caFormulas{\Sigma}{S_i}\dtVar\caVar\enspace\land\\
			&\qqquad\caFMinMax{i}{n}{m}\in\caFormulas{\Sigma}{S_i}\dtVar\caVar~[\textrm{for some }i\in \isId]\enspace,\\
			v\in \caVar_i &\implies \caFActive v \in\caFormulas{\Sigma}{S_i}{\dtVar}{\caVar}~[\textrm{for some }i\in \isId]\enspace,\\
			v\in \caVar_i\land v'\in \caVar_j \land p\in \cp{\isInt_i}{2} \land p'\in \cp{\isInt_j}{3}&\implies \caFConn{v}{p}{v'}{p'} \in\caFormulas{\Sigma}{S_i}{\dtVar}{\caVar}~[\textrm{for some }i,j\in \isId]\enspace,\\
			p\in \cp{\isInt_i}{2} \land p'\in \cp{\isInt_j}{3}&\implies \caIRConn{i}{p}{j}{p'} \in\caFormulas{\Sigma}{S_i}{\dtVar}{\caVar}~[\textrm{for some }i,j\in \isId]\enspace.
		\end{mymathbox}
		\caption{Inductive definition of \glsentryplural{gls:cnfassert} $\caFormulas{\Sigma}{S_i}{\dtVar}{\caVar}$ over signature $\Sigma$, \glsentrytext{gls:ifspec} $S_i=\ifspec$, \glsentryplural{gls:dt:var} $\dtVar$, and \glsentryplural{gls:ca:var} $\caVar=(\caVar_i)_{i\in \isId}$\enspace.}\label{fig:CA:syn:f}
	\end{figure*}
\end{definition}
Note that the following predicates can be used for the specification of \glsentryplural{gls:cnfassert}:
\begin{itemize}
	\item \emph{\Glspl{gls:cns:act}} can be used to denote activation and deactivation of components. With $\caFActive c$, e.g., we denote the activation of component $c$. With $\caFMin{i}{n}$, $\caFMax{i}{n}$ we denote the constraint that at least $n$ or at most $n$ components with interface $i$ are active at each point in time.	
	\item \emph{\Glspl{gls:cns:conn}} can be used to denote constraints over the connection between components. With $\caFConn{c}{p}{c'}{p'}$, e.g., we denote a constraint that port $p$ of component $c$ has to be connected to port $p'$ of component $c'$.
	Finally, with $\caIRConn{i}{p}{j}{p'}$ we denote that port $p$ of every component with interface $i$ is connected to port $p'$ of every component with interface $j$.
\end{itemize}



The semantics of \glsentryplural{gls:cnfassert} is defined over an \glsentrytext{gls:alg} with corresponding \glsentrytext{gls:dt:va} and an \glsentrytext{gls:if:int} with corresponding \glsentrytext{gls:ca:va}.
\begin{sloppypar}
\begin{definition}[\Glsentrytext{gls:ca:sem}]~\label{def:cnf:term:sem}
	The \emph{\gls{gls:ca:sem}} for \emph{\glsentryplural{gls:cnfassert}} $\caFormulas{\Sigma}{S_i}\dtVar\caVar$ over \glsentrytext{gls:alg} $A\in\algAll{\Sigma}$ with corresponding \glsentrytext{gls:dt:va} $\dtVI\in\dtVAll{\dtVar}{A}$, \glsentrytext{gls:if:int} $\cint\in\cintAll{S_i}{A}$ with corresponding \glsentryplural{gls:ca:va} $\caVI\in\caVAll{\caVar}{\cint}$ is the relation $\camod[\dtVI][\caVI]{\_}{\_}{A}{J}\subseteq \cnfAll{\cicomp{\cint}}\times\caFormulas{\Sigma}{S_i}\dtVar\caVar$ characterized by the equations in Fig.~\ref{fig:CA:sem:f}.
	\begin{figure*}
		\begin{mymathbox}[1cm]{\Glsentryplural{gls:cnfassert}: semantics}
			\camod[\dtVI][\caVI]{k}{b}{A}{J} &\iff \gamma(b)~\textrm{[for $b\in\sigPred^0$]}\enspace,\\	
			\camod[\dtVI][\caVI]{k}{b(t_1,\cdots,t_n)}{A}{J} &\iff \gamma(b)(\dtsem{t_1}{\dtVI}{A}, \cdots, \dtsem{t_n}{\dtVI}{A})~\textrm{[for $b\in\sigPred^{n+1}$]}\enspace,\\
			\camod[\dtVI][\caVI]{k}{\dteq{t}{t'}}{A}{J} &\iff \dtsem{t}{\dtVI}{A}=\dtsem{t'}{\dtVI}{A}\enspace,\\
			\camod[\dtVI][\caVI]{k}{\dtland{E}{E'}}{A}{J} &\iff \camod[\dtVI][\caVI]{k}{E}{A}{J} \land \camod[\dtVI][\caVI]{k}{E'}{A}{J}\enspace,\\
			\camod[\dtVI][\caVI]{k}{\dtlor{E}{E'}}{A}{J} &\iff \camod[\dtVI][\caVI]{k}{E}{A}{J} \lor \camod[\dtVI][\caVI]{k}{E'}{A}{J}\enspace,\\
			\camod[\dtVI][\caVI]{k}{\dtimplies{E}{E'}}{A}{J} &\iff \camod[\dtVI][\caVI]{k}{E}{A}{J} \implies \camod[\dtVI][\caVI]{k}{E'}{A}{J}\enspace,\\			
			\camod[\dtVI][\caVI]{k}{E \dtiff E'}{A}{J} &\iff \camod[\dtVI][\caVI]{k}{E}{A}{J} \iff \camod[\dtVI][\caVI]{k}{E'}{A}{J}\enspace,\\
			\camod[\dtVI][\caVI]{k}{\dtexists{x}~E}{A}{J} &\iff \exists x'\in \alpha(s)\colon \camod[\fupdate{\dtVI}{s}{x}{x'}][\caVI]{k}{E}{A}{J}~\textrm{[for $s\in\sigSort$ and $x\in\dtVar_s$]}\enspace,\\
			\camod[\dtVI][\caVI]{k}{\dtforall{x}~E}{A}{J} &\iff \forall x'\in \alpha(s)\colon \camod[\fupdate{\dtVI}{s}{x}{x'}][\caVI]{k}{E}{A}{J}~\textrm{[for $s\in\sigSort$ and $x\in\dtVar_s$]}\enspace,\\			
			\camod[\dtVI][\caVI]{k}{\caexists{x}~E}{A}{J} &\iff
			\exists x'\in \cmpName{\cicomp[i]{J}}\colon \camod[\dtVI][\fupdate{\caVI}{i}{x}{x'}]{k}{E}{A}{J}~\textrm{[for $i\in \isId$ and $x\in \caVar_i$]}\enspace,\\
			\camod[\dtVI][\caVI]{k}{\caforall{x}~E}{A}{J} &\iff
			\forall x'\in \cmpName{\cicomp[i]{J}}\colon \camod[\dtVI][\fupdate{\caVI}{i}{x}{x'}]{k}{E}{A}{J}~\textrm{[for $i\in \isId$ and $x \in\caVar_i$]}\enspace,\\
			\camod[\dtVI][\caVI]{k}{\caFMin{i}{n}}{A}{J} &\iff \card{\cicomp[i]{J}\cap\cnfComp}\ge n~\textrm{[for $k=\configuration$ and $i\in \isId$]}\enspace,\\
			\camod[\dtVI][\caVI]{k}{\caFMax{i}{m}}{A}{J} &\iff \card{\cicomp[i]{J}\cap\cnfComp}\le m~\textrm{[for $k=\configuration$ and $i\in \isId$]}\enspace,\\
			\camod[\dtVI][\caVI]{k}{\caFMinMax{i}{n}{m}}{A}{J} &\iff \camod[\dtVI][\caVI]{k}{\caFMin{i}{n}}{A}{J} \enspace\land\enspace \camod[\dtVI][\caVI]{k}{\caFMax{i}{m}}{A}{J}\enspace,\\			
			\camod[\dtVI][\caVI]{k}{\caFActive{v}}{A}{J} &\iff \caVI_i(v)\in\cmpName{\cnfComp}~\textrm{[for $k=\configuration$, $i\in \isId$, and $v\in \caVar_i$]}\enspace,\\
			\camod[\dtVI][\caVI]{k}{\caFConn{v}{p}{v'}{p'}}{A}{J} &\iff (\caVI(v'),p')\in\cnfConn(\caVI(v),p)~\textrm{[for $k=\configuration$]}\enspace,\\
			\camod[\dtVI][\caVI]{k}{\caIRConn{i}{p}{j}{p'}}{A}{J} &\iff \camod[\dtVI][\caVI]{k}{\caforall{v,v'}~\big(\caFActive{v}\land\caFActive{v'}\implies\caFConn{v}{p}{v'}{p'}\big)}{A}{J}~\textrm{[for $v\in\caVar_i$, $v'\in\caVar_j$]}\enspace.
		\end{mymathbox}
		\caption{Recursive definition of models relation for \glsentryplural{gls:cnfassert} $\caFormulas{\dtVar}{\caVar}{\Sigma}{S_i}$ with \glsentrytext{gls:ifspec} $S_i=\ifspec$, algebra $A\in\algAll{\Sigma}$ with corresponding \glsentryplural{gls:dt:va} $\dtVI=(\dtVI_s)_{s\in\sigSort}$, and \glsentrytext{gls:ifspec:int} $\cint\in\cintAll{S_i}{A}$ with corresponding \glsentrytext{gls:ca:va} $\caVI=(\caVI_i)_{i\in\isId}$\enspace.}\label{fig:CA:sem:f}
	\end{figure*}
	A \glsentrytext{gls:cnfassert} $\gamma$ is \emph{valid} for \glsentrytext{gls:aconf} $k\in\cnfAll{\cicomp{\cint}}$ under \glsentrytext{gls:alg} $A\in\algAll{\Sigma}$ and \glsentrytext{gls:if:int} $\cint\in\cintAll{S_i}{A}$ iff there exists corresponding \glsentrytext{gls:dt:va} $\dtVI\in\dtVAll{\dtVar}{A}$ and \glsentrytext{gls:ca:va} $\caVI\in\caVAll{\caVar}{\cint}$, such that $\camod[\dtVI][\caVI]{k}{\gamma}{A}{J}$.
	Configuration $k$ is a \emph{model} of $\gamma$, written $\camod{k}{\gamma}{A}{J}$ iff for each corresponding \glsentrytext{gls:dt:va} $\dtVI$ and \glsentrytext{gls:ca:va} $\caVI$ we have $\camod[\dtVI][\caVI]{k}{\gamma}{A}{J}$.
	Configuration $k$ is a \emph{model} of a set of \glsentryplural{gls:cnfassert} $\Gamma$, written $\camod{k}{\Gamma}{A}{J}$ iff $\camod{k}{\gamma}{A}{J}$ for each $\gamma\in\Gamma$.
\end{definition}
\end{sloppypar}
Note that the semantics of \glsentryplural{gls:cnfassert} is given in terms of a relation, parametrized by an \glsentrytext{gls:if:int}. It determines all valid \glsentryplural{gls:aconf} (over the components provided by the \glsentrytext{gls:if:int}) for a given \glsentrytext{gls:cnfassert}.

\subsubsection{\glsentrytext{gls:cnftraceassert}}\label{sec:const:ta}
\Glsentryplural{gls:cnftraceassert} are a means to directly specify sets of \glsentryplural{gls:cnf:trace}.
They are formulated as \glsentrytext{gls:templog} formulas over \glsentryplural{gls:cnfassert} and consist of the common temporal operators and so-called rigid variables for \glsentryplural{gls:datatype} and components. Compared to ``normal'' variables, which are newly interpreted at each point in time, these variables are interpreted only once for the whole execution trace.

Thus, we assume the existence of a set of \emph{\glspl{gls:ct:dtvar}} (a family of disjoint sets of variables $\ctaDTVar=(\ctaDTVar_s)_{s\in\sigSort}$ with $\ctaDTVar_s$ denoting a set of variables for each sort $s\in\sigSort$) and \emph{\glspl{gls:ct:cvar}} (a family of disjoint sets of variables $\ctaCVar=(\ctaCVar_i)_{i\in\isId}$ with $\ctaCVar_i$ denoting a set of variables for each interface identifier $i\in\isId$).
\begin{definition}[\Glsentrytext{gls:cnftraceassert}]
	The set of all \glspl{gls:cnftraceassert} over \glsentrytext{gls:sig} $\Sigma$, \glsentrytext{gls:ifspec} $S_i\in\IS{S_p}$ over \glsentrytext{gls:if:pspec} $S_p$, \glsentryplural{gls:dt:var} $\dtVar$, \glsentryplural{gls:ct:dtvar} $\ctaDTVar$, \glsentryplural{gls:ca:var} $\caVar$, and \glsentryplural{gls:ct:cvar} $\ctaCVar$ is the smallest set $\ctFormulas{\Sigma}{S_i}{\dtVar}{\caVar}{\ctaDTVar}{\ctaCVar}$ satisfying the equations in Fig.~\ref{fig:CTA:syn}.
	\begin{figure*}
		\begin{mymathbox}[2.5cm]{\Glsentryplural{gls:cnftraceassert}: syntax}
			\phi \in \caFormulas{\Sigma}{S_i}{\dtVar\cup\ctaDTVar}{\caVar\cup\ctaCVar} &\implies \phi \in \ctFormulas{\Sigma}{S_i}{\dtVar}{\caVar}{\ctaDTVar}{\ctaCVar}\enspace,\\
			\gamma \in \ctFormulas{\Sigma}{S_i}{\dtVar}{\caVar}{\ctaDTVar}{\ctaCVar} &\implies \ctnext{\gamma},~\cteventually{\gamma},~\ctglobally{\gamma} \in \ctFormulas{\Sigma}{S_i}{\dtVar}{\caVar}{\ctaDTVar}{\ctaCVar}\enspace,\\
			\gamma,\gamma' \in \ctFormulas{\Sigma}{S_i}{\dtVar}{\caVar}{\ctaDTVar}{\ctaCVar} &\implies\ctuntilS{\gamma}{\gamma'},~\ctuntilW{\gamma}{\gamma'}\in \ctFormulas{\Sigma}{S_i}{\dtVar}{\caVar}{\ctaDTVar}{\ctaCVar}\enspace,\\
			x\in \ctaDTVar_s \land \gamma \in \ctFormulas{\Sigma}{S_i}{\dtVar}{\caVar}{\ctaDTVar}{\ctaCVar} &\implies\ctforallD{x}~\gamma\in \ctFormulas{\Sigma}{S_i}{\dtVar}{\caVar}{\ctaDTVar}{\ctaCVar} \enspace\land\\
			&\qqquad\ctexistsD{x}~\gamma\in \ctFormulas{\Sigma}{S_i}{\dtVar}{\caVar}{\ctaDTVar}{\ctaCVar}~\textrm{[for $s\in S$]}\enspace,\\
			x\in \ctaCVar_i \land \gamma \in \ctFormulas{\Sigma}{S_i}{\dtVar}{\caVar}{\ctaDTVar}{\ctaCVar} &\implies\ctforallC{x}~\gamma\in \ctFormulas{\Sigma}{S_i}{\dtVar}{\caVar}{\ctaDTVar}{\ctaCVar}\enspace \land\\
			&\qqquad\ctexistsC{x}~\gamma\in \ctFormulas{\Sigma}{S_i}{\dtVar}{\caVar}{\ctaDTVar}{\ctaCVar}~\textrm{[for $i\in \isId$]}\enspace.
		\end{mymathbox}
		\caption{Inductive definition of \glsentryplural{gls:cnftraceassert} $\ctFormulas{\Sigma}{S_i}{\dtVar}{\caVar}{\ctaDTVar}{\ctaCVar}$ over signature $\Sigma=\signature$, interface specification $S_i=\ifspec$, \glsentryplural{gls:dt:var} $\dtVar=(\dtVar_s)_{s\in \sigSort}$, \glsentryplural{gls:ca:var} $\caVar=(\caVar_i)_{i\in \isId}$, \glsentryplural{gls:ct:dtvar} $\ctaDTVar=(\ctaDTVar_s)_{s\in \sigSort}$, and \glsentryplural{gls:ct:cvar} $\ctaCVar=(\ctaCVar_i)_{i\in \isId}$\enspace.}\label{fig:CTA:syn}
	\end{figure*}
\end{definition}

The semantics of \glsentryplural{gls:cnftraceassert} is given according to~\cite{Manna2012}. It is defined over an algebra, an \glsentrytext{gls:if:int}, \emph{\glspl{gls:ct:dtva}} (a family of mappings $\ctaDTVI=(\ctaDTVI_s)_{s\in\sigSort}$ with $\ctaDTVI_s\colon \dtVar_s\to \alpha(s)$ for each sort $s\in\sigSort$) and \emph{\glspl{gls:ct:cva}} (a family of mappings $\ctaCVI=(\ctaCVI_i)_{i\in\isId}$ with $\ctaCVI_i\colon \ctaCVar_i\to \cmpName{\cicomp[i]{\cint}}$ for each interface identifier $i\in\isId$, where $\cint$ is the corresponding \glsentrytext{gls:ifspec:int}).
It relates each \glsentrytext{gls:cnftraceassert} with a \glsentrytext{gls:cnf:trace} and point in time such that the trace satisfies the assertion at the corresponding time point.
\begin{sloppypar}
\begin{definition}[\Glsentrytext{gls:cta:sem}]
	The \emph{\gls{gls:cta:sem}} for \emph{\glsentryplural{gls:cnftraceassert}} $\ctFormulas{\Sigma}{S_i}{\dtVar}{\caVar}{\ctaDTVar}{\ctaCVar}$ over \glsentrytext{gls:alg} $A\in\algAll{\Sigma}$ with corresponding \glsentrytext{gls:dt:va} $\dtVI\in\dtVAll{\dtVar}{A}$ and \glsentrytext{gls:ct:dtva} $\ctaDTVI\in\ctaDTVAll{\dtVar}{A}$, \glsentrytext{gls:if:int} $\cint\in\cintAll{S_i}{A}$, with corresponding \glsentrytext{gls:ca:va} $\caVI\in\caVAll{\caVar}{\cint}$ and \glsentrytext{gls:ct:cva} $\ctaCVI\in\ctaCVAll{\caVar}{\cint}$ is the relation $\ctmod[\ctaDTVI][\ctaCVI]{\_}{\_}{A}{J}\subseteq(\cnfAll{\cicomp{\cint}} \times \NN) \times \ctFormulas{\Sigma}{S_i}{\dtVar}{\caVar}{\ctaDTVar}{\ctaCVar}$ characterized by the equations in Fig.~\ref{fig:CTA:sem}.
	\begin{figure*}
		\begin{mymathbox}[1cm]{\Glsentryplural{gls:cnftraceassert}: semantics}
			\ctmod[\ctaDTVI][\ctaCVI]{(t,n)}{\phi}{A}{J} &\iff \exists \dtVI\in\dtVAll{\dtVar}{A} ,\caVI\in\caVAll{\caVar}{J}\colon\camod[\fmerge{\dtVI}{\ctaDTVI}][\fmerge{\caVI}{\ctaCVI}]{t(n)}{\phi}{A}{J}\enspace,\\
			\ctmod[\ctaDTVI][\ctaCVI]{(t,n)}{\ctnext{\gamma}}{A}{J} &\iff\ctmod[\ctaDTVI][\ctaCVI]{(t,n+1)}{\gamma}{A}{J}\enspace,\\
			\ctmod[\ctaDTVI][\ctaCVI]{(t,n)}{\cteventually{\gamma}}{A}{J} &\iff \exists n'\geq n\colon \ctmod[\ctaDTVI][\ctaCVI]{(t,n')}{\gamma}{A}{J}\enspace,\\
			\ctmod[\ctaDTVI][\ctaCVI]{(t,n)}{\ctglobally{\gamma}}{A}{J} &\iff \forall n'\geq n\colon \ctmod[\ctaDTVI][\ctaCVI]{(t,n')}{\gamma}{A}{J}\enspace,\\
			\ctmod[\ctaDTVI][\ctaCVI]{(t,n)}{\ctuntilS{\gamma}{\gamma'}}{A}{J} &\iff \exists n'\geq n\colon \ctmod[\ctaDTVI][\ctaCVI]{(t,n')}{\gamma'}{A}{J}\enspace\land\enspace\forall n\leq m < n'\colon \ctmod[\ctaDTVI][\ctaCVI]{(t,m)}{\gamma}{A}{J}\enspace,\\
			\ctmod[\ctaDTVI][\ctaCVI]{(t,n)}{\ctuntilW{\gamma}{\gamma'}}{A}{J} &\iff \ctmod[\ctaDTVI][\ctaCVI]{(t,n)}{\ctuntilS{\gamma}{\gamma'}}{A}{J}\enspace\lor\enspace\ctmod[\ctaDTVI][\ctaCVI]{(t,n)}{\ctglobally{\gamma}}{A}{J}\enspace,\\
			\ctmod[\ctaDTVI][\ctaCVI]{(t,n)}{\ctexistsD{x}~\gamma}{A}{J}&\iff \exists x'\in \alpha(X)\colon\ctmod[\fupdate{\ctaDTVI}{X}{x}{x'}][\ctaCVI]{(t,n)}{\gamma}{A}{J}~\textrm{[for $X\in \sigSort$]}\enspace,\\
			\ctmod[\ctaDTVI][\ctaCVI]{(t,n)}{\ctforallD{x}~\gamma}{A}{J}&\iff \forall x'\in \alpha(X)\colon\ctmod[\fupdate{\ctaDTVI}{X}{x}{x'}][\ctaCVI]{(t,n)}{\gamma}{A}{J}~\textrm{[for $X\in \sigSort$]}\enspace,\\
			\ctmod[\ctaDTVI][\ctaCVI]{(t,n)}{\ctexistsC{x}~\gamma}{A}{J}&\iff \exists x'\in \cint^{-1}(X)\colon\ctmod[\ctaDTVI][\fupdate{\ctaCVI}{X}{x}{x'}]{(t,n)}{\gamma}{A}{J}~\textrm{[for $X\in \isId$]}\enspace,\\
			\ctmod[\ctaDTVI][\ctaCVI]{(t,n)}{\ctforallC{x}~\gamma}{A}{J}&\iff \forall x'\in \cint^{-1}(X)\colon\ctmod[\ctaDTVI][\fupdate{\ctaCVI}{X}{x}{x'}]{(t,n)}{\gamma}{A}{J}~\textrm{[for $X\in \isId$]}\enspace.
		\end{mymathbox}
		\caption{Recursive definition of models relation for \glsentryplural{gls:cnftraceassert} $\ctFormulas{\Sigma}{S_i}{\dtVar}{\caVar}{\ctaDTVar}{\ctaCVar}$ over signature $\Sigma=\signature$, \glsentrytext{gls:ifspec} $S_i=\ifspec$, algebra $A=\algebra$ with corresponding \glsentrytext{gls:ct:dtva} $\ctaDTVI=(\ctaDTVI_s)_{s\in\sigSort}$, and \glsentrytext{gls:ifspec:int} $J=\cint\in\cintAll{S_i}{Q}$ with corresponding \glsentrytext{gls:ct:cva} $\ctaCVI=(\ctaCVI_i)_{i\in\isId}$.}\label{fig:CTA:sem}
	\end{figure*}

	A \glsentrytext{gls:cnftraceassert} $\gamma$ is \emph{valid} for \glsentrytext{gls:cnf:trace} $t\in\allTraces{\cicomp{\cint}}$ under \glsentrytext{gls:alg} $A\in\algAll{\Sigma}$ and \glsentrytext{gls:if:int} $J=\cint\in\cintAll{S_i}{A}$ iff there exists corresponding \glsentrytext{gls:ct:dtva} $\ctaDTVI\in\ctaDTVAll{\dtVar}{A}$ and \glsentrytext{gls:ct:cva} $\ctaCVI\in\ctaCVAll{\caVar}{\cint}$, such that $\ctmod[\ctaDTVI][\ctaCVI]{(t,0)}{\gamma}{A}{J}$.
	Trace $t$ is a \emph{model} of $\gamma$, written $\ctmod{t}{\gamma}{A}{J}$ iff for each corresponding \glsentrytext{gls:ct:dtva} $\ctaDTVI$ and \glsentrytext{gls:ct:cva} $\ctaCVI$ we have $\ctmod[\ctaDTVI][\ctaCVI]{(t,0)}{\gamma}{A}{J}$.
	Trace $t$ is a \emph{model} of a set of \glsentryplural{gls:cnftraceassert} $\Gamma$, written $\ctmod{t}{\Gamma}{A}{J}$ iff $\ctmod{t}{\gamma}{A}{J}$ for each $\gamma\in\Gamma$.
\end{definition}
\end{sloppypar}
Note the existential quantification for \glsentryplural{gls:dt:va} and \glsentryplural{gls:ca:va} meaning that these variables are interpreted at each point in time, compared to the rigid once.

\subsection{Specifying \glsentryplural{gls:cnftraceassert}}
\Glsentryplural{gls:cnftraceassert} can be specified by means of \glspl{gls:cta:temp} (Fig.~\ref{fig:ctspec}).
\begin{figure}
	\centering
	\begin{ctstmp}{Name}{ifSpec}
		\ctvar{var1, var2}{$\mathtt{Sort1}$}
		\ctvar{var3}{$\mathtt{Sort2}$}			
		\ctline[dashed]
		\ctaxiom{$\mathit{assertion1}(\mathrm{var1},\mathrm{var2},\mathrm{var3})$}
		\ctaxiom{$\mathit{assertion2}(\mathrm{var1},\mathrm{var2},\mathrm{var3})$}		
	\end{ctstmp}
	\caption{Configuration trace specification template}
	\label{fig:ctspec}
\end{figure}
Each template has a name and can import \glsentryplural{gls:ifspec:temp} by means of their name.
Then, a list of variables for the different sorts/interfaces are defined.
Finally, a list of configuration trace assertions are formulated over the variables and interfaces specified by the corresponding \glsentryplural{gls:ifspec}.

\subsection{\glsentrytext{gls:bb:p}: \glsentrytext{gls:const:spec}}
In the following we provide an \glsentrytext{gls:const:spec} for \glsentrytext{gls:bb:p} architectures.
First, we specify constraints regarding the behavior of \glsentryplural{BB} and \glsentryplural{KS}, respectively.
Then, we provide activation and connection constraints for such architectures by means of \glsentryplural{gls:cta:temp}.

\subsubsection{\glsentrytext{gls:bb:bb} behavior}
A \glsentrytext{BB} provides the \emph{current state} towards solving the original problem and forwards problems and solutions from \glsentryplural{KS}.
Fig.~\ref{fig:bb:bhv:bb} provides a specification of the \glsentryplural{BB} behavior in terms of a \glsentrytext{gls:cta:temp} consisting of three \glsentryplural{gls:cnftraceassert}:
\begin{itemize}
	\item if a solution to a subproblem is received on its input, then it is eventually provided at its output (Eq.~\ref{eq:bb:b:bb:1}).
	\item if solving a problem requires a set of subproblems to be solved first, those problems are eventually provided at its output (Eq.~\eqref{eq:bb:b:bb:2}).
	\item a problem is provided as long as it is not solved (Eq.~\eqref{eq:bb:b:bb:3}).
\end{itemize}
\begin{figure*}
	\centering
	\begin{ctstmp}[300]{Blackboard\_Behavior}{Blackboard}
		\ctvar{$\mathit{bb}$}{BB}
		\ctvar{$p,~p'$}{$\mathtt{PROB}$}
		\ctvar{$P$}{$\mathtt{PROB~SET}$}		
		\ctvar{$s$}{$\mathtt{SOL}$}
		\ctline[dashed]
		\ctspace[5pt]		
		\ctaxiom[eq:bb:b:bb:1]{$\ctglobally{\Big(\dtimplies{(p, s)\in \caTVal{\mathit{bb}}{\bbis}}{\cteventually{\big((p,s)\in \caTVal{\mathit{bb}}{\bbos}\big)}}\Big)}$}
		\ctspace[5pt]
		\ctaxiom[eq:bb:b:bb:2]{$\ctglobally{\dtimplies{\Big((p, P)\in \caTVal{\mathit{bb}}{\bbip}}{\big(\dtforall{p'\in P}~\left(\cteventually{p'\in\caTVal{\mathit{bb}}{\bbop}}\right)}\big)\Big)}$}
		\ctspace[5pt]		
		\ctaxiom[eq:bb:b:bb:3]{$\ctglobally{\Big(\dtimplies{p \in \caTVal{\mathit{bb}}{\bbop}}{\ctuntilW{p\in \caTVal{\mathit{bb}}{\bbop}}{(p,\mathit{solve}(p))\in \caTVal{\mathit{bb}}{\bbis}}}}\Big)$}
		\ctspace[5pt]		
	\end{ctstmp}
	\caption{\Glsentrytext{gls:const:spec}: \glsentrytext{BB} behavior.}
	\label{fig:bb:bhv:bb}
\end{figure*}%

\subsubsection{\glsentrytext{gls:bb:ks} behavior}
A \glsentrytext{KS} receives open problems via $\ksip$ and solutions for other problems via $\ksis$. It might contribute to the solution of the original problem by solving subproblems.
Fig.~\ref{fig:bb:bhv:ks} provides a specification of the \glsentryplural{KS} behavior in terms of a \glsentrytext{gls:cta:temp} consisting of three \glsentryplural{gls:cnftraceassert}:
\begin{itemize}
	\item if a \glsentrytext{KS} gets correct solutions for all the required subproblems, then it solves the problem eventually (Eq.~\eqref{eq:bb:b:ks:1}).
	\item in order to solve a problem, a \glsentrytext{KS} requires solutions only for smaller problems (Eq.~\eqref{eq:bb:b:ks:2}).
	\item if a \glsentrytext{KS} is able to solve a problem it will eventually communicate this (Eq.~\eqref{eq:bb:b:ks:3}).
\end{itemize}
\begin{figure*}
	\centering
	\begin{ctstmp}[400]{Knowledgesource\_Behavior}{Blackboard}
		\ctvar{$\mathit{ks}$}{KS}
		\ctvar{$p,~q$}{$\mathtt{PROB}$}
		\ctvar{$P$}{$\mathtt{PROB~SET}$}		
		\ctline[dashed]
		\ctspace[5pt]		
		\ctaxiom[eq:bb:b:ks:1]{$\Box\Big(\dtforall{(p,P)\in\caTVal{\mathit{ks}}{\ksop}}~\big(\dtimplies{\left(\dtforall{q\in P}~
						\cteventually{(q,\mathit{solve}(q))\in \caTVal{\mathit{ks}}{\ksis}}\right)}{\cteventually{(p,\mathit{solve}(p))\in \caTVal{\mathit{ks}}{\ksos}}}\big)\Big)$}
		\ctspace[5pt]					
		\ctaxiom[eq:bb:b:ks:2]{$\ctglobally{\Big(\dtforall{(p,P)\in\caTVal{\mathit{ks}}{\ksop}}~\dtforall{q\in P}~q \prec p\Big)}$}
		\ctspace[5pt]		
		\ctaxiom[eq:bb:b:ks:3]{$\ctglobally{\Big(p\in \caTVal{\mathit{ks}}{\mathit{prob}}\land p\in \caTVal{\mathit{ks}}{\ksip}\implies \cteventually{(\dtexists{P}~(p,P)\in \caTVal{\mathit{ks}}{\ksop})}\Big)}$}
		\ctspace[5pt]
	\end{ctstmp}
	\caption{\Glsentrytext{gls:const:spec}: \glsentrytext{KS} behavior.}
	\label{fig:bb:bhv:ks}
\end{figure*}%

\subsubsection{Activation constraints}\label{sec:runexa:actprop}
Activation constraints for the \glsentrytext{gls:bb:p} pattern are described by two \glsentryplural{gls:cnftraceassert} provided in the \glsentrytext{gls:cta:temp} in Fig.~\ref{fig:bb:act}:
\begin{itemize}
	\item Eq.~\eqref{eq:bb:a:1} denotes the conditions that there is a unique \glsentrytext{BB} component which is always activated (in contrast to \glsentryplural{KS} components which can be activated and deactivated arbitrarily).
	\item Eq.~\eqref{eq:bb:a:2} requires that whenever a \glsentrytext{KS} component offers to solve some problem, it is always activated when solutions to the required subproblems are provided.
\end{itemize}
\begin{figure*}
	\centering
	\begin{ctstmp}[450]{Blackboard\_Activation}{Blackboard}
		\ctvar{$\mathit{bb},~\mathit{bb'}$}{BB}
		\ctvar{$p,~q$}{$\mathtt{PROB}$}
		\ctvar{$P$}{$\mathtt{PROB~SET}$}		
		\ctline[dashed]
		\ctaxiom[eq:bb:a:1]{$\ctglobally{(\caFActive{\mathit{bb}}\land\ctforallC{\mathit{bb'}}~\mathit{bb'}=\mathit{bb})}$}
		\ctspace[2pt]		
		\ctaxiom[eq:bb:a:2]{$\Box\Big(\dtforall{(p,P)\in\caTVal{\mathit{ks}}{\ksop}}~\big(\dtforall{q\in P}~\dtimplies{(\cteventually{(q,\mathit{solve}(q))\in \caTVal{\mathit{bb}}{\bbos}})}{\cteventually{((q,\mathit{solve}(q))\in \caTVal{\mathit{bb}}{\bbos} \land \caFActive{\mathit{ks}})}}\big)\Big)$}
		\ctspace[5pt]
	\end{ctstmp}
	\caption{\Glsentrytext{gls:const:spec}: activation specification.}
	\label{fig:bb:act}
\end{figure*}%

\subsubsection{Connection constraints}\label{sec:runexa:conprop}
Connection constraints are also specified by a \glsentrytext{gls:cta:temp} provided in Fig.~\ref{fig:bb:conn}.
It consists of two \glsentryplural{gls:cnftraceassert}:
Eq.~\eqref{eq:bb:c:1} describes the required connections for all executions while Eq.~\eqref{eq:bb:c:2} describes all connections which are not allowed.
Roughly speaking the specification requires that for each point in time, input port $\ksip$ of a \glsentrytext{KS} is connected (only) to output port $\bbop$ of the \glsentrytext{BB} component, input port $\ksis$ of a \glsentrytext{KS} is connected (only) to output port $\bbos$ of the \glsentrytext{BB} component, output port $\ksop$ of a \glsentrytext{KS} is connected (only) to input port $\bbip$ of the \glsentrytext{BB} component, and output port $\ksos$ of a \glsentrytext{KS} is connected (only) to input port $\bbis$ of the \glsentrytext{BB} component.
\begin{figure*}
	\centering
	\begin{ctstmp}[400]{Connection}{Blackboard}
		\ctvar{$\mathit{bb}$}{BB}
		\ctvar{$\mathit{ks}$}{KS}
		\ctline[dashed]
		\ctspace[5pt]		
		\ctaxiom[eq:bb:c:1]{$\ctglobally{\Big(\caFConn{\mathit{ks}}{\ksip}{\mathit{bb}}{\bbop}\land\caFConn{\mathit{ks}}{\ksis}{\mathit{bb}}{\bbos}\land\caFConn{\mathit{bb}}{\bbip}{\mathit{ks}}{\ksop}\land\caFConn{\mathit{bb}}{\bbis}{\mathit{ks}}{\ksos}\Big)}$}
		\ctspace[5pt]		
		\ctaxiom[eq:bb:c:2]{$\ctglobally{\Big(\neg\big(\caFConn{\mathit{ks}}{\ksip}{\mathit{bb}}{\bbos}\big)\land\neg\big(\caFConn{\mathit{ks}}{\ksis}{\mathit{bb}}{\bbop}\big)\land\neg\big(\caFConn{\mathit{bb}}{\bbip}{\mathit{ks}}{\ksos}\big)\land \neg\big(\caFConn{\mathit{bb}}{\bbis}{\mathit{ks}}{\ksop}\big)\Big)}$}
		\ctspace[5pt]
	\end{ctstmp}
	\caption{\Glsentrytext{gls:const:spec}: connection specification.}
	\label{fig:bb:conn}
\end{figure*}%

\subsection{specifying \glsentryplural{gls:cns:arch}}
As stated in the introduction of this section, \glsentryplural{gls:cns:arch} are specified in three steps by specifying \glsentryplural{gls:datatype}, interfaces and \glsentryplural{gls:cnf:trace}.

\begin{sloppypar}
	\begin{definition}[\Glsentrytext{gls:const:spec}]
	An \emph{\gls{gls:const:spec}} over \glsentryplural{gls:dt:var} $\dtVar$, $\dtVar'$ and \glsentryplural{gls:ca:var} $\caVar$, \glsentryplural{gls:ct:dtvar} $\ctaDTVar$ and \glsentryplural{gls:ct:cvar} $\ctaCVar$ is a 6-tuple $\acspec$, consisting of:
	\begin{itemize}
		\item a \glsentrytext{gls:sig} $\Sigma$,
		\item a \glsentrytext{gls:dt:spec} $\dtspec\subseteq\dtFormula{\Sigma}{\dtVar'}$,
		\item a \glsentrytext{gls:if:pspec} $\specP\in\pspecAll{\Sigma}$,		
		\item an \glsentryplural{gls:ifspec} $\acIf=\ifspec\in\IS{\specP}$,
		\item \glsentryplural{gls:ifspec} $(\acIA_i)_{i\in\isId}$, where $\acIA_i\subseteq\ifFormulas{\Sigma}{\isInt_i}{\dtVar}$ for each interface identifier $i\in\isId$, and
		\item a set of \glsentryplural{gls:cnftraceassert} $\acCTA\subseteq\ctFormulas{\Sigma}{\acIf}{\dtVar}{\caVar}{\ctaDTVar}{\ctaCVar}$.
	\end{itemize}
\end{definition}
\end{sloppypar}

The semantics of an \glsentrytext{gls:const:spec} is given in terms of a set of \glsentryplural{gls:cnf:trace}.
Algorithm~\ref{alg:ac:sem} describes how to systematically derive the semantics of an \glsentrytext{gls:const:spec}.
\renewcommand{\algorithmicrequire}{\textbf{Input:}}
\renewcommand{\algorithmicensure}{\textbf{Output:}}
\begin{algorithm}
	\caption{Calculate semantics of \Glsentrytext{gls:const:spec}}
	\label{alg:ac:sem}
	\begin{algorithmic}
		\REQUIRE architecture constraint specification $S$ with:
		\begin{itemize}
			\item \glsentrytext{gls:sig} $\Sigma$,
			\item \glsentrytext{gls:dt:spec} $\dtspec\subseteq\dtFormula{\Sigma}{\dtVar'}$,
			\item \glsentrytext{gls:if:pspec} $\specP\in\pspecAll{\Sigma}$,		
			\item \glsentryplural{gls:ifspec} $\acIf=\ifspec\in\IS{\specP}$,
			\item \glsentryplural{gls:ifspec} $(\acIA_i)_{i\in\isId}$, where $\acIA_i\subseteq\ifFormulas{\Sigma}{\isInt_i}{\dtVar}$ for each interface identifier $i\in\isId$, and
			\item set of \glsentryplural{gls:cnftraceassert} $\acCTA\subseteq\ctFormulas{\Sigma}{\acIf}{\dtVar}{\caVar}{\ctaDTVar}{\ctaCVar}$.
		\end{itemize}
		\ENSURE a set of configuration traces $T$ satisfying $S$
		\STATE {$A \Leftarrow \algAll{\Sigma}$, such that $\dtmod{A}{\dtspec}$}
		\STATE{$J \Leftarrow \cintAll{S_i}{A}$, such that $\forall i\in \isId, j \in J_i \colon \ifmod{j}{\Omega_i}{A}$}
		\STATE {$T\Leftarrow t\in\allTraces{\cicomp{\cint}}$, such that $\ctmod{t}{\Gamma}{A}{\cint}$}
		\RETURN $T$
	\end{algorithmic}
\end{algorithm}

\begin{sloppypar}
A set of \glsentryplural{gls:cnf:trace} $T\subseteq\allTraces{\cicomp{\cint}}$ fulfills an \glsentrytext{gls:const:spec} $\acspec$ over \glsentryplural{gls:dt:var} $\dtVar$, $\dtVar'$ and \glsentryplural{gls:ca:var} $\caVar$, \glsentryplural{gls:ct:dtvar} $\ctaDTVar$ and \glsentryplural{gls:ct:cvar} iff there exists an \glsentrytext{gls:alg} $A\subseteq\algAll{\Sigma}$ and \glsentrytext{gls:ifspec:int} $J$, such that:
\end{sloppypar}
\begin{itemize}
	\item $A$ is a model of the \glsentrytext{gls:dt:spec}:	$\dtmod{A}{\dtspec}$\enspace,
	\item $J_i$ satisfies the corresponding \glsentrytext{gls:if:assert}: $\ifmod{j}{\Omega_i}{A}$\enspace,
	\item each trace $t\in T$ is a model of $\acCTA$: $\ctmod{t}{\acCTA}{A}{J}$\enspace. 
\end{itemize}	

Fig.~\ref{fig:spec} depicts the relationship of the syntactic and semantic concepts to specify architecture properties. First, a \glsentrytext{gls:dt:spec} determines an \glsentrytext{gls:alg} with corresponding sets of messages and operations on those sets. Later on, an \glsentrytext{gls:ifspec} determines a set of components valuated by messages from the corresponding \glsentrytext{gls:alg}.
Finally, the set of \glsentryplural{gls:cnftraceassert} determine a set of \glsentrytext{gls:cnf:trace} over those components. Note that \glsentryplural{gls:cnftraceassert} may use certain operations specified in the corresponding \glsentrytext{gls:dt:spec} which is why \glsentryplural{gls:cnf:trace} depend on the concrete interpretation of those operations.
\begin{figure}\centering
	\begin{tikzpicture}
	\node (sem) at (0,1) {Semantic Domains};
	\node (syn) at (4.5,1) {Syntactic Domains};	
	
	\node (alg) at (0,0) {Algebra};
	\node[align=center] (mess) at (1,-1) {Sets of \\ Messages};
	\node (op) at (-1,-1.5) {Operations};
	\node (cmp) at (1,-2.5) {Components};	
	\node (ct) at (0,-5) {Configuration Traces};
	
	\node[align=center] (ds) at (4.5,0) {Datatype \\ Specification};
	\node[align=center] (is) at (4.5,-2.5) {Interface \\ Specification};
	\node[align=center] (cta) at (4.5,-5) {Configuration \\ Trace \\ Assertions};

	\draw[-latex] (is.north) -- (ds.south);
	\draw[-latex] (cta.north) -- (is.south);
	
	\draw[dotted] (2.5,1) -- (2.5,-5.5);
	\draw (-2,0.5) -- (6,0.5);
	
	\draw[dashed] (alg.south) -- (mess.north);
	\draw[dashed] (alg.south) -- (op.north);
	\draw[-latex,dashed] (ds.west) -- (alg.east);
	\draw[-latex,dashed] (is.west) -- (cmp.east);
	\draw[-latex,dashed] (cta.west) -- (ct.east);
	\draw[-latex] (cmp.north) -- (mess.south);
	\draw[-latex] (ct.north) -- (cmp.south);
	\draw[-latex] (ct.north) -- (op.south);
	\end{tikzpicture}
	\caption{Relationship between syntactic and semantic concepts to specify \glsentryplural{gls:cns:arch}.}
\end{figure}\label{fig:spec}

\subsection{Summary}
Table~\ref{tab:cta} provides an overview of the concepts introduced in this section.
For each concept it provides a brief description and related notation.
\begin{table*}
	\centering
	\caption{Overview of concepts to specify \glsentryplural{gls:cns:arch}.\label{tab:cta}}
	\begin{tabular}{r@{\hspace{10pt}}p{7cm}@{\hspace{15pt}}p{2.6cm}}
		\toprule
		\textbf{Concept} & \textbf{\textit{Description}} & \textbf{\textit{Related Notation}} \\
		\toprule
		\glsentrytext{gls:ca:var}&\glsentrydesc{gls:ca:var}&\glsentrysymbol{gls:ca:var}\\
		\cmidrule{2-3}		
		\glsentrytext{gls:ca:term} &\glsentrydesc{gls:ca:term}&\glsentrysymbol{gls:ca:term}\\
		\emph{\glsentrytext{gls:cns:act}}&\emph{\glsentrydesc{gls:cns:act}}&\\
		\emph{\glsentrytext{gls:cns:conn}}&\emph{\glsentrydesc{gls:cns:conn}}&\\
		\cmidrule{2-3}		
		\glsentrytext{gls:ca:va}&\glsentrydesc{gls:ca:va}&\glsentrysymbol{gls:ca:va}\\
		\cmidrule{2-3}		
		\glsentrytext{gls:ca:term:sem}&\glsentrydesc{gls:ca:term:sem}&\glsentrysymbol{gls:ca:term:sem}\\
		\cmidrule{2-3}		
		\glsentrytext{gls:cnfassert}&\glsentrydesc{gls:cnfassert}&\glsentrysymbol{gls:cnfassert}\\
		\cmidrule{2-3}		
		\glsentrytext{gls:ca:sem}&\glsentrydesc{gls:ca:sem}&\glsentrysymbol{gls:ca:sem}\\
		\cmidrule{2-3}		
		\glsentrytext{gls:ct:dtvar}&\glsentrydesc{gls:ct:dtvar}&\glsentrysymbol{gls:ct:dtvar}\\
		\cmidrule{2-3}		
		\glsentrytext{gls:ct:cvar}&\glsentrydesc{gls:ct:cvar}&\glsentrysymbol{gls:ct:cvar}\\
		\cmidrule{2-3}		
		\glsentrytext{gls:cnftraceassert}&\glsentrydesc{gls:cnftraceassert}&\glsentrysymbol{gls:cnftraceassert}\\
		\cmidrule{2-3}		
		\glsentrytext{gls:ct:dtva}&\glsentrydesc{gls:ct:dtva} &\glsentrysymbol{gls:ct:dtva}\\
		\cmidrule{2-3}		
		\glsentrytext{gls:ct:cva}&\glsentrydesc{gls:ct:cva} &\glsentrysymbol{gls:ct:cva}\\	
		\cmidrule{2-3}		
		\glsentrytext{gls:cta:sem}&\glsentrydesc{gls:cta:sem}&\glsentrysymbol{gls:cta:sem}\\		
		\cmidrule{2-3}		
		\glsentrytext{gls:cta:temp}&\glsentrydesc{gls:cta:temp}&\glsentrysymbol{gls:cta:temp}\\
		\cmidrule{2-3}		
		\glsentrytext{gls:const:spec}&\glsentrydesc{gls:const:spec}&\glsentrysymbol{gls:const:spec}\\
		\bottomrule
	\end{tabular}
\end{table*}

%% file: cnfdiag.tex
\section{\glsentryplural{gls:cnfdiag}}\label{sec:cnfd}
\Glsentryplural{gls:cnftraceassert} are actually sufficient to specify each property of dynamic architectures.
However, sometimes, certain common constraints are better expressed by a so-called \glsentrytext{gls:cnfdiag}.
\Glsentryplural{gls:cnfdiag} complement \glsentryplural{gls:cnftraceassert} and are well-suited to specify interfaces and introduce certain common connection and activation constraints in one graphical notation.

\subsection{Simple \glsentryplural{gls:cnfdiag}}
In its simplest form, a \gls{gls:cnfdiag} is just a graphical representation of an \glsentrytext{gls:ifspec}. It consists of boxes and small circles, representing an interface identifier and corresponding local, input, and output ports. Thereby, transparent circles inside a component represent local ports, while white and black circles on the border of a component represent input and output ports, respectively. In addition, the ports are annotated with their name. The diagram is surrounded by a box which adds a name to the specification, a reference to imported \glsentryplural{gls:if:pspec}, and a set of \glsentryplural{gls:if:assert}.

Fig.~\ref{fig:cnfdiag} shows a graphical representation of a \glsentrytext{gls:cnfdiag} $\mathit{Name}$ corresponding to an \glsentrytext{gls:ifspec} $S_i=\ifspec$ and family of \glsentryplural{gls:if:assert} $\Omega$, with:
\begin{itemize}
	\item interface identifiers	$\isId=\{\mathit{If1},\mathit{If2}\}$,
	\item interfaces $\isInt_{\mathit{If1}}=(\{i_0\},\{o_1\})$ and $\isInt_{\mathit{If2}}=(\emptyset,\{o_0\})$, and
	\item \glsentryplural{gls:if:assert} $\Omega_{\mathit{If1}}=\{\mathit{InterfaceAssertion}\}$.
\end{itemize}
\begin{figure}
	\centering
	\input{img/cnfd.tex}
	\caption{Simple \glsentrytext{gls:cnfdiag} as a graphical means to specify component types.}
	\label{fig:cnfdiag}	
\end{figure}%
As for \glsentryplural{gls:ifspec} a \glsentrytext{gls:cnfdiag} may include also a corresponding port specification.
Fig.~\ref{fig:cnfdiag}, e.g., includes a port specification declaring ports $l_0, i_0, o_0, o_1$ and corresponding types $\pstype(i_0)=\pstype(o_1)=\mathit{Sort1}$ and $\pstype(o_0)=\pstype(l_0)=\mathit{Sort2}$.

\subsection{Advanced \glsentryplural{gls:cnfdiag}}
Sometimes, it is convenient to annotate \glsentryplural{gls:cnfdiag} by certain activation and connection constraints.
These annotations are actually graphical synonyms for certain \glsentryplural{gls:cnftraceassert} and they can be separated into activation and connection annotations.
\subsection{Activation annotations}
Activation annotations enhance a \glsentrytext{gls:cnfdiag} by constraints regarding the activation and deactivation of certain components.
Thus, they are modeled by predicates or mappings over interface identifiers.

\subsubsection{Min-Max annotations}
\Glsentryplural{gls:act:minmax} restrict the number of active components of a certain type.
\begin{definition}[\Glsentrytext{gls:act:minmax}]
	A \emph{\gls{gls:act:minmax}} for an interface specification $\ifspec$ is a pair of mappings $\cdMinMax$, with $\mmMin$, $\mmMax\colon\pFun{\isId}{\NN}$.
\end{definition}
Note that not every interface needs to be annotated.

A min-max annotation can be easily specified by adding the two numbers to an interface identifier in a \glsentrytext{gls:cnfdiag}.
Fig.~\ref{fig:cd:act}, for example, depicts a constraint that at least $n$ but at most $m$ components of type $\mathit{If}$ are active at each point in time.
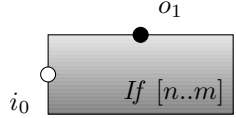
\begin{figure}
	\centering
	\input{img/active.tex}
	\caption{Min-max annotation requiring at least $n$ but at most $m$ active components of type $\mathit{If}$ at each point in time.}
	\label{fig:cd:act}	
\end{figure}%

The semantics of \glsentryplural{gls:act:minmax} is given by means of corresponding \glsentryplural{gls:cnftraceassert}.
\begin{definition}[\Glsentrytext{gls:act:minmax} semantics]
	The \emph{semantics} of a \emph{\glsentrytext{gls:act:minmax}} $\cdMinMax$, for \glsentrytext{gls:ifspec} $\ifspec$ is given by the following \glsentrytext{gls:cnftraceassert}:
	\begin{equation}
		\ctglobally{\Big(\bigwedge_{i\in\domain{\mmMin}}\caFMin{i}{\mmMin(i)}\enspace\land\enspace\bigwedge_{i\in\domain{\mmMax}}\caFMax{i}{\mmMax(i)}\Big)}\enspace.
	\end{equation}
\end{definition}
Note that if $\mmMin(i)=\mmMax(i)$, then one number can be omitted and only one is to be annotated in the corresponding \glsentrytext{gls:cnfdiag}.

\subsubsection{Rigid Annotations}
A \glsentrytext{gls:act:minmax} does only constrain the number of components for a certain interface at a certain point in time.
It does not say anything about which components these are.
Assume, for example, we want to specify that a unique component of type $i$ is active at each point in time. If we put a min-max constraint of $1$ for interface $i$, than this means that exactly one component of type $i$ is active at each point in time. However, it can be that at some point only component $c_1$ is active while at another time $c_2$ is active.

To specify that at each point in time the same components have to be activated we can use so-called rigid annotations.
\begin{definition}[\Glsentrytext{gls:act:rigid}]
	A \emph{\gls{gls:act:rigid}} for an \glsentryplural{gls:ifspec} $\ifspec$ and \glsentryplural{gls:ct:cvar} $\ctaCVar=(\ctaCVar_i)_{i\in\isId}$ is a mapping $\cdRig\colon$ $\isId\to\pset{\ctaCVar}$, such that $\forall i\in\isId\colon \cdRig(i)\subseteq\ctaCVar_i$\enspace.
\end{definition}

A rigid annotation is specified by a list of variables for each interface. Note, however, that we require the use of \emph{rigid} component variables here.

Fig.~\ref{fig:cd:rig} depicts a constraint that only components $c_1$ and $c_2$ are activated throughout system execution.
\begin{figure}
	\centering
	\input{img/rigid.tex}
	\caption{Rigid annotation requiring that only components $c_1$ and $c_2$ are activated throughout system execution.}
	\label{fig:cd:rig}	
\end{figure}
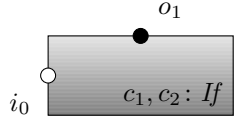%

The semantics of rigid annotations is given by means of configuration trace assertions.
\begin{definition}[\Glsentrytext{gls:act:rigid} semantics]
	The \emph{semantics} of a \emph{\glsentrytext{gls:act:rigid}} $\cdRig$ for interface specification $\ifspec$ is given by the following configuration trace assertion:
	\begin{equation}
		\ctglobally{\Big(\bigwedge_{i\in\isId}\big(\forall v\colon \bigvee_{c \in A_r(i)} (c=v)\big)\Big)}\enspace,
	\end{equation}
	where $v\in\caVar$ is a (non-rigid) \glsentrytext{gls:ca:var}.
\end{definition}

\subsection{Connection annotations}
Connection annotations enhance a \glsentrytext{gls:cnfdiag} by constraints regarding the connection of certain components.
Thus, they are modeled by predicates or mappings over relations over interface ports.

\begin{definition}[\Glsentrytext{gls:conn:req}]
	A \emph{\gls{gls:conn:req}} for an \glsentrytext{gls:ifspec} $S_i$ is a relation $\cdRConn\colon\pFun{\ifIn{S_i}}{\ifOut{S_i}}$.
\end{definition}

A required connection annotation is expressed by solid connections between the corresponding ports.
Figure~\ref{fig:cd:rc} denotes a constraint that a component of type $\mathit{If1}$ is always connected to a component of type $\mathit{If2}$ through ports $i$ and $o$, respectively.
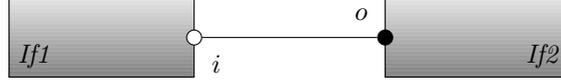
\begin{figure}%
	\centering%
	\input{img/rcon}%
	\caption{Required connection annotation requiring components of type $\mathit{If1}$ to be always connected to a component of type $\mathit{If2}$ through ports $i$ and $o$, respectively.}%
	\label{fig:cd:rc}%
\end{figure}%

\begin{definition}[\Glsentrytext{gls:conn:req} semantics]\label{def:ann:conn}
	The \emph{semantics} of a \emph{\glsentrytext{gls:conn:req}} $\cdRConn$ for \glsentrytext{gls:ifspec} $S_i\in\ifAll{S_p}$ over \glsentrytext{gls:if:pspec} $S_p\in\pspecAll{\Sigma}$ is given by the following \glsentrytext{gls:cnftraceassert}:
	\begin{equation}\label{eq:ann:conn}
		\ctglobally{\Big(\bigwedge_{((j,i),(k,o))\in \cdRConn} \caFConn{j}{i}{k}{o}\enspace\land \bigwedge_{((j,i),(k,o))\in\mathit{rest}(\cdRConn)}\neg \big(\caFConn{j}{i}{k}{o}\big)\Big)}\enspace,
	\end{equation}
	where $\mathit{rest}(\cdRConn)=\ifIn{S_i}\times \ifOut{S_i}\setminus\cdRConn$.
\end{definition}

Note that a required connection annotation induces a \emph{full} homomorphism (a homomorphism preserving non-edge) between a configuration diagram and corresponding architecture configuration.

\begin{property}[Required connection induces full homomorphism]
	Let $\cdRConn$ be a \glsentrytext{gls:conn:req} for \glsentrytext{gls:ifspec} $S_i=\ifspec$ with induced \glsentrytext{gls:cnftraceassert} $\varphi$. Moreover, let $\cint\in\cintAll{S_i}{A}$ be a corresponding \glsentrytext{gls:ifspec:int} under an algebra $A=\algebra\in\algAll{\Sigma}$.
	Finally, let $\delta\colon\cmpName{\cicomp{\cint}}\to\isId$ denote the interface for each component identifier and $\intIn=(\intIn_d)_{d\in\cmpName{\cicomp{\cint}}}$ and $\intOut=(\intOut_d)_{d\in\cmpName{\cicomp{\cint}}}$ denoting the corresponding input and output port interpretations for component identifier $c$.
	
	Then, $(\intIn,\intOut)$ form a homomorphism from each \glsentrytext{gls:aconf} of each trace $t\in \allTraces{\cicomp{\delta}}$ satisfying $\varphi$ (for each point in time) to the corresponding \glsentrytext{gls:conn:feasible}:
	\begin{equation}
		\begin{split}
			&\forall t\in \big\{t\in\allTraces{\cicomp{\cint}}\mid \ctmod{t}{\varphi}{A}{J}\big\}, n\in \NN,c\in\cp{t(n)}{1},\\
			&c'\in\cp{t(n)}{1},d=\cp{c}{1},d'=\cp{c'}{1},p\in\cp{c}{3},p'\in\cp{c}{4}\colon\\
			&\qquad((d,p),(d',p'))\in \cp{t(n)}{2}\iff\\
			&\qquad\Big((\delta(d),\intIn_{\delta(d)}(p)),(\delta(d'),\intOut_{\delta(d')}(p'))\Big)\in \cdRConn\enspace.
		\end{split}
	\end{equation}
\end{property}
\begin{proof}
	Let $t\in\big\{t\in\allTraces{\cicomp{\cint}}\mid \ctmod{t}{\varphi}{A}{J}\big\}$, $n\in\NN$, $c\in\cp{t(n)}{1}$, $c'\in\cp{t(n)}{1}$, $d=\cp{c}{1}$, $d'=\cp{c'}{1}$, $p\in\cp{c}{3}$, and $p'\in\cp{c}{4}$.
	
	($\implies$):
		Assume $((d,p),(d',p'))\in \cp{t(n)}{2}$ and show $\Big((\delta(d),\intIn_{\delta(d)}(p)),(\delta(d'),\intOut_{\delta(d')}(p'))\Big)\in \cdRConn$ by contradiction.
		Thus, assume $\neg\Big((\delta(d),\intIn_{\delta(d)}(p)),(\delta(d'),\intOut_{\delta(d')}(p'))\Big)\in \cdRConn$.
		The, from Def.~\ref{def:ann:conn}, have that $\varphi$ contains conjunction $\neg \big(\caFConn{j}{i}{k}{o}\big)$.
		Thus, since $\ctmod{t}{\varphi}{A}{J}$, have $((d,p),(d',p'))\notin \cp{t(n)}{2}$ by Def.~\ref{def:cnf:term:sem} which is in contradiction with the assumption.
		
	($\impliedby$):
		Assume $\Big((\delta(d),\intIn_{\delta(d)}(p)),(\delta(d'),\intOut_{\delta(d')}(p'))\Big)\in \cdRConn$.
		Thus, from Def.~\ref{def:ann:conn}, have that $\varphi$ contains conjunction $\caFConn{j}{i}{k}{o}$.
		Thus, since $\ctmod{t}{\varphi}{A}{J}$, have $((d,p),(d',p'))\in \cp{t(n)}{2}$ by Def.~\ref{def:cnf:term:sem}.
\end{proof}

\subsection{\glsentrytext{gls:bb:p}: \glsentrytext{gls:cnfdiag}}
The \glsentrytext{gls:ifspec} of the \glsentrytext{gls:bb:p} pattern as well as the activation and connection constraints of Eq.~\eqref{eq:bb:a:1} and Eq.~\eqref{eq:bb:c:1}, respectively, could have been also expressed by the \glsentrytext{gls:cnfdiag} in Fig.~\ref{fig:BB:CD}:
\begin{figure}%
	\centering%
	\input{img/blackboard}%
	\caption{\Glsentrytext{gls:cnfdiag} for \glsentrytext{gls:bb:p} architectures.}%
	\label{fig:BB:CD}%
\end{figure}%
\begin{itemize}
	\item The interface specification is given by the two interfaces $\emph{KK}$ and $\emph{BB}$, respectively.
	\item Eq.~\eqref{eq:bb:a:1} is addressed by adding variable $\mathit{bb}$ and the corresponding min-max annotation. 
	\item Eq.~\eqref{eq:bb:c:1} is addressed by the solid connections between the ports.
\end{itemize}

\subsection{Refining \glsentryplural{gls:cnfdiag}}
\Glsentryplural{gls:cnfdiag} are well-suited to specify interfaces and certain common activation and connection constraints.

However, not all constraints can be specified only by means of \glsentryplural{gls:cnfdiag} which is why \glsentryplural{gls:cnfdiag} are usually refined by adding further constraints by means of \glsentryplural{gls:cnftraceassert}.

\subsection{Summary}
Table~\ref{tab:cnfdiag} provides an overview of the concepts introduced in this section.
For each concept a brief description as well as related notation is provided.
\begin{table*}
	\centering
	\caption{Overview of concepts for \glsentryplural{gls:cnfdiag}.\label{tab:cnfdiag}}
	\begin{tabular}{r@{\hspace{10pt}}p{7.5cm}@{\hspace{15pt}}p{3cm}}
		\toprule
		\textbf{Concept} & \textbf{\textit{Description}} & \textbf{\textit{Related Notation}} \\
		\midrule
		\glsentrytext{gls:cnfdiag} &\glsentrydesc{gls:cnfdiag}&\glsentrysymbol{gls:cnfdiag}\\
		\cmidrule{2-3}			
		activation annotations & annotations to constraint the activation/deactivation of components &\\
		\emph{\glsentrytext{gls:act:minmax}} & \emph{\glsentrydesc{gls:act:minmax}} & \emph{\glsentrysymbol{gls:act:minmax}}\\
		\emph{\glsentrytext{gls:act:rigid}} & \emph{\glsentrydesc{gls:act:rigid}} & \emph{\glsentrysymbol{gls:act:rigid}}\\
		\cmidrule{2-3}		
		connection annotations & annotations to constrain the connection between components & \\
		\emph{\glsentrytext{gls:conn:req}} & \emph{\glsentrydesc{gls:conn:req}} &
		\begin{tikzpicture}
			\draw (0,0) -- (1,0);
		\end{tikzpicture}\\
		\bottomrule
	\end{tabular}
\end{table*}

%% file: img/cnfd.tex
\begin{tikzpicture}[scale=1, every node/.style={transform shape},lport/.style={circle,draw,inner sep=2pt},iport/.style={circle,draw,fill=white,inner sep=2pt},oport/.style={circle,draw,fill,inner sep=2pt}, component/.style={draw,minimum height=30pt,minimum width=70pt, bottom color=black!50, top color=black!10}, sComp/.style={draw,minimum height=30pt,minimum width=70pt, bottom color=black!50, top color=black!10}, conn/.style={dotted}, sconn/.style={draw}]
		\node[draw, minimum height=162pt, minimum width=9cm] at (0,-0.3) {};
		\draw[double, double distance = 2pt] (-4.5,55pt) -- (4.5,55pt);
		\node[anchor=base west, text width=3cm] at (-4.5,61pt) {\textbf{Diagram} Name};
		\node[anchor=base east, text width=6cm, align=right] at (4.5,61pt) {\textbf{based on} PSpec \textbf{uses} DTSpec};
		\draw (-4.5,30pt) -- (4.5,30pt);
		\node[anchor=base west, text width=7.5cm] at (-4.5,43pt) {$i_0,o_1\colon Sort1$};
		\node[anchor=base west, text width=7.5cm] at (-4.5,33pt) {$o_0,l_0\colon Sort2$};
		
		iffalse Component C1------------------------------------------------------------- fi
		\node[component] (c1) at (-2,-0.3) {};
		\node[above left] at (c1.south east) {$\mathit{If1}$};
		\node [lport] (c1l0) at ($(c1)+(-20pt,5pt)$) {};
		\node [right = 1pt of c1l0]{$l_0=\{A\}$};	
		\node [iport] (c1i0) at ($(c1)+(-35pt,0pt)$) {};
		\node [below left = 1pt and 1pt of c1i0]{$i_0$};
		\node [oport] (c1o1) at ($(c1)+(0pt,15pt)$) {};
		\node [above right = 1pt and 1pt of c1o1]{$o_1$};
		iffalse Component C2------------------------------------------------------------- fi
		\node[component] (c2) at (2,-0.3) {};
		\node[above left] at (c2.south east) {$\mathit{If2}$};		
		\node [oport] (c2o0) at ($(c2)+(0pt,15pt)$) {};
		\node [above left = 1pt and 1pt of c2o0]{$o_0$};

		\draw[thin] (-4.5,-40pt) -- (4.5,-40pt);
		\node[anchor=base west, text width=7.5cm] at (-4.5,-50pt) {\textbf{var}};
		\node[anchor=base west, text width=7.5cm] at (-3.5,-50pt) {$v_1\colon$};
		\node[anchor=base west, text width=7.5cm] at (-2.5,-50pt) {$\mathit{Sort1}$};
		\node[anchor=base west, text width=7.5cm] at (-3.5,-60pt) {$v_2\colon$};
		\node[anchor=base west, text width=7.5cm] at (-2.5,-60pt) {$\mathit{Sort2}$};
		\node [anchor=base east, align=right, text width=7.5cm] at (4.5,-70pt) {$\mathit{If1}$};
		\draw[dashed] (-4.5,-67pt) -- (3.8,-67pt);
		\draw[dashed] (4.3,-67pt) -- (4.5,-67pt);		
		\node[anchor=base west, text width=7.5cm] at (-4.5,-83pt) {$\mathit{InterfaceAssertion}(\mathit{If1}.i_0,\mathit{If1}.o_1,\mathit{If1}.l_0,\mathit{If2}.o_0)$};
\end{tikzpicture}

%% file: img/active.tex
\begin{tikzpicture}[scale=1, every node/.style={transform shape},iport/.style={circle,draw,fill=white,inner sep=2pt},oport/.style={circle,draw,fill,inner sep=2pt}, component/.style={draw,minimum height=30pt,minimum width=70pt, bottom color=black!50, top color=black!10}, sComp/.style={draw,minimum height=30pt,minimum width=70pt, bottom color=black!50, top color=black!10}, conn/.style={dotted}, sconn/.style={draw}]

	iffalse Component C1------------------------------------------------------------- fi
	\node[component] (c1) at (-2,0) {};
	\node[above left] at (c1.south east) {$\mathit{If}$ $[n..m]$};
	\node [iport] (c1i0) at ($(c1)+(-35pt,0pt)$) {};
	\node [below left = 1pt and 1pt of c1i0]{$i_0$};
	\node [oport] (c1o1) at ($(c1)+(0pt,15pt)$) {};
	\node [above right = 1pt and 1pt of c1o1]{$o_1$};
\end{tikzpicture}

%% file: img/rigid.tex
\begin{tikzpicture}[scale=1, every node/.style={transform shape},iport/.style={circle,draw,fill=white,inner sep=2pt},oport/.style={circle,draw,fill,inner sep=2pt}, component/.style={draw,minimum height=30pt,minimum width=70pt, bottom color=black!50, top color=black!10}, sComp/.style={draw,minimum height=30pt,minimum width=70pt, bottom color=black!50, top color=black!10}, conn/.style={dotted}, sconn/.style={draw}]

	iffalse Component C1------------------------------------------------------------- fi
	\node[component] (c1) at (-2,0) {};
	\node[above left] at (c1.south east) {$c_1,c_2\colon\mathit{If}$};
	\node [iport] (c1i0) at ($(c1)+(-35pt,0pt)$) {};
	\node [below left = 1pt and 1pt of c1i0]{$i_0$};
	\node [oport] (c1o1) at ($(c1)+(0pt,15pt)$) {};
	\node [above right = 1pt and 1pt of c1o1]{$o_1$};
\end{tikzpicture}

%% file: img/rcon.tex
\begin{tikzpicture}[scale=1, every node/.style={transform shape}, iport/.style={circle,draw,fill=white,inner sep=2pt},oport/.style={circle,draw,fill,inner sep=2pt}, component/.style={draw,minimum height=30pt,minimum width=70pt, bottom color=black!50, top color=black!10}, conn/.style={dotted}, sConn/.style={draw}, fConn/.style={dashed}]
		iffalse Component C1------------------------------------------------------------- fi
		\node[component] (c1) at (-2.5,0) {};
		\node[above right] at (c1.south west) {$\mathit{If1}$};		
		\node [iport] (c1i0) at ($(c1)+(35pt,0pt)$) {};
		\node [below right = 1pt and 1pt of c1i0]{$i$};

		iffalse Component C2------------------------------------------------------------- fi
		\node [component] (c2) at (2.5,0) {};
		\node[above left] at (c2.south east) {$\mathit{If2}$};		
		\node [oport] (c2o0) at ($(c2)-(35pt,0pt)$) {};
		\node [above left = 1pt and 1pt of c2o0]{$o$};				

		\draw [sConn] (c1i0) edge (c2o0);
		
\end{tikzpicture}

%% file: img/blackboard.tex
\begin{tikzpicture}[scale=1, every node/.style={transform shape}, lport/.style={circle,draw,inner sep=2pt}, iport/.style={circle,draw,fill=white,inner sep=2pt},oport/.style={circle,draw,fill,inner sep=2pt}, component/.style={draw,minimum height=40pt,minimum width=85pt, bottom color=black!50, top color=black!10}, conn/.style={dotted}, sConn/.style={draw}, fConn/.style={dashed}]
		
		\node[draw, minimum height=245pt, minimum width=10cm] at (0,7pt) {};
		\draw[double, double distance = 2pt] (-5,110pt) -- (5,110pt);
		\node[anchor=base west, text width=4cm] at (-5,118pt) {\textbf{Diagram} Blackboard};
		\node[anchor=base east, text width=6cm, align=right] at (5,118pt) {\textbf{based on} Blackboard \textbf{uses} ProbSol};
		\node [anchor=base west, text width=5cm] at (-5,3.4) {$\mathit{prob}:\pset{\mathtt{PROB}}$};		
		\draw[thin] (-5,90pt) -- (5,90pt);
		
		iffalse Component C1------------------------------------------------------------- fi
		\node[component] (c1) at (0,1.7) {};
		\node[below right] at (c1.base west) {$\mathit{KS}$};
		\node [lport] (c1l0) at ($(c1)+(-15pt,5pt)$) {};
		\node [right = 1pt of c1l0]{$\mathit{prob}$};
		\node [iport] (c1i0) at ($(c1)-(30pt,20pt)$) {};
		\node [below left = 1pt and 1pt of c1i0]{$\ksip$};
		\node [iport] (c1i1) at ($(c1)-(10pt,20pt)$) {};
		\node [below left = 1pt and 1pt of c1i1]{$\ksis$};
		\node [oport] (c1o0) at ($(c1)+(10pt,-20pt)$) {};
		\node [below left = 1pt and 1pt of c1o0]{$\ksop$};
		\node [oport] (c1o1) at ($(c1)+(30pt,-20pt)$) {};
		\node [below left = 1pt and 1pt of c1o1]{$\ksos$};
		iffalse Component C2------------------------------------------------------------- fi
		\node [component] (c2) at (0,-1) {};
		\node[above right] at (c2.south west) {$\mathit{bb}\colon\mathit{BB}~[1]$};		
		\node [oport] (c2o0) at ($(c2)+(-30pt,20pt)$) {};
		\node [above right = 1pt and 1pt of c2o0]{$\bbop$};
		\node [oport] (c2o1) at ($(c2)+(-10pt,20pt)$) {};	
		\node [above right = 1pt and 1pt of c2o1]{$\bbos$};		
		\node [iport] (c2i0) at ($(c2)+(10pt,20pt)$) {};			
		\node [above right = 1pt and 1pt of c2i0]{$\bbip$};		
		\node [iport] (c2i1) at ($(c2)+(30pt,20pt)$) {};			
		\node [above right = 1pt and 1pt of c2i1]{$\bbis$};				

		\draw [sConn] (c1i0) edge (c2o0);
		\draw [sConn] (c1i1) edge (c2o1);		
		\draw [sConn] (c2i0) edge (c1o0);
		\draw [sConn] (c2i1) edge (c1o1);
		
		\draw [thin] (-5,-2.5) -- (5,-2.5);

		\node [anchor=base west, text width=4cm] at (-5,-2.8) {\textbf{var}};
		\node [anchor=base west, text width=4cm] at (-4,-2.8) {p$\colon$};
		\node [anchor=base west, text width=4cm] at (-3,-2.8) {$\mathtt{PROP}$};
		\node [anchor=base west, text width=4cm] at (-4,-3.2) {P};
		\node [anchor=base west, text width=4cm] at (-3,-3.2) {$\pset{\mathtt{PROP}}$};
		
		\draw [dashed] (-5,-3.4) -- (4.2,-3.4);
		\draw [dashed] (4.7,-3.4) -- (5,-3.4);
		\node [anchor=base west] at (4.15,-3.5) {$\mathit{bb}$};		
		\node [anchor=base west, text width=4cm] at (-5,-3.8) {$\ksop=(p,P)\implies p\in\mathit{prob}$};
\end{tikzpicture}

%% file: verifying.tex
\section{Verifying a specification}\label{sec:analysis}
As demonstrated by the example, the approach allows for formal specification of patterns of dynamic architectures.
Such a specification is useful, for example, to check pattern conformance of an architecture, i.e., whether a concrete architecture implements a certain pattern.
On the other hand, having a formal specification of a pattern allows to formally analyze the specification.

This section demonstrates how a specification can be used to formally reason about it.
Therefore, we specify a characteristic guarantee of \glsentrytext{gls:bb:p} architectures by means of \glsentryplural{gls:cnftraceassert} and prove it from the specification developed so far.

\subsection{Specifying properties}
First, a property is specified over the architecture.
The property can be formally specified by applying the techniques presented so far.
As stated in the introduction of this article, one characteristic property of a \glsentrytext{gls:bb:p} architecture is its ability to (collaboratively) solve a complex problem even if no single \glsentrytext{KS} exists which is able to solve the problem on its own.

\begin{theorem}\label{thm:bb}
	Assuming that \glsentryplural{KS} are active when required:
	\begin{equation}\label{eq:bb:a}
		\ctglobally{\Big(\dtforall{p\in\caTVal{\mathit{bb}}{\bbop}}~\cteventually{\big(\dtexists{\mathit{ks}}~p\in\caTVal{\mathit{ks}}{\mathit{prob}}\big)}\Big)}\enspace,
	\end{equation}
	a \glsentrytext{gls:bb:p} architecture guarantees to solve the original problem:
	\begin{align}\label{eq:bb:c}
	\ctglobally{\Big(\dtimplies{p\in \caTVal{\mathit{bb}}{\bbip}}{\cteventually{(p,\mathit{solve(p)})\in \caTVal{\mathit{bb}}{\bbos}}}\Big)\enspace.}
	\end{align}
\end{theorem}

\subsection{Verifying the specification}
Then, the specification is verified w.r.t. the identified property by proving it from the specification.
Therefore, the constraints introduced in the specification of the pattern serve as the major arguments throughout the proof.
In the following we prove Thm.~\ref{thm:bb} by applying the different constraints specified for the pattern.

\begin{proof}
	The proof is by well-founded induction over the problem relation $\prec$:
	First, by Eq.~\eqref{eq:bb:a:1} or Fig.~\ref{fig:BB:CD} there exists a unique blackboard component $\mathit{bb}$ which is always activated.
	Then, by Eq.~\eqref{eq:bb:a}, we are sure that for each problem eventually a \texttt{KnowledgeSource} $\mathit{ks}$ exists which is capable to solve the problem. By Eq.~\eqref{eq:bb:b:ks:3} the $\mathit{ks}$ will eventually communicate the subproblems $p'$ it requires to solve the original problem $p$ on its port $\ksop$ and this information is then transferred to port $\bbip$ of $\mathit{bb}$ by the connection constraints imposed by Eq.~\eqref{eq:bb:c:1} or Fig.~\ref{fig:BB:CD}.
	By Eq.~\eqref{eq:bb:b:bb:2}, $\mathit{bb}$ will provide these subproblems $p'$ eventually on its output port $\bbop$ and publish it as long as it is not solved (as required by Eq.~\eqref{eq:bb:b:bb:3}).
	Since the subproblems $p'$ provided to $\mathit{bb}$ are strictly less than the original problem $p$ (due to Eq.~\eqref{eq:bb:b:ks:2}), they will eventually be solved and its solutions $s'$ provided on port $\bbos$ of $\mathit{bb}$ by the induction hypothesis.
	$\mathit{ks}$ will eventually be activated for each solution $s'$ (Eq.~\eqref{eq:bb:a:2}) and due to Eq.~\eqref{eq:bb:c:1} or Fig.~\ref{fig:BB:CD} the solutions $s'$ are transferred to the corresponding port $\ksis$ of $\mathit{ks}$.
	Thus, $\mathit{ks}$ eventually has all solutions $s'$ to its subproblems $p'$ and will then solve the original problem $p$ by Eq.~\eqref{eq:bb:b:ks:1} and publish the solution $s$ on its port $\ksos$.
	Solution $s$ is received eventually by $\mathit{bb}$ on its port $\bbis$ due to Eq.~\eqref{eq:bb:c:1} or Fig.~\ref{fig:BB:CD} and is finally provided by $\mathit{bb}$ on its port $\bbos$ due to Eq.~\eqref{eq:bb:b:bb:1}.
\end{proof}

%% file: discussion.tex
\section{Discussion}\label{sec:discussion}
The approach presented in this article is characterized by the following properties:
\begin{itemize}
	\item \emph{Formal}: The approach is based on a formal foundation with a formal semantics for each specification technique.
	\item \emph{Uniform}: Each specification technique is based on a uniform model for dynamic architectures.
	\item \emph{Abstract}: The approach is based on a rather abstract notion of architecture.
	\item \emph{Model-theoretic semantics}: The semantics of each technique is given in terms of models which satisfy a corresponding specification.
\end{itemize}

These properties induce several benefits as well as some drawbacks which we will briefly discuss in the following.

\subsection{Unambiguous interpretation}
Due to its formal nature, specifications can be interpreted as mathematical models.
As demonstrated in Sect.~\ref{sec:analysis}, this enables formal analyses and verifications of the specifications.

\subsection{Consistent specification techniques}
Since the semantics of each technique is given in terms of a uniform model, inconsistencies can be detected more easily since the impact of different specification assertions can be directly related to each other.

\subsection{Generality}
Due to the abstract nature of the underlying model, the approach is very general.
Its specifications and corresponding verification results can be interpreted for different, concrete architecture instances.

\subsection{Extensibility}
Since the semantics is given in terms of model-theory, this makes the approach easily extensible.
New constructs can be easily integrated by describing its impact of the underlying model.

\subsection{Stepwise refinement}
Another implication of the model-theoretic semantics is the possibility for stepwise refinement of specifications.
After specifying a set of interfaces we can add more and more constraints.
Thus, gradually lessen the space of possible architectures satisfying the specification.

\subsection{Potential limitations}
Of course, the approach imposes some limitations which we will briefly discuss in the following.

\subsubsection{Generality}
While generality is listed as a benefit of the approach, it can also be seen as a drawback.
Due to the abstract nature of the approach, it omits several details regarding component instantiation and communication between components.
Thus, the approach is not well-suited in situations in which it is important to reason about these, more detailed aspects.

\subsubsection{Consistency of specifications}
Although the approach is based on a uniform model of architectures, this does not ensure consistency of specifications developed with the approach.
Constraints are expressed in temporal logic formulas which can be complex and induce inconsistencies of specifications.

\subsubsection{Practical evaluation}
In this article we provided the theoretical foundations of our approach and demonstrated the concepts by means of a small example.
To demonstrate the power of the approach it should be applied to more real world specifications.

%% file: bgrw.tex
\section{Related Work}\label{sec:bgrw}
In this article we described an approach to specify constraints for dynamic architectures.
Thus, related work can be found in three different areas.
\begin{inparaenum}[(i)]
	\item specification of dynamic architectures,
	\item specification of architectural constraints, and
	\item specification of dynamic reconfiguration.
\end{inparaenum}
In the following we briefly discuss each of them in more detail.

\subsection{Specification of dynamic architectures}
Over the last decades, several so-called Architecture Description Languages (ADLs) emerged to support in the formal specification of architectures. Some of them also support the specification of dynamic aspects~\cite{Inverardi1995,Luckham1995,Magee1996,Allen1997,Allen1998,Oquendo2004,Dashofy2001,Garlan2003}.

While ADLs support in the formal specification of (dynamic) architectures, they were developed with the aim to specify individual architecture instances, rather than architecture constraints which requires more abstract specification techniques.

By providing a language to specify such constraints, we actually complement these approaches.
Our language can be used to specify architectural constraints and verify them against the concrete architectures specified in one of those languages.

\subsection{Specification of Architectural Constraints}
Attempts to formalize architectural styles and patterns required more abstract specification techniques and focused on the specification of architectural constraints, rather than concrete architectures.

Such constraints are either specified by a general specification language such as Z~\cite{Spivey1992} (as, for example, by Abowd \etal~\cite{Abowd1995}), algebraic specifications (as, for example, by Moriconi \etal~\cite{Moriconi1995} and Penix \etal~\cite{Penix1997}), graph grammars (as, for example, by Le~Métayer~\cite{LeMetayer1998}) or by the use of process algebras (as, for example, by Bernardo \etal~\cite{Bernardo2000}) or directly from architectural primitives (as, for example, by Mehta and Medvidovic~\cite{Mehta2003}).

While these approaches focus on the specification of architectural constraints rather than architectures, they do usually not allow for the specification of dynamic architectural constraints which is the focus of this work.

\subsection{Specification of dynamic reconfigurations}
Recently, some attempts were made to model dynamic reconfigurations in an abstract, language independent manner.

\subsubsection{Stateless reconfiguration}
The first approaches in this area focused on plain structural evolution.
Examples include the work of Le~M\'{e}tayer~\cite{LeMetayer1998}, Hirsch and Montanari~\cite{Hirsch2002}, and Mavridou \etal~\cite{Mavridou2015}.
While these approaches focus on the specification of constraints for dynamic architectures, similar as for ADLs, the relation of behavioral and structural aspects is not considered.

\subsubsection{State-full reconfiguration}
More recent approaches focus on the interrelation of behavior and configuration and are probably most closely related to our work.

One prominent example here is the work of Wermelinger \etal~\cite{Wermelinger2001,Wermelinger2002}.
The authors introduce a graph based architecture reconfiguration language based on the unity language~\cite{Chandy1989}.
The authors recognize that the interplay between topology and run-time behavior is important and so their language also allows for the specification of reconfiguration constraints formulated over run-time behavior.

Bruni \etal~\cite{Bruni2008} provide a graph-based approach to dynamic reconfiguration.
Reconfigurations are modeled as typed graph grammars.
Also here, the authors provide a mechanism to express architectural constraints.

Another approach in this area is the one of Batista \etal~\cite{Batista2005} where reconfiguration is specified as a set of reconfiguration rules.
In a reconfiguration rule, a predicate is specified to trigger a reconfiguration and the result of a reconfiguration is specified in terms of attach and detach operations.

While these works actually recognize the relationship of behavior and state in the specification of dynamic reconfigurations, they usually focus on the specification of concrete architecture instances, rather than architecture constraints.
The specification of constraints is only of secondary rule which is why they usually focus on the specification of static architecture constraints rather than dynamic once.

\subsubsection{Dynamic reconfiguration constraints}
Work in this area focus on the specification of dynamic reconfiguration constraints and is most closely related to our work.
However, to the best of our knowledge there exist only three approaches in this area.
In the following we are going to discuss each of them in more detail.

One of the first approaches in this area is from Dormoy \etal~\cite{Dormoy2010} who provide a temporal logic for dynamic reconfiguration called FTPL.
FTPL allows for the specification of component architecture evolution which is modeled by a transition system over architecture configurations and so-called evolution operations.
While FTPL is very promising, it focuses on the temporal aspect. Thus, we complement their work by providing explicit specification techniques for data-types, interfaces and architecture configurations as well.

Castro \etal~\cite{Castro2010} provides a categorical approach to model dynamic architecture reconfigurations in terms of institutions.
While the approach provides fundamental insights into the specification of dynamic architecture properties, their model remains implicitly in the categorical constructions. Thus, we complement their work by providing an explicit model of dynamic architecture properties.

Another example is the one of Fiadeiro and Lopes~\cite{Fiadeiro2013} who provide an approach similar to ours.
In their approach they use a rather abstract notion of state and configuration. While this makes the approach widely applicable, it has to be specialized for different domains. Indeed, our work can actually be seen as a specialization of their model by providing a concrete notion of state (as ports valuated by messages) and configuration (as connection of component ports).

%% file: conclusion.tex
\section{Conclusion}\label{sec:conc}
With this article we provide a formal approach for the specification of properties for dynamic architectures by means of \glsentryplural{gls:cns:arch}.

To this end, we first introduce an abstract model for dynamic architectures (Sect.~\ref{sec:model}).
Thereby, an architecture is modeled as a set of \glsentryplural{gls:cnf:trace} which are sequences over \glsentryplural{gls:aconf}.
An \glsentrytext{gls:aconf}, on the other hand, is a set of active components and connections between their ports.
A component consists of input, output, and local ports and a valuation of its ports with messages. Components are not allowed to change their interface over time, nor are they allowed to change the valuation of their local ports over time (since they act as a kind of configuration parameter).

In Sect.~\ref{sec:ds}-Sect.~\ref{sec:cnfd} we then describe the details of our approach to specify constraints for dynamic architectures:
\begin{itemize}
	\item First, a signature is specified defining the basic sorts, function, and predicate symbols.
	\item Then, \glsentryplural{gls:datatype} are specified in terms of algebraic specification techniques over the signature (Sect.~\ref{sec:ds}).
	\item Also interfaces are specified over the signature. An \glsentrytext{gls:ifspec} defines a set of interfaces which consist of an identifier and corresponding ports.
	An \glsentrytext{gls:ifspec} allows for the specification of component types by associating interfaces with invariants formulated as over its ports (Sect.~\ref{sec:ifspec}).
	\item Finally, architecture constraints are formulated over the \glsentrytext{gls:ifspec} by means of \glsentryplural{gls:cnfassert}, i.e., linear temporal formulas over the interfaces (Sect.~\ref{sec:cnfd}). To this end, activation and connection predicates are introduced to express activation and connection constraints, respectively.
\end{itemize}
For each specification technique a formal description of its syntax as well as its semantics in terms of our model introduced in Sect.~\ref{sec:model} was provided.

To support in the specification process the notion of \glsentrytext{gls:cnfdiag} was introduced in Sect.~\ref{sec:cnfd} as a graphical notation to specify interfaces and certain common activation and connection constraints. To this end, the notion of activation and connection annotations was introduced to easily express certain common activation and connection constraints.

The approach allows to specify constraints for dynamic architectures.
Therefore, it is well-suited for the specification of patterns for such architectures and enables formal analyses of such patterns as discussed in Sect.~\ref{sec:analysis}.
This is demonstrated by a running example in which we specify the \glsentrytext{gls:bb:p} architecture pattern and verify one of its key characteristic properties.
Therefore, with our work we complement existing approaches for the specification of dynamic architectures which focus on the specification of concrete architecture instances rather than properties.

Future work arises in three major directions:
\begin{inparaenum}[(i)]
	\item To support in the verification of specifications we are currently implementing of our approach for the interactive theorem prover Isabelle/HOL.
	\item On the theoretic side we are interested in a calculus of dynamic architectures to support the reasoning of dynamic architectures.
	\item Finally, we are working on an integration of our approach into current ADLs to support the specification of architecture constraints for those ADLs.
\end{inparaenum}

\section{Acknowledgements}
We would like to thank Veronika Bauer, Mario Gleirscher, Vasileios Koutsoumpas, Xiuna Zhu, and all the anonymous reviewers for their comments and helpful suggestions on earlier versions of the paper.

%% file: conventions.tex
\section{Appendix A: Conventions}\label{sec:conv}
In the following we introduce some conventions used throughout the paper.

\begin{sloppypar}
	\begin{definition}[Inverse function]
		For a function $f\colon A\to B$, we denote by $\inverse{f}\colon B\to\pset{A}$, the \emph{inverse} function of $f$.
	\end{definition}
\end{sloppypar}

\begin{sloppypar}
	\begin{definition}[Bijective function]
		With $A\leftrightarrow B$ we denote a \emph{bijective} function from $A$ to $B$.
	\end{definition}
\end{sloppypar}

\begin{sloppypar}
	\begin{definition}[Projection]
		For an n-tuple $C=(c_1, \dots, c_n)$, we denote by $\cp{c}{i}=c_i$ with $1\leq i \leq n$ the \emph{projection} to the $i$-th component of $C$.
	\end{definition}
\end{sloppypar}

\begin{sloppypar}
	\begin{definition}[Partial function]
		We denote by $\pFun{X}{Y}$, the set of \emph{partial} functions from a set $X$ to a set $Y$:
		\begin{align*}
			\pFun{X}{Y} & \defeq \{f\subseteq X{\times}Y\mid \forall x\in X,~y_1,y_2\in Y\colon\\
			&\quad((x,y_1)\in f\wedge(x,y_2)\in f)\implies y_1=y_2\}\enspace.
		\end{align*}
		For a partial function $f\colon \pFun{X}{Y}$, we denote by:
		\begin{itemize}
			\item $\domain{f}=\{x\in X \mid \exists y\in Y\colon f(x)=y\}$, the set of elements for which $f$ is defined, and with
			\item $\range{f}=\{y\in Y \mid \exists x\in X\colon f(x)=y\}$, the set of elements returned by $f$.
			\item By $f \rest {X'}\colon \pFun{X'}{Y}$, such that $\forall x\in X'\colon f \rest {X'}(x)=f(x)$, the restriction of $f$ to the set $X'\subseteq X$.
		\end{itemize}	
	\end{definition}
\end{sloppypar}

\begin{sloppypar}
	\begin{definition}[Cartesian power]
		For a set $S$ and number $n\in\NN$ we denote with $S^n$ the \emph{cartesian power} of $S$ to $n$:
		\begin{equation}
			\cProd{S}{n}=\{(s_1,\dots, s_n)\mid s_i\in S \text{ for all } i=1, \dots, n\}\enspace.
		\end{equation}
	\end{definition}
\end{sloppypar}

\begin{sloppypar}
	\begin{definition}[Function update]
		For a function $f\colon D\to R$ and elements $d\in D$ and $r\in R$, we denote with $\update{f}{d}{r}\colon D\to R$ a function which is equal to $f$ but maps $d$ to $r$:
		\begin{equation}
			\update{f}{d}{r}(x)\defeq
			\begin{cases}
				r & \text{if } x=d\enspace,\\
				f(x) & \text{else .}\\
			\end{cases}
		\end{equation}
		
		For a family of functions $F=(F_i)_{i\in I}$ with index set $I$, index $j\in I$, function $F_j\colon D\to R$, elements $d\in D$ and $r\in R$, we denote by $\fupdate{F}{j}{d}{r}$ a family where function $F_j$ is updated to $\update{F_j}{d}{r}$.
		\begin{equation}
			\fupdate{F}{j}{d}{r}_i\defeq
			\begin{cases}
				\update{F_i}{d}{r} & \text{if } i=j\enspace,\\
				F_i & \text{else .}\\
			\end{cases}
		\end{equation}
	\end{definition}
\end{sloppypar}

\begin{sloppypar}
	\begin{definition}[Function merge]
		Given functions $f\colon D_f\to R_f$ and $g\colon D_g\to R_g$, with disjoint $D$s and $R$s, we denote with $\fmerge fg\colon (D_f\cup D_g)\to (R_f\cup R_g)$ the \emph{merge} of the two functions:
		\begin{equation}
			(\fmerge fg) (x) \defeq
			\begin{cases}
				f(x) & \text{if } x\in D_f\enspace,\\
				g(x) & \text{else .}\\
			\end{cases}
		\end{equation}
	\end{definition}
\end{sloppypar}